\newenvironment{comment}[1]{}{}
\newcommand{\bco}{\begin{comment}}
\newcommand{\eco}{\end{comment}}

\newcommand{\co}{\footnotesize\textBlack}
\newcommand{\oc}{\normalsize\textBlack}
\newcommand{\ex}{\sf\textBlack}
\newcommand{\xe}{\rm\textBlack}

\newcommand{\sm}{mo\-d\`e\-le stan\-dard}
\newcommand{\cm}{cen\-tre de mas\-se}
\newcommand{\cmf}{re\-p\`e\-re du \cm}
\newcommand{\xs}{sec\-tion ef\-fi\-ca\-ce}
\newcommand{\xss}{sec\-tions ef\-fi\-ca\-ces}

\newcommand{\bc}{con\-di\-tion ini\-tia\-le}
\newcommand{\bcs}{con\-di\-tions ini\-tia\-les}
\newcommand{\lc}{col\-li\-sioneur li\-n\'eaire}

\newcommand{\irrep}{re\-pr\'e\-sen\-ta\-tion ir\-r\'e\-duc\-ti\-ble}
\newcommand{\irreps}{re\-pr\'e\-sen\-ta\-tions ir\-r\'ed\-uc\-ti\-bles}
\newcommand{\susy}{su\-per\-sy\-m\'e\-trie}
\newcommand{\susic}{su\-per\-sy\-m\'e\-tri\-que}
\newcommand{\cl}{niveau de confiance}
\newcommand{\cls}{niveaux de confiance}
\newcommand{\rg}{grou\-pe de re\-nor\-ma\-li\-sa\-tion}
\newcommand{\br}{rap\-port de bran\-che\-ment}
\newcommand{\brs}{rap\-ports de bran\-che\-ment}
\newcommand{\dof}{de\-gr\'e de li\-ber\-t\'e}
\newcommand{\dofs}{de\-gr\'es de li\-be\-rt\'e}
\newcommand{\qft}{th\'eorie quan\-ti\-que des champs}
\newcommand{\vev}{valeur moyenne dans le vide}
\newcommand{\vevs}{valeurs moyennes dans le vide}
\newcommand{\ssb}{bri\-sure spon\-ta\-n\'ee de sy\-m\'e\-trie}
\newcommand{\EW}{\'electro\-fai\-ble}
\newcommand{\EM}{\'electro\-ma\-gn\'e\-ti\-que}
\newcommand{\wma}{angle de m\'elange \EW}
\newcommand{\qn}{nom\-bre quan\-ti\-que}
\newcommand{\qns}{nom\-bres quan\-ti\-ques}
\newcommand{\gu}{gran\-de uni\-fi\-ca\-tion}
\newcommand{\gut}{th\'eorie de gran\-de uni\-fi\-ca\-tion}
\newcommand{\guts}{th\'eories de gran\-de uni\-fi\-ca\-tion}
\newcommand{\lrsm}{mo\-d\`e\-le sy\-m\'e\-tri\-que gauche-droite}

\newcommand{\etc}{{\em etc.}}
\newcommand{\ie}{c'est-\`a-di\-re}
\newcommand{\Ie}{C'est-\`a-di\-re}
\newcommand{\eg}{par exem\-ple}
\newcommand{\Eg}{Par exem\-ple}
\newcommand{\pe}{\mbox{$e^+e^-$}}
\newcommand{\ee}{\mbox{$e^-e^-$}}

\newcommand{\sw}{\mbox{$\sin\theta_w$}}
\newcommand{\cw}{\mbox{$\cos\theta_w$}}
\newcommand{\st}{\mbox{$\sin\theta_s$}}
\newcommand{\ct}{\mbox{$\cos\theta_s$}}
\newcommand{\swt}{\mbox{$\sin^2\theta_w$}}
\newcommand{\cwt}{\mbox{$\cos^2\theta_w$}}
\newcommand{\swf}{\mbox{$\sin^4\theta_w$}}

\def\beq{\begin{equation}}
\def\eeq{\end{equation}}
\def\bea{\begin{eqnarray}}
\def\eea{\end{eqnarray}}
\def\barr{\begin{array}}
\def\earr{\end{array}}

\def\lsim{\mathrel{\vcenter{\hbox{$<$}\nointerlineskip\hbox{$\sim$}}}}

\documentstyle[dina4,axodraw,colordvi,german,twoside,12pt]{book}
\begin{document}
\background{White}
\textBlack
\setcounter{tocdepth}{5}
\selectlanguage{\french}
\mdqoff
\parindent0em
\parskip2ex plus.5ex minus.5ex
\headsep10mm
\footskip15mm
\textheight220mm
\renewcommand{\baselinestretch}{1.05}
\addtolength{\evensidemargin}{-20pt}

\pagenumbering{roman}
\pagestyle{empty}

\title{\LARGE\bf Au-del\`a du Mod\`ele Standard}

\bigskip

\author{\Large Frank Cuypers}

\bigskip\bigskip

\date{mai 1997}
%\date{\today}

\maketitle

\clearpage
\thispagestyle{empty}
\setcounter{page}{2}
\mbox{}

\clearpage
\thispagestyle{empty}

\begin{flushright}
PSI Report 97-03\\
May 1997
%\today
\end{flushright}

\vfill
\begin{center}

{\LARGE\bf Au-del\`a du Mod\`ele Standard}

\bigskip\bigskip
{\Large Frank Cuypers}

\bigskip
{\tt cuypers@psi.ch}

{Paul Scherrer Institute,
CH-5232 Villigen PSI,
Suisse}

\bigskip\bigskip\bigskip
\begin{minipage}[t]{12cm}
Ces notes de cours
visent \`a d\'ecrire de mani\`ere p\'edagogique
certaines extensions populaires
du \sm\ des interactions fortes et \EW s.
Les sujets abord\'es
comprennent
le secteur de Higgs,
le mod\`ele sym\'etrique gauche-droit,
la grande unification
et la supersym\'etrie.
L'accent est mis sur les cons\'equences ph\'enom\'enologiques
et les m\'ethodes d'investigation.
\end{minipage}

\end{center}
\vfill

\clearpage
\thispagestyle{empty}
\mbox{}

\pagestyle{headings}

\chapter*{Pr\'eface}

Ce fascicule est inspir\'e
des th\`emes abord\'es 
lors d'un cours du troisi\`eme cycle
donn\'e \`a Nantes du 21 au 25 avril 1997.
Il s'adresse \`a des chercheurs
en formation doctorale ou au-del\`a,
qui ont au pr\'ealable eu l'occasion de se familiariser
avec le \sm\ des interactions fortes et \EW s,
et qui sont int\'eress\'es par l'\'etude de ses extensions.
La connaissance des techniques perturbatives de la \qft,
\ie\ en particulier le calcul de diagrammes de Feynman,
est un pr\'erequis utile.

Le cours peut \^etre lu \`a trois niveaux diff\'erents,
caract\'eris\'es par trois \'ecritures %et couleurs\footnote{
%Si la lecture est effectu\'ee \`a l'aide d'un moniteur en couleur.}
diff\'erentes:
\begin{itemize}
\item
Le texte principal est \'ecrit en caract\`eres \`a s\'erifs %noirs
de taille normale.
\item
\co
Certains commentaires,
bien que je les estime tr\`es int\'eressants,
ne sont pas n\'ecessaires \`a la compr\'ehension du reste du texte
et peuvent \^etre omis en premi\`ere lecture.
Ces commentaires sont \'ecrits en caract\`eres \`a s\'erifs %bleus
de plus petite taille.
\oc
\item
\ex
Certaines parties plus techniques
peuvent \^etre consid\'er\'ees comme des exercices.
Leur lecture est particuli\`erement recommand\'ee
\`a ceux qui sont int\'eress\'es 
\`a effectuer de la recherche
dans le domaine de la ph\'enom\'enologies des particules \'el\'ementaires.
Ces calculs sont \'ecrits en caract\`eres sans s\'erifs %verts
de taille normale.
\xe
\end{itemize}

Un certain nombre de r\'ef\'erences sont cit\'ees au fil du texte.
Elles sont divis\'ees en deux classes:
\begin{itemize}
\item
Les ouvrages de r\'ef\'erences
\`a caract\`ere p\'edagogique ou de compilation.
Leur lecture est recommand\'ee
\`a ceux qui sont int\'eress\'es 
de mani\`ere g\'en\'erale
par la physique des hautes \'energies.
Ces r\'ef\'erences sont \'ecrites en caract\`eres gras,
comme \eg~{\bf\cite{pdg}}.
\item
Les articles dont la lecture ne s'impose absolument pas,
surtout en premi\`ere lecture!
Je ne les mentionne que pour rendre justice \`a mes sources
et pour permettre \`a ceux 
qui d\'esireraient en savoir plus
sur un sujet bien particulier
de satisfaire leur curiosit\'e.
Ces r\'ef\'erences sont \'ecrites en caract\`eres normaux,
comme \eg~\cite{mmc}.
\end{itemize}

Je remercie vivement Jan Govaerts et Milan Locher
pour leurs nombreuses remarques et suggestions
quant au contenu de ce texte,
ainsi que pour leur m\'eticuleux d\'epistage
de mes innombrables fautes d'haurtocrafe.%orthographe.

Une copie de ce texte
peut \^etre obtenue 
en format Postscript
\`a l'adresse WWW 
{\tt http://www.hep.psi.ch/cuypers/bsm.ps}
et tous commentaires, suggestions ou corrections
envoy\'ees 
\`a l'adresse email
{\tt cuypers@psi.ch}
seront fort appr\'eci\'es.

\tableofcontents 
\listoffigures 
\listoftables 

\clearpage
\mbox{}
\clearpage

\pagenumbering{arabic}
\chapter{Introduction}

Si l'on veut aujourd'hui formuler une th\'eorie
des particules \'el\'ementaires et de leurs interactions,
aucun chemin ne passe plus \`a c\^ot\'es 
du \sm\ des interactions fortes et \EW s.
En d\'epit de ses d\'efauts de principe,
le \sm\ pr\'edit de mani\`ere si pr\'ecise
la foule immense de donn\'ees exp\'erimentales
que nous avons jusqu'\`a pr\'esent obtenues,
qu'il ne fait plus aucun doute 
qu'il constitue la description ad\'equate
des ph\'enom\`enes ayant lieu
jusqu'\`a une \'echelle d'\'energies 
de l'ordre de quelques centaines de GeV.

\co
De ce succ\`es fantastique du \sm\ 
a r\'esult\'e la triste situation
dans laquelle se trouve la physique de particules:
depuis voil\`a vingt ans,
aucune nouvelle d\'ecouverte n'est venu r\'evolutionner 
l'image que nous nous faisons du monde microscopique.
Les d\'ecouvertes des bosons interm\'ediaires $W^\pm,Z^0$
et du quark $t$,
ne sont d'un point de vue \'epist\'emologique 
que des anecdotes,
puisque l'existence de ces particules avait \'et\'e pr\'edite
bien auparavent.
Depuis la d\'ecouverte du quark $b$,
aucune exp\'erience n'est parvenue 
\`a mettre en \'evidence un ph\'enom\`ene impr\'evu!
\oc

\co
Nous vivons en ce moment une \'epoque malsaine de la physique des particules,
o\`u la th\'eorie est en avance sur l'exp\'erience,
o\`u les pr\'edictions du \sm\ 
dictent la marche \`a suivre.
Ceci n'est pas sans rappeler la situation de la fin du si\`ecle pass\'e,
avant la d\'ecouverte de la radioactivit\'e 
(Becquerel en 1896)
et de l'\'electron
(Thomson en 1997).
\`A cette \'epoque beaucoup consid\'eraient d\'ej\`a la physique
comme une science en d\'eclin:
tout avait \'et\'e d\'ecouvert
et il ne suffisait plus que de ``boucher les trous''.
On se souviendra qu'il fut conseill\'e \`a Max Planck
de ne pas entamer des \'etudes de physique 
pour cette raison!\dots
\oc

\co
Un si\`ecle pus tard,
les conditions aux limites sont analogues,
mais la r\'eaction de la communaut\'e scientifique 
est radicalement oppos\'ee.
Plus personne ne songerait \`a affirmer aujourd'hui
que notre description actuelle des ph\'enom\`enes physiques
constitue une explication compl\`ete de la r\'ealit\'e.
Bien au contraire,
l'absence de guide exp\'erimental
a permis et provoqu\'e
un impressionant foisonnement de conjectures
d\'ecrivant l'univers des distances inf\'erieures \`a l'atto-m\`etre
(1 am = $10^{-18}$ m).
Tous ces mod\`eles sont bas\'es sur le \sm,
tout comme la relativit\'e g\'en\'erale 
est bas\'ee sur la gravitation de Newton,
mais au-del\`a d'une \'energie d'environs 100 GeV
leurs pr\'edictions divergent du tout au tout.
\oc

Le but de ce cours
est de donner au lecteur une base de travail
pour l'\'etude de quelques unes
des nombreuses th\'eories
{\em au-del\`a du \sm}.
Le choix des sujets abord\'es
est bien s\^ur avant tout 
dict\'e par mon go\^ut personel.
Mais il y a aussi une certaine logique \`a la base.
Je commencerai par une description relativement d\'etaill\'ee
des propri\'et\'es du boson de Higgs du \sm\
et montrerai qu'il est difficile de modifier radicalement ce secteur.
J'examinerai ensuite le \lrsm\
qui \'elargit de mani\`ere tr\`es naturelle et \'el\'egante
la sym\'etrie de jauge du \sm,
tout en \'evitant les \'ecueils du secteur de Higgs.
La lacune principale de ce mod\`ele
est de n'\^etre pas unifi\'e,
le sujet du chapitre suivant
qui traite des \guts.
Ayant montr\'e que ces derni\`eres
souffrent de mani\`ere end\'emique
d'un probl\`eme majeur,
dit de {\em hi\'erarchie},
j'indiquerai finalement 
comment la supersym\'etrie peut palier \`a ce d\'efaut.
Finalement,
je compilerai une petite liste 
(bien incompl\`ete)
d'autres th\'eories et mod\`eles populaires,
sans toutefois entrer dans les d\'etails.

Les deux principaux outils d'exp\'erimentation
qui nous permettront 
de falsifier la plupart de ces extensions du \sm,
voire d'en s\'electionner une,
verrons le jour
au d\'ebut du si\`ecle prochain.
Il s'agit
du collisionneur hadronique LHC du CERN
et du \lc\ de leptons 
projet\'e 
aux \'Etats-Unis (NLC),
au Japon (JLC) 
et en Europe (TESLA).
Alors que le LHC est d'ores et d\'ej\`a en construction,
aucun des projets de \lc\ n'a jusqu'\`a pr\'esent \'et\'e approuv\'e
par les instances politiques.
Il y a peu de doutes,
toutefois,
qu'au moins un d'entre eux sera concr\'etis\'e sous peu,
probablement dans le cadre d'une coop\'eration intercontinentale.
J'appellerai cette machine le FLC,
pour {\em Future Linear Collider}.

\bco{
Je passerai sous silence 
les exp\'eriences
du TEVATRON au Fermilab
et du LEP2 au CERN.
Au niveau d'un cours d'introduction 
ces machines sont similaires respectivement au LHC et au FLC,
mais en moins performant.
}\eco

Il existe d'autres projets
tr\`es int\'eressants,
tel les collisionneurs de muons~\cite{mmc} ou de photons~\cite{plc},
mais leur nivau de maturit\'e
est encore beaucoup moins avanc\'e.
Je ne les mentionnerai donc qu'en passant.

\chapter[Le \sm]
{Le \sm\ des interactions fortes et \'electrofaibles}

Je suppose que les lignes g\'en\'erales du \sm\ 
vous sont famili\`eres.
Toutefois,
surtout afin de d\'efinir mes notations,
je r\'esume dans la table~\ref{tsm}
un petit br\'eviaire de sa faune.
Je conseille vivement 
\`a ceux qui seraient int\'eress\'es
\`a effectuer de la recherche 
dans la ph\'enom\'enologie du \sm\ et de ses extensions,
de se familiariser avec les propri\'et\'es 
de ces particules
en jetant un 
(voire de nombreux!)
coup d'\oe il 
dans une des compilation bisannuelles
du Particle Data Group~{\bf\cite{pdg}}

\setlength{\arraycolsep}{.5em}
\renewcommand{\arraystretch}{1.6}
\newcommand{\sdt}[2]{\raisebox{ 1.5ex}[-1.5ex]{\shortstack[l]{$#1$\\$#2$}}}
\newcommand{\sdl}[2]{\raisebox{-1.5ex}{\shortstack[l]{$#1$\\$#2$}}}
\newcommand{\sdr}[2]{\raisebox{-1.5ex}{\shortstack[r]{$#1$\\$#2$}}}
\newcommand{\sdc}[2]{\raisebox{-1.5ex}{\shortstack[c]{$#1$\\$#2$}}}
\begin{table}[htb]
$$
\begin{array}{||l|c||c|c|r@{\hspace{2em}}|r@{\hspace{1em}}r@{\hspace{1.5em}}||}
\hline\hline
\multicolumn{2}{||r||}{\sdr{\mbox{\large nombres}}{\mbox{\large quantiques}}}
& {SO(1,3)}
& {SU(3)_c}
& \multicolumn{1}{c|}{SU(2)_L}
& \multicolumn{1}{c}{U(1)_Y}
& \multicolumn{1}{c||}{U(1)_{\rm EM}}
\\
\multicolumn{2}{||l||}{\begin{picture}(0,0)\put(3,15){\line(3,-2){40}}\end{picture}}
& {\mbox{spin}}
& {\mbox{couleur}}
& \multicolumn{1}{c|}{\sdc{\mbox{projection}}{\mbox{d'isospin}}}
& \multicolumn{1}{c}{\sdc{\mbox{hyper-}}{\mbox{charge}} }
& \multicolumn{1}{c||}{\sdc{\mbox{charge}}{\mbox{\'electrique}}}
\\
\multicolumn{2}{||l||}{\mbox{\large particules}}
& {J}
& 
& \multicolumn{1}{c|}{T^3}
& \multicolumn{1}{c}{Q_Y}
& \multicolumn{1}{c||}{Q_{\rm EM}}
\\
\hline\hline
\mbox{\Red{boson de Higgs}}
& \Red{H} 
& \Red{0} 
& \Red{{\bf1}} 
& \Red{-1/2} 
& \Red{1/2} 
& \Red{0} 
\\
\hline\hline
\mbox{leptons droits}
& \ell_R 
& 
& 
& 0 
& -1
& -1 
\\
\cline{1-2}\cline{5-7}
& \nu_L 
& 
& {\bf1} 
& 1/2 
& 
& 0 
\\
\sdt{\mbox{leptons gauches}}{}
& \ell_L
& 
& 
& -1/2
& \sdt{-1/2}{}
& -1
\\
\cline{1-2}\cline{4-7}
\mbox{quarks up droits}
& u_R 
& 1/2 
& 
& 0 
& 2/3
& 2/3 
\\
\cline{1-2}\cline{5-7}
\mbox{quarks down droits}
& d_R 
& 
& 
& 0 
& -1/3
& -1/3 
\\
\cline{1-2}\cline{5-7}
& u_L 
& 
& \raisebox{2.5ex}[-2.5ex]{\bf3} 
& 1/2
& 
& 2/3
\\
\sdt{\mbox{quarks gauches}}{}
& d_L
& 
& 
& -1/2
& \sdt{1/6}{}
& -1/3
\\
\hline\hline
\mbox{photon}
& \gamma
& 
& 
& 0
& 0 
& 0 
\\
\cline{1-2}\cline{5-7}
& W^+
& 
& 
&  1
&  0
&  1  
\\
\sdl{\mbox{bosons de jauge}}{\mbox{\EW s}}
& Z^0
& 1
& \raisebox{2.5ex}[-2.5ex]{\bf1} 
& 0
& 0
& 0 
\\
& W^-
& 
& 
& -1
& 0
& -1 
\\
\cline{1-2}\cline{4-7}
\mbox{gluons}
& g
& 
& {\bf8} 
& 0
& 0
& 0 
\\
\hline\hline
\end{array}
$$
\caption{
Spectre des particules du \sm.
L'hypercharge $Y$ est d\'efinie de mani\`ere \`a ce que la charge \EM\ $Q$
soit donn\'ee par la relation
$Q=T^3+Y$.
}
\label{tsm}
\end{table}
\renewcommand{\arraystretch}{1}

Je n'ai pas indiqu\'e dans la table~\ref{tsm}
la triplication des g\'en\'erations de fermions,
car mise-\`a-part la l\'eg\`ere diff\'erence
entre les \'etats propres de jauge
et les \'etats propres de masse des quarks,
param\'etris\'ee en terme de la matrice Cabbibo-Kobayashi-Maskawa,
il s'agit-l\`a d'un cl\^onage trivial.
Dans le cadre de ce cours d'introduction
j'ignorerai le m\'elange qui en r\'esulte 
et traiterai les \'etats propres de jauge des fermions
comme des \'etats propres de masse.

Cette approximation ne se justifie \'evidemment pas
lorsqu'on traite des bosons de jauge eux-m\^emes.
Leur m\'elange est sp\'ecifi\'e par l'\wma\
et j'\'ecrirai leurs \'etats propres de jauge
en fonction de leurs \'etats propres de masse
de la mani\`ere suivante:

\beq
\label{mixing}
\left\{
\begin{array}{rcl}
W_1 & = & \displaystyle {1\over\sqrt{2}}~  \left( W^+ + W^- \right) \\\\
W_2 & = & \displaystyle {1\over i\sqrt{2}} \left( W^+ - W^- \right) 
\end{array}
\right.
\qquad\qquad
\left\{
\begin{array}{rcl}
B   & = & \cw~ A ~-~ \sw~ Z \\\\
W_3 & = & \sw~ A ~+~ \cw~ Z
\end{array}
\right.
\eeq

Les constantes de couplage $g$, $g'$ et $e$
des sym\'etries $SU(2)_L$, $U(1)_Y$ et \EM\
sont reli\'ees par les \'egalit\'es

\beq
\label{cc}
g \sw ~=~ g' \cw ~=~ e~
\eeq

et je d\'efinis la normalisation de l'hypercharge $Y$
de mani\`ere \`a ce que la charge \EM\ 
soit donn\'ee par 
$Q=T^3+Y$,
o\`u $T^3$ est la projection d'isospin.

Les masses des vecteurs de jauge
sont li\'ees par la relation

\beq
\label{rhosm}
\rho~
=~
{m_W^2 \over m_Z^2 \cwt}~
=~
1~
~.
\eeq

Bien que le \sm\ domine victorieusement
la sc\`ene exp\'erimentale
depuis pr\`es d'un quart de si\`ecle,
il n'y a aucun doute
qu'il ne constitue qu'une approximation
\`a ``basses'' \'energies
d'une r\'ealit\'e 
plus complexe,
plus sym\'etrique,
plus belle,
qui ne se manifestera
qu'\`a des \'energies plus \'elev\'ees.

En effet,
un grand nombre de questions restent encore sans r\'eponses,
telles que:

\begin{itemize}
\item
Quel est le m\'ecanisme
qui explique les nombreux ordres de grandeur
s\'eparant les masses des trois g\'en\'erations de fermions?
\item
Pourquoi, d'ailleurs,
n'y a-t-il que trois g\'en\'erations?
Ou bien y en aurait-il plus?
\item
Quelle est la raison pour laquelle
le potentiel de Higgs acquiert
sa singuli\`ere forme de 
{\em sombr\'ero}?
\item
Pourquoi y a-t-il trois constantes de couplage ind\'ependantes
($g$, $g'$ et $g_s$)?
N'oublions pas que la th\'eorie \EW\ 
n'est pas le moins du monde une th\'eorie unifi\'ee!
\item
Et m\^eme si l'on unifie les interactions \EW s et fortes,
nous verrons plus loin que cette unification ne peut s'effectuer
qu'\`a une \'echelle d'\'energies immens\'ement plus \'elev\'ee
($\sim10^{15}$ GeV)
que celle de la \ssb\ \EW\
($\sim10^{2}$ GeV)!
Quel m\'ecanisme physique est \`a l'origine de cette hi\'erarchie
ahurissante?
\item
Finalement,
nous ne disposons m\^eme pas d'une th\'eorie quantique de la gravitation.
Quelles que soient donc nos \'elucubrations les plus brillantes,
elles deviendront fatalement caduques
\`a l'\'echelle de la masse de Planck
($\sim10^{19}$ GeV).
\`A cette \'echelle,
la vision que nous avons de la structure de notre espace-temps
devra probablement \^etre modifi\'ee de mani\`ere dramatique.
Peut-\^etre les supercordes
apporteront-elles un jour une description satisfaisante.
Mais cela est d\'ej\`a une toute autre histoire\dots
\end{itemize}

Il existe un grand nombre de th\'eories
qui tentent
tant bien que mal 
de r\'esoudre ces \'enigmes.
Certaines de ces th\'eories 
sont plus s\'eduisantes ou populaires que d'autres,
mais toutes ont un point en commun:
elles contiennent le \sm\
comme th\'eorie effective \`a basse \'energie.

H\'el\`as,
comme nous le verrons bient\^ot,
les r\'eponses apport\'ees par ces th\'eories
aux questions pos\'ees plus haut
engendrent \`a chaque fois de nouvelles questions!
Nous sommes encore loin
du r\^eve platonicien
d'une th\'eorie unique expliquant le monde,
et seuls de nouvelles exp\'eriences
ont une chance de nous sortir de l'impasse
dans laquelle nous a men\'e le succ\`es du \sm.

\chapter{Le cha\^\i non manquant: le boson de Higgs}

Le boson de Higgs est la v\'eritable clef de vo\^ute du \sm\footnote{
D'apr\`es Sheldon Glashow,
c'est aussi son pot de chambre!
}.
C'est en effet gr\^ace au m\'ecanisme de Higgs
qu'\`a la fois les fermions
et les bosons de jauge \EW s 
acqui\`erent leur masses,
tout en pr\'eservant la renormalisabilit\'e de la th\'eorie.
Il reste n\'eanmoins
que jusqu'\`a pr\'esent
aucune exp\'erience n'a pu mettre en \'evidence 
le moindre indice convaincant
quant \`a son existence.
Il n'est donc pas \'etonnant
que la d\'ecouverte du boson de Higgs
est l'un des buts majeurs
de toutes les exp\'eriences pr\'evues prochainement
en physique des hautes \'energies.

Bien s\^ur, 
pour beaucoup de nos coll\`egues,
le boson Higgs fait partie int\'egrale du \sm\
et ne devrait donc pas faire l'objet 
d'un cours consacr\'e \`a la physique 
{\em au-del\`a du \sm}.
Mais comme il s'agit-l\`a d'un sujet si fondamental,
je d\'ecris dans les pages qui suivent 
quelques propri\'et\'es importantes du boson de Higgs
ainsi que les principales strat\'egies 
visant \`a sa d\'ecouverte.
J'examinerai aussi les puissantes contraintes exp\'erimentales
auxquelles doivent satisfaire d'\'eventuelles extensions
du secteur de Higgs.

Pour ceux qui sont int\'eress\'es \`a en savoir plus \`a ce sujet,
je recommande la lecture du
{\em Higgs Hunter's Guide}~{{\bf\cite{hhg}},
qui constitue aujourd'hui encore
la compilation la plus importante
de ce que nous (ne) savons (pas) du Higgs.

\section{Les d\'esint\'egrations du Higgs}

Les trois m\'ecanismes les plus importants
qui m\`enent \`a la d\'esint\'egration du Higgs
sont d\'ecrits par les diagrammes de Feynman 
de la figure~\ref{fhdecfeyn}.
Il s'agit des d\'esint\'egrations 
en une paire de fermions,
en une paire de bosons de jauge interm\'ediaires et
en une paire de gluons ou photons.
Pour la d\'esint\'egration en deux photons
il y a encore d'autres diagrammes 
impliquant les bosons de jauge $W$
que je n'ai pas repr\'esent\'es,
mais qui jouent un r\^ole crucial.

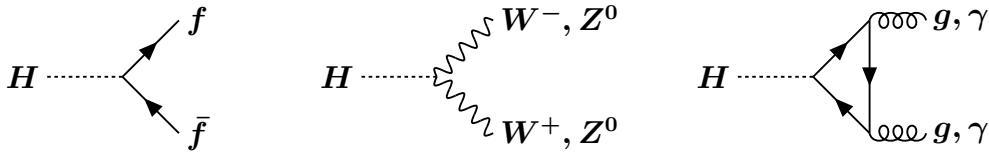
\begin{figure}[htb]
\unitlength.5mm
\SetScale{1.418}
\begin{boldmath}
\begin{center}
\begin{picture}(45,40)(0,0)
\DashLine(00,15)(20,15){1}
\ArrowLine(35,00)(20,15)
\ArrowLine(20,15)(35,30)
\Text(-2,15)[r]{$H$}
\Text(37,00)[l]{$\bar f$}
\Text(37,30)[l]{$f$}
\end{picture}
\hspace{4em}
\begin{picture}(45,40)(0,0)
\DashLine(00,15)(20,15){1}
\Photon(35,00)(20,15){2}{4.5}
\Photon(20,15)(35,30){2}{4.5}
\Text(-2,15)[r]{$H$}
\Text(37,00)[l]{$W^+,Z^0$}
\Text(37,30)[l]{$W^-,Z^0$}
\end{picture}
\hspace{6em}
\begin{picture}(60,40)(0,0)
\DashLine(00,15)(20,15){1}
\ArrowLine(20,15)(35,30)
\ArrowLine(35,30)(35,00)
\ArrowLine(35,00)(20,15)
\Gluon(35,00)(50,00){-2}{3.5}
\Gluon(35,30)(50,30){2}{3.5}
\Text(-2,15)[r]{$H$}
\Text(52,00)[l]{$g,\gamma$}
\Text(52,30)[l]{$g,\gamma$}
\end{picture}
\end{center}
\end{boldmath}
\caption{
Diagrammes de Feynman \`a l'ordre le plus bas
responsables de la d\'esint\'egration du boson de Higgs en une paire 
de fermions, de bosons interm\'ediaires et de gluons ou photons.
Il existe encore d'autres diagrammes du m\^eme ordre
pour la production de photons.
}
\label{fhdecfeyn}
\end{figure}

Ces m\'ecanismes de d\'esint\'egration du Higgs
d\'ependent de mani\`ere cruciale 
de la masse de ce dernier.
Il y a trois raisons pour cela:

\begin{enumerate}
\item
Primo,
plus le Higgs est lourd,
plus il a de modes permis cin\'ematiquement 
pour se d\'esint\'egrer.
\c Ca, c'est \'evident!
\item
Ce qui est moins trivial,
c'est que le couplage du Higgs \`a une autre particule
est proportionnel \`a la masse de celle-ci,
et par cons\'equent le Higgs aura tendance \`a se d\'esint\'egrer
pr\'ef\'erentiellement en des particules lourdes,
pour autant que cela lui est permis
par conservation de l'\'energie.
\item
Finalement,
et ceci est loin d'\^etre \'evident,
comme les composantes longitudinales des bosons de jauge massifs 
$W^\pm$ et $Z^0$
sont en essence 
les bosons de Goldstone
qui constituent les trois composantes du doublet scalaire complexe
absorb\'ees par les bosons de jauge dans le m\'ecanisme de Higgs,
leur couplage au boson de Higgs physique
est proportionnel \`a la masse de ce dernier
et non pas \`a la masse du boson de jauge.\footnote{
Si cela vous para\^\i t Chinois,
peut-\^etre le calcul explicite
de la d\'esint\'egration $H \to W^+W^-$,
que j'effectue plus loin,
rendra ce ph\'enom\`ene plus clair.
}
\end{enumerate}

\begin{figure}[htb]
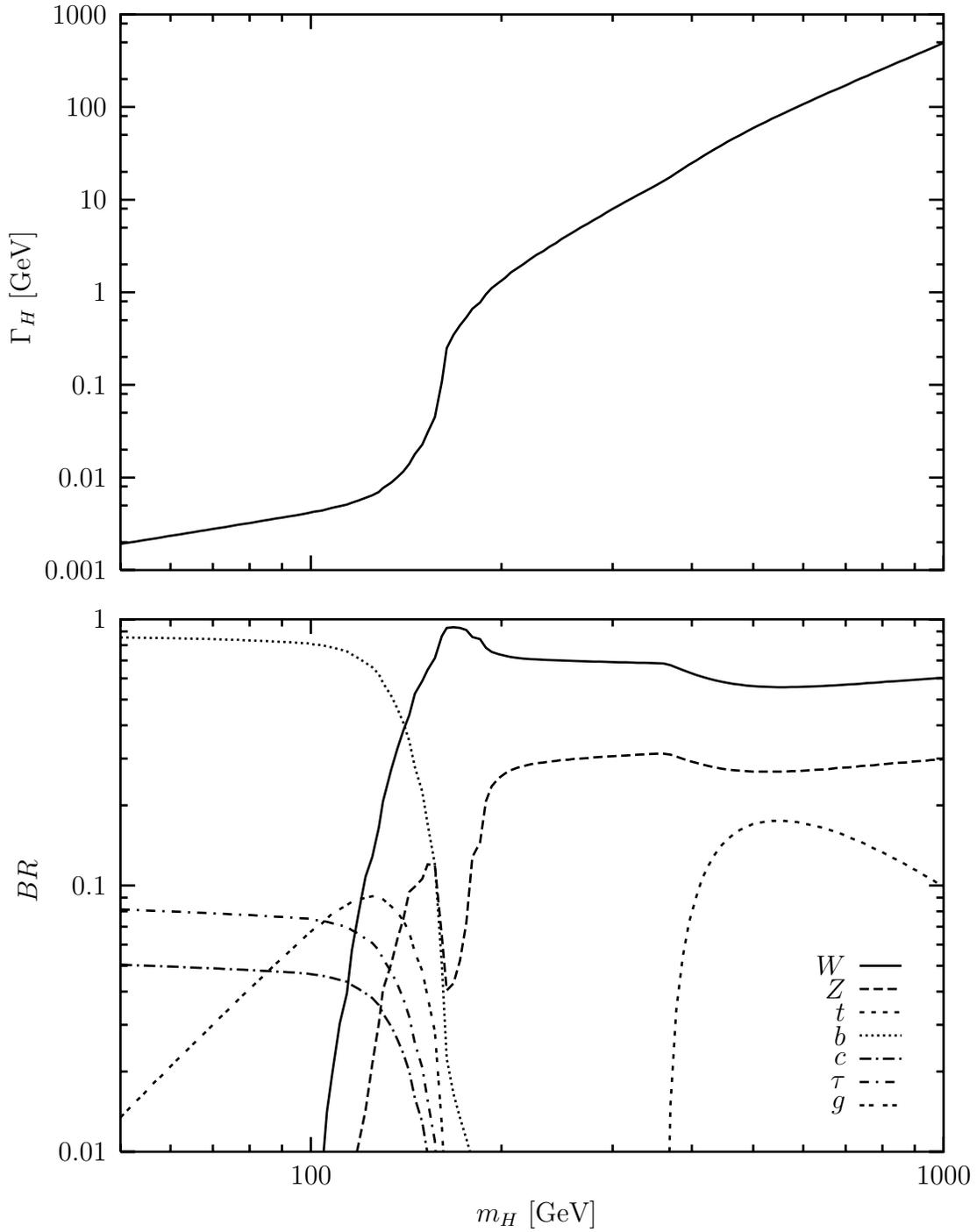

\centerline{\input{hdec.pstex}}
\centerline{\input{hbr.pstex}}
\bigskip
\caption{
Largeur de d\'esint\'egration du Higgs
et ses principaux \brs\
en fonction de sa masse.
}
\bigskip
\label{fhdec}
\end{figure}

\clearpage

J'ai repr\'esent\'e la largeur de d\'esint\'egration du Higgs
et ses principaux \brs\
dans la figure~\ref{fhdec}
en fonction de la masse du Higgs.
Puisque l'existence d'un Higgs de masse inf\'erieure \`a environ 60 GeV
a \'et\'e pr\'eclue par les exp\'eriences du LEP,
je ne consid\'ere donc ici 
que les masses sup\'erieures \`a 50 GeV.
Examinons bri\`evement 
les cons\'equences de ces pr\'edictions.

On voit clairement 
que l'on peut s\'eparer le domaine de masses
en deux r\'egions distinctes.
Un Higgs l\'eger\footnote{
Pour des raisons historiques,
on l'appelle aussi parfois
``de masse interm\'ediaire''.
},
d'une masse de moins de 140 GeV,
et un Higgs lourd 
d'une masse exc\'edant ces 140 GeV.
Il s'agit-l\`a du seuil 
au-del\`a duquel la d\'esint\'egration en une paire de $W$
devient dominante.

\subsection{Le Higgs l\'eger}

Un Higgs l\'eger
se d\'esint\`egre pr\'ef\'erentiellement en une paire de quarks $b$.
Les autres modes de d\'esint\'egration importants
sont ceux en une paire de lepton $\tau$ et de quarks $c$,
dont les \brs\ ne d\'epassent pas les 8\% et 5\%,
respectivement.

\ex
Calculons plus en d\'etails
cet important (et simple!) m\'ecanisme de d\'esint\'egration
en une paire de fermions.
En toute g\'en\'eralit\'e,
la largeur de d\'esint\'egration diff\'erentielle d'une particule
en deux particules $a$ et $b$
est donn\'ee par~\cite{iz}

\bea
d\Gamma
~=~
\underbrace%
{|{\cal M}|^2}_%
{\mbox{dynamique}}
&\times&
\underbrace%
{1 \over 2m_H}_%
{\mbox{flux}}
\\\nonumber
\\\nonumber
&\times&
\underbrace%
{{d^3p_{a} \over (2\pi)^3~2E_{a}}~{d^3p_{b} \over (2\pi)^3~2E_{b}}~(2\pi)^4~\delta^4(p_{H}-p_{a}-p_{b})}_%
{\mbox{espace des phases}}
~.
\eea

Dans le rep\`ere du \cm,
o\`u $p_H=m_H$,
ceci devient:

\beq
d\Gamma
~=~
|{\cal M}|^2
~\times~
{1 \over 2m_H}
~\times~
{k_{a} \over m_H}~{d\Omega_{a} \over 16\pi^2}
~,
\eeq

o\`u $k_{a} = \sqrt{(m_H/2)^2 - m_{a}^2}$
est le moment cin\'etique de la particule $a$.

Comme le Higgs est un scalaire,
il se d\'esint\`egre de mani\`ere isotropique
dans son rep\`re propre
et l'int\'egration sur l'angle solide du fermion
fournit un simple facteur $4\pi$.
La largeur partielle
est donc

\beq
\label{hw}
\Gamma~
=~
{\beta \over 16\pi m_H}~
|{\cal M}|^2~
~,
\eeq

o\`u $\beta$ est la vitesse des produits de d\'esint\'egration
(dont je suppose ici qu'ils ont la m\^eme masse)

\beq
\beta ~=~ \sqrt{1 - {4m_{a}^2 \over m_H^2}}~
~.
\eeq

Reste maintenant \`a calculer le carr\'e de la norme
de l'\'el\'ement de matrice $|{\cal M}|^2$
pour le cas o\`u $a$ et $b$ sont des quarks $b$
(ou tout autre fermion).
%et puis le cas o\`u il s'agit de bosons $W$
%(ou tout autre vecteur).
En appliquant au premier graphe de la figure~\ref{fhdecfeyn}
les r\`egles de Feynman
avec les couplages appropri\'es du Higgs,
on obtient pour l'\'el\'ement de matrice

\beq
{\cal M} 
~=~
{e m_b \over 2\sw m_W}~
 \bar u_{b} v_{\bar b} ~
\eeq

et son hermitien conjugu\'e

\beq
{\cal M}^{\dag}
~=~
{e m_b \over 2\sw m_W}~
 \bar v_{\bar b} u_{b} ~
~,
\eeq

o\`u $e$ est la charge de l'\'electron,
\sw\ est le sinus de l'\wma\
et $m_W$ et $m_b$ sont les masses du boson $W$ et du quark $b$.
En sommant sur les polarisations de l'\'etat final
on obtient pour le carr\'e de la norme
de l'\'el\'ement de matrice

\beq
|{\cal M}|^2~
=~
N_c~
\left({e m_b \over 2\sw m_W}\right)^2~
\underbrace{tr[ (p_{b}+m_b) (p_{\bar b}-m_b) ]}_%
{\displaystyle
4~( p_{b} \cdot p_{\bar b} - m_b^2) ~=~ 2~( m_H^2 - 4m_b^2 )
}~
~,
\eeq

o\`u $N_c$ est le nombre de couleurs port\'ees par le quark,
\ie\ $N_c=3$.
(S'il s'\'etait agit d'un lepton
on aurait bien s\^ur $N_c=1$.)
Si l'on exprime la charge de l'\'electron $e$
en fonction de la constante de structure fine $\alpha = e^2 / 4\pi$,
la largeur (\ref{hw}) devient

\beq
\Gamma_{H \to b\bar b}~
=~
N_c~
{\alpha \over 8\swt}~
\left( {m_b \over m_W} \right)^2~
\beta^3~
m_H~
\qquad\qquad
\beta = \sqrt{1 - {4m_{b}^2 \over m_H^2}}~
~.
\eeq

Comme il se doit,
cette largeur est bien proportionnelle au carr\'e de la masses du quark.
\xe

La d\'esint\'egration en une paire de gluons
est un autre mode particuli\`erement int\'eressant.
En effet,
contrairement aux autres principaux m\'ecanismes,
il s'op\`ere au travers d'une boucle,
comme celle repr\'esent\'ee dans la figure~\ref{fhdecfeyn}.
C'est le quark $t$ qui domine largement 
la contribution \`a cette boucle
en raison de sa grande masse,
qui gouverne son couplage au Higgs.
En fait, 
dans ce processus il y a une compensation partielle
des effets de propagateurs
(qui vont dans le sens d'une diminution pour des masses croissantes)
et des effets de couplage
(qui vont dans le sens contraire):
en fin de compte 
les effets de couplage dominent.
Bien que ce mode de d\'esint\'egration 
n'ait pas beaucoup d'influence sur la largeur totale du Higgs
dans le cadre du \sm,
ce qui le rend si int\'eressant
est le fait 
que s'il existe d'autres quarks non observ\'es
encore plus lourds que le top,
ils se manifesteront \`a ``basse'' \'energie
par ce m\'ecanisme!
Il s'agit-l\`a d'un ph\'enom\`ene de 
{\em non-d\'ecouplage},
qui est intimement li\'e \`a la \ssb.

Quoique tr\`es rares,
les d\'esint\'egrations en une paire de photons
sont aussi d'une importance majeure.
Dans le cadre du \sm\
elles sont {\em grosso modo} supprim\'ees
d'un facteur $(\alpha/\alpha_s)^2\approx.01$
par rapport aux d\'esint\'egrations gluoniques.
Mais elles sont sensibles
\`a l'existence de toute particule massive{\footnote{
Qui, bien s\^ur,
acquiert sa masse par son couplage \`a ce Higgs en particulier.
}
charg\'ee,
et non seulement aux particules color\'ees.
Autre aspect important,
il existe en g\'en\'eral
peu de bruit de fond imittant
ce signal de d\'esint\'egration en deux photons de hautes \'energies.
Nous reviendrons \`a cette caract\'eristique
au sous-chapitre suivant,
quand je traiterai de la d\'etection de Higgs l\'egers
dans les collisions hadroniques.

\subsection{Le Higgs lourd}

D\'ej\`a en-de\c c\`a du seuil de production de $W$,
les d\'esint\'egrations en bosons de jauge massifs
r\'eels ({\em e.g.} $H \to W^+W^-$) 
ou virtuels ({\em e.g.} $H \to W^+W^{-*} \to W^+e^-\nu_e$)
dominent enti\`erement la sc\`ene
et d\'ecroissent dramatiquement la dur\'ee de vie du Higgs.
Au-del\`a d'une masse de 200 GeV
le Higgs se d\'esint\`egre deux fois plus souvent 
en une paire de $W$ 
qu'en une paire de $Z$.
Le \br\ en une paire de quarks $t$
n'exc\`ede jamais les 20\%.

\ex
Calculons plus en d\'etails
l'importante (et simple!) largeur de d\'esint\'egration
en une paire de vecteurs.
L'\'el\'ement de matrice $|{\cal M}|^2$
pour le cas o\`u $a$ et $b$ sont des bosons $W$
(ou tout autre vecteur)
est obtenu
en appliquant au deuxi\`eme graphe de la figure~\ref{fhdecfeyn}
les r\`egles de Feynman
avec les couplages appropri\'es du Higgs:

\beq
{\cal M} 
~=~
{e m_W \over \sw}~
\varepsilon_{W^+} \cdot \varepsilon_{W^-}~
~.
\eeq

Les coordonn\'ees des vecteurs de polarisation 
transverse ($L$ et $R$) et longitudinale ($l$) 
peuvent \^etre choisies comme suit:

\bea
\label{polvec}
&&
\varepsilon_{W^+}(L)~
\equiv~
{1\over\sqrt{2}}\left(\begin{array}{c}0\\1\\i\\0\end{array}\right)
\qquad\qquad
\varepsilon_{W^-}(L)~
\equiv~
{1\over\sqrt{2}}\left(\begin{array}{c}0\\1\\-i\\0\end{array}\right)
\\\nonumber\\\nonumber
&&
\varepsilon_{W^+}(R)~
\equiv~
{1\over\sqrt{2}}\left(\begin{array}{c}0\\1\\-i\\0\end{array}\right)
\qquad\qquad
\varepsilon_{W^-}(R)~
\equiv~
{1\over\sqrt{2}}\left(\begin{array}{c}0\\1\\i\\0\end{array}\right)
\\\nonumber\\\nonumber
&&
\varepsilon_{W^+}(l)~
\equiv~
{1\over m_W}\left(\begin{array}{c}k_W\\0\\0\\E_W\end{array}\right)
\qquad\qquad
\varepsilon_{W^-}(l)~
\equiv~
{1\over m_W}\left(\begin{array}{c}k_W\\0\\0\\-E_W\end{array}\right)
~,
\eea

o\`u $k_W$ et $E_W$ sont le moment cin\'etique et l'\'energie du $W$.
En sommant sur ces trois polarisations de l'\'etat final
on obtient pour le carr\'e de la norme
de l'\'el\'ement de matrice

\bea
|{\cal M}|^2
&=&
\left({e m_W \over \sw}\right)^2~
\times
\\\nonumber
&&\quad
\underbrace{
\left(~
  |\varepsilon_{W^+}(L) \cdot \varepsilon_{W^-}(L)|^2~
+ |\varepsilon_{W^+}(R) \cdot \varepsilon_{W^-}(R)|^2~
+ |\varepsilon_{W^+}(l) \cdot \varepsilon_{W^-}(l)|^2~
\right)
}_%
{\displaystyle
1^2 + 1^2 + (2E_W^2 - m_W^2)^2/m_W^4 
~=~ 
3 - {m_H^2 \over m_W^2} + {m_H^4 \over 4m_W^4}
}
\eea

o\`u j'ai utilis\'e le fait que dans le rep\`ere du \cm\ $E_W = m_H/2$.
Si l'on \'ecrit la charge de l'\'electron $e$
en fonction de la constante de structure fine $\alpha = e^2 / 4\pi$,
la largeur (\ref{hw}) devient

\bea
&&
\Gamma_{H \to W^+W^-}~
=~
{\alpha \over 64\swt}~
\left( {m_H \over m_W} \right)^2~
\beta~
(3 - 2\beta^2 + 3\beta^4)~
m_H~
\\\nonumber
&&\qquad\qquad\qquad\qquad\qquad\qquad\qquad\qquad
\beta = \sqrt{1 - {4m_{W}^2 \over m_H^2}}~
~.
\eea

Cette largeur n'est pas proportionnelle au carr\'e de la masse du $W$
comme on aurait pu s'y attendre na\"\i vement.
Cela est d\^u \`a la contibution du vecteur de polarisation longitudinale
$\varepsilon_{W}(l)$,
qui correspond au \dof\ du Higgs
que le $W$ a absorb\'e pour aqu\'erir sa masse.
Comme il s'agit donc en fait d'un Higgs ``d\'eguis\'e'',
le couplage de cette composante du $W$ au Higgs physique
est proportionnelle a la masse du Higgs et non celle du $W$.

Similairement,
en tenant compte du diff\'erent vertex $HZ^0Z^0$ 
et du facteur $1/2$ pour des particules identiques dans l'\'etat final,
on trouve pour 
la largeur partielle de d\'esint\'egration du Higgs en une paire de $Z$

\bea
&&
\Gamma_{H \to Z^0Z^0}~
=~
{\alpha \over 128\swt}~
\left( {m_H \over m_W} \right)^2~
\beta~
(3 - 2\beta^2 + 3\beta^4)~
m_H~
\\\nonumber
&&\qquad\qquad\qquad\qquad\qquad\qquad\qquad\qquad
\beta = \sqrt{1 - {4m_{Z}^2 \over m_H^2}}~
~.
\eea

Un Higgs tr\`es lourd
($m_H \gg m_{W,Z}$)
se d\'esint\`egrera donc deux fois plus souvent en une paire de $W$
qu'en une paire de $Z$,
comme on peut le v\'erifier
sur la figure~\ref{fhdec}.
\xe

%\co
Au-del\`a d'une masse de 1.4 TeV
on obtient m\^eme une largeur du Higgs exc\'edant sa masse
$\Gamma_H>m_H$.
\`A ce stade,
cela n'a plus beaucoup de sens
de traiter ce champ comme une vraie particule.
Comme en outre un Higgs aussi lourd
aurait des auto-interactions tr\`es fortes,
nous p\'en\'etrons-l\`a un domaine particuli\`erement sp\'eculatif
de la physique du Higgs,
o\`u les m\'ethodes perturbatives ne sont plus applicables.
%\oc

\section{Production du Higgs}

La d\'ecouverte de bosons de Higgs
n'est pas une t\^ache ais\'ee,
en raison du faible couplage du Higgs 
aux fermions l\'egers,
\ie\ les electrons ou les quarks $u$ et $d$
(confin\'es \`a l'int\'erieur du proton) 
qui constituent les faisceaux de nos acc\'el\'erateurs.
La production de Higgs
ne s'effectue donc principalement par l'interm\'ediaire
de particules lourdes,
et est donc \'etouff\'ee en cons\'equence.

Je me concentre ici uniquement sur les deux types de collisionneurs
susceptibles de produire de bosons de Higgs
dans l'avenir le plus proche,
\ie\ le LHC et le FLC. 
Comme nous le verrons,
ces deux machines sont compl\'ementaires
et on peut grossi\`erement r\'esumer la situation de la mani\`ere suivante:
le FLC permettra de d\'ecouvrir un Higgs l\'eger
et le LHC permettra de d\'ecouvrir un Higgs lourd.

\co
Notons n\'eanmoins qu'il existe deux projets d'acc\'el\'erateurs
particuli\`erement bien adapt\'es 
\`a la production et l'\'etude de bosons de Higgs:
un collisionneur de muon~\cite{mmc}
et un collisionneur de photon~\cite{plc}.
Ces deux types originaux de collisions
produiraient des Higgs en grande abondance
par les processus inverses
de ceux d\'ecrits respectivement 
par le premier et le dernier diagramme de Feynman 
de la figure~\ref{fhdecfeyn}.
\oc

\co
Le principe fondamental du collisionneur de muons
est bien s\^ur de tout faire tr\`es vite!
\Ie,
les produires,
les collecter,
les refroidir,
les acc\'el\'erer
et finalement les collisionner quelques centaines de fois
dans un anneau de stockage.
Tout un programme!
Heureusement,
la dur\'ee de vie du muon de quelques microsecondes
combin\'ee \`a la dilatation relativiste
en fait des Mathusalems 
aux \'echelles de temps courrantes d'un acc\'el\'erateur!
\oc

\co
Le collisionneur de photons
peut \^etre construit \`a partir d'un \lc\ d'\'electrons,
en \'eclairant les faisceaux d'\'electrons
peu avant leur points d'interaction
par un puissant laser.
Le rayonnement Compton qui en r\'esulte
poss\`ede toutes les caract\'eristiques 
d'un faisceau de photons tr\`es d\'ecent.
Facile?
En fait,
il y a beaucoup de probl\`emes techniques,
mais les experts sont d'accord pour affirmer 
que la r\'ealisation de cette exp\'erience est possible.
\oc

\subsection{Collisions hadroniques}

Le principal m\'ecanisme de production de Higgs
dans les collisions $pp$ ou $p\bar p$ 
est d\'ecrit par le premier graphe de Feynman 
de la figure~\ref{fhhadfeyn}.
Il s'agit de la {\em fusion de gluons}
$gg \to H$,
la r\'eaction inverse de la d\'esint\'egration en une paire de gluons.
Comme je l'avais remarqu\'e plus haut,
l'effet de la boucle fermionique
est amplifi\'e par l'importante masse du top.

\co
Ce n'est qu'en premi\`ere approximation
qu'un proton est form\'e de deux quarks $u$ et d'un quark $d$.
En fait,
outre ces {\em quarks de valence},
il est aussi form\'e d'une {\em mer} infinie
de paires de quark anti-quark
et de gluons.
Une description de cette mer
est inaccessible par nos m\'ethodes perturbatives,
mais son contenu peut \^etre explor\'e
dans les collisions lepton-proton,
comme \eg\ \`a l'anneau HERA du DESY.
Il en r\'esulte
que les \xss\ de diffusion hadroniques aux hautes \'energies
peuvent toujours se calculer \`a l'aide de la convolution
d'une \xs\ microscopique calcul\'ee en th\'eorie des perturbations
et d'une fonction de structure du proton.
Cette derni\`ere,
qui est mesur\'ee dans les collisions lepton-proton,
fournit l'information quant au contenu {\em partonique} du proton.
\Eg,
la \xs\ de production d'un Higgs par fusion de gluon
au d\'epart d'une collision proton anti-proton $p\bar p \to HX$,
o\`u le symbole $X$ 
repr\'esente les fragments qui subsistent du proton et de
l'anti-proton,
est donn\'ee par
\oc

\co
\beq
\label{sf}
\sigma(p\bar p \to HX)~
=~
\int\limits^1_0 dx_1~
\int\limits^1_0 dx_2~
f^g_p(x_1)~
f^g_{\bar p}(x_2)~
\sigma(gg \to H)
~,
\eeq
\oc

\co
o\`u $f^g_p(x)$
est la fonction de structure
qui donne la probabilit\'e de trouver un gluon 
dont la fraction de moment est $x$
dans un proton,
et $\sigma(gg \to H)$
est la \xs\ partonique.
\oc

\begin{figure}[htb]
\unitlength.5mm
\SetScale{1.418}
\begin{boldmath}
\begin{center}
\begin{picture}(80,40)(0,-5)
\Gluon(00,00)(20,00){-2}{3.5}
\Gluon(00,30)(20,30){2}{3.5}
\ArrowLine(40,15)(20,30)
\ArrowLine(20,30)(20,00)
\ArrowLine(20,00)(40,15)
\DashLine(40,15)(60,15){1}
\Text(28,15)[c]{$t$}
\Text(62,15)[l]{$\Red{H}$}
\Text(-2,00)[r]{$g$}
\Text(-2,30)[r]{$g$}
\end{picture}
\hspace{6em}
\begin{picture}(55,50)(0,0)
\ArrowLine(00,35)(20,35)
\ArrowLine(20,35)(40,40)
\ArrowLine(00,05)(20,05)
\ArrowLine(20,05)(40,00)
\Photon(20,35)(25,20){2}{3.5}
\Photon(20,05)(25,20){-2}{3.5}
\DashLine(25,20)(50,20){1}
\Text(-2,05)[r]{$q'$}
\Text(-2,35)[r]{$q$}
\Text(42,40)[l]{$q''$}
\Text(42,00)[l]{$q'''$}
\Text(52,20)[l]{$\Red{H}$}
\Text(20,12)[r]{$W$}
\Text(20,28)[r]{$W$}
\end{picture}
\end{center}
\end{boldmath}
\caption{
Diagrammes de Feynman d\'ecrivant la production de bosons de Higgs
par fusion de gluons et par fusion de $W$
dans des collisions $pp$ ou $p\bar p$.
}
\label{fhhadfeyn}
\end{figure}
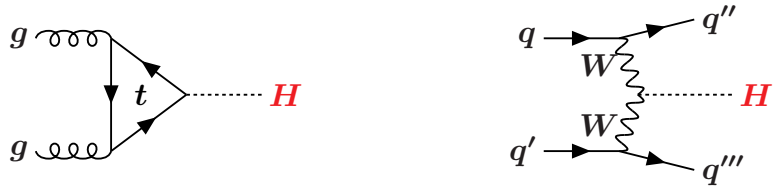

Si le Higgs est lourd
et se d\'esint\`egre donc pour la majeure partie 
en une paire de $W$ ou de $Z$,
les d\'esint\'egrations leptoniques de ces derniers
fournissent des signatures tr\`es significatives
qui souffrent de peu de bruit de fond.
Le LHC ne devrait donc avoir aucun probl\`eme 
\`a d\'ecouvrir un Higgs lourd.

\co
Toutefois,
si la masse du Higgs exc\`ede environ 800 GeV,
ses auto-interactions commencent \`a devenir fortes
et des effets non-perturbatifs peuvent se manifester.
En cons\'equence nos pr\'edictions concernant 
la production d'un Higgs aussi ob\`ese
sont tr\`es incertaines 
et d\'ependent fortement de mod\`eles.
Mais il est certain
que dans ce cas
{\em la fusion de $W$},
$W^+W^- \to H$,
\ie\ la r\'eaction inverse de la d\'esint\'egration en une paire de $W$,
joue un r\^ole dominant.
Cette r\'eaction est d\'ecrite par le deuxi\`eme graphe de Feynman 
de la figure~\ref{fhhadfeyn}.
\oc

Par contre,
si le Higgs est l\'eger
la situation est beaucoup moins claire pour le LHC.
En effet,
bien que les \xss\ de production de Higgs 
soient tr\`es \'elev\'ees,
leur d\'esint\`egration en une paire de quarks $b$
ne donne pas lieu \`a un signal particuli\`erement rare.
Bien au contraire,
le signal du Higgs sera enti\`erement submerg\'e 
par les innombrables \'ev\`enements \`a deux jets
qui r\'esultent des fusions de gluons 
ou des annihilations de paires quark-antiquark.

Le seul moyen d'explorer le domaine des Higgs l\'egers
est de se concentrer sur les d\'esint\'egrations rares,
en particulier la d\'esint\'egration en une paire de photons.
Le prix \`a payer est bien s\^ur
une r\'eduction dramatique dans le nombre d'\'ev\`enements Higgs.
Il n'est pas encore clair \`a l'heure actuelle
si les performances des d\'etecteurs 
(ATLAS et CMS)
du LHC
permettront de d\'ecouvrir un Higgs l\'eger.

Il est donc prudent d'affirmer 
que le LHC pourra certainement d\'ecouvrir le boson de Higgs
pr\'edit par le \sm,
si sa masse est comprise entre 
140 GeV et 800 GeV.

\subsection{Collisions \'electrons-positrons}

Les deux principaux m\'ecanismes de production de Higgs
dans les collisions \pe\ 
sont d\'ecrits par les graphes de Feynman de la figure~\ref{fhprodfeyn}.
Il s'agit de la {\em Higgsstrahlung}
$\pe\to ZH$
(\ie\ de la production d'un Higgs et d'un $Z$ \`a partir d'un $Z$ virtuel)
et de la {\em fusion de $W$}
$\pe\to\bar\nu\nu H$
(\ie\ la r\'eaction inverse de la d\'esint\'egration en une paire de $W$).

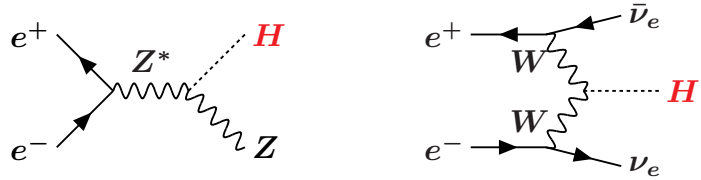
\begin{figure}[htb]
\unitlength.5mm
\SetScale{1.418}
\begin{boldmath}
\begin{center}
\begin{picture}(55,50)(0,0)
\ArrowLine(15,20)(00,35)
\ArrowLine(00,05)(15,20)
\Photon(15,20)(35,20){2}{4.5}
\Photon(35,20)(50,05){-2}{4.5}
\DashLine(35,20)(50,35){1}
\Text(-2,05)[r]{$e^-$}
\Text(-2,35)[r]{$e^+$}
\Text(52,05)[l]{$Z$}
\Text(52,35)[l]{$\Red{H}$}
\Text(25,25)[bc]{$Z^*$}
\end{picture}
\hspace{6em}
\begin{picture}(55,50)(0,0)
\ArrowLine(20,35)(00,35)
\ArrowLine(40,40)(20,35)
\ArrowLine(00,05)(20,05)
\ArrowLine(20,05)(40,00)
\Photon(20,35)(30,20){2}{3.5}
\Photon(20,05)(30,20){-2}{3.5}
\DashLine(30,20)(50,20){1}
\Text(-2,05)[r]{$e^-$}
\Text(-2,35)[r]{$e^+$}
\Text(42,40)[l]{$\bar\nu_e$}
\Text(42,00)[l]{$\nu_e$}
\Text(52,20)[l]{$\Red{H}$}
\Text(21,12)[r]{$W$}
\Text(21,28)[r]{$W$}
\end{picture}
\end{center}
\end{boldmath}
\caption{
Diagrammes de Feynman d\'ecrivant la production de bosons de Higgs
par Higgsstrahlung et par fusion de $W$
dans des collisions \pe.
}
\label{fhprodfeyn}
\end{figure}

Comme on peut le voir
sur le deuxi\`eme graphe de la figure~\ref{fhprod},
typiquement
la Higgsstrahlung domine la production de Higgs
aux basses \'energies
alors que la fusion de $W$ domine aux hautes \'energies.
\'Evidemment,
ce que sont des ``basses'' et des ``hautes'' \'energies
d\'epend de la masse du Higgs.
Il s'av\`ere
que l'\'energie pour laquelle
ces deux \xss\ sont \'egales
est donn\'ee approximativement par
$\sqrt{s} \approx 0.6 m_H + 400$ GeV.

\begin{figure}[htb]
\unitlength1mm
\makebox(0,60)[bl]{\includegraphics{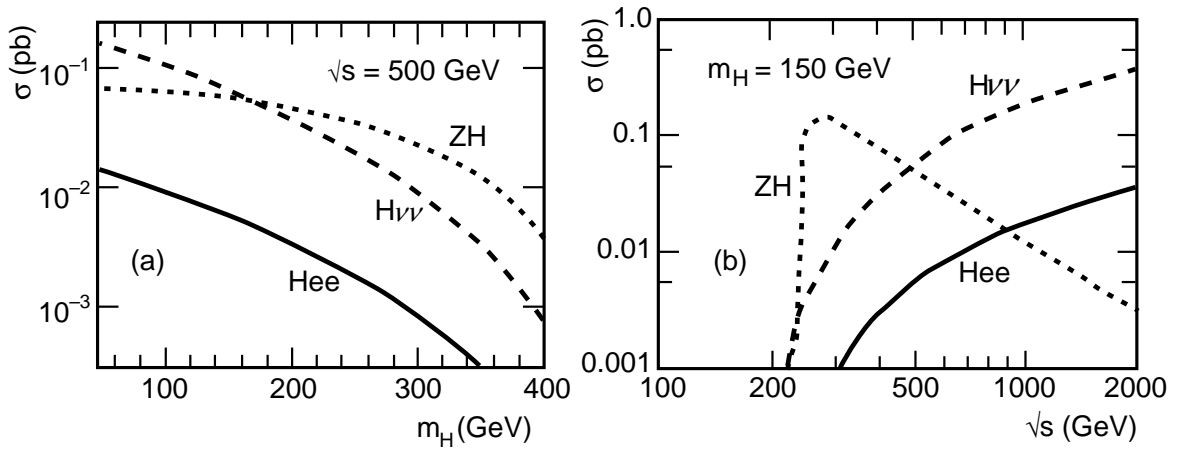}}
\caption{
Sections efficaces de production de bosons de Higgs au FLC
par Higgsstrahlung ($ZH$),
et par fusion de $W$ ($H\nu\nu$)
et par fusion de $Z$ ($Hee$).
Cette figure est adapt\'ee de la r\'ef\'erence~\protect\cite{nlc}.
}
\label{fhprod}
\end{figure}

\co
Les graphes de la figure~\ref{fhprod}
montrent aussi les \xss\ de production de Higgs
par fusion de $Z$,
$\pe\to\pe H$.
Elles sont environ dix fois moins importantes
que celles de la fusion de $W$
en raison du plus faible couplage des \'electrons au $Z$.
N\'eanmoins,
ce mode de production peut jouer un r\^ole important
pour d\'eterminer la nature du Higgs,
si nous avons affaire \`a un secteur de Higgs non-minimal.
\oc

\begin{figure}[htb]
\unitlength1mm
\makebox(0,50)[bl]{\includegraphics{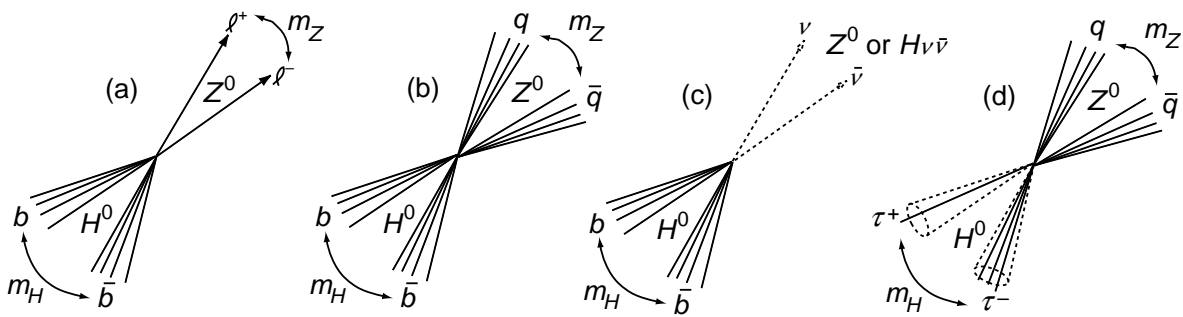}}
\caption{
Quelques unes des topologies possibles
pour des \'ev\`enements 
o\`u un boson de Higgs a \'et\'e produit au FLC.
Cette figure est adapt\'ee de la r\'ef\'erence~\protect\cite{nlc}.
}
\label{fhtopo}
\end{figure}

Puisque les luminosit\'es projet\'es au FLC
sont de l'ordre de 10--100 fb$^{-1}$,
ces \xss\ de l'ordre de 100 fb
permettraient d'obtenir un \'echantillon d'environ $10^3$ \`a $10^4$
\'ev\`enements contenant un Higgs.
Mais \`a quoi ressemblent ces \'ev\`enements?
Cela d\'epend fortement du mode de d\'esint\'egration du Higgs
et \'eventuellement de celui du $Z$.
Cela peut varier d'\'ev\`enements \`a 6 jets
(\eg\
dans le cas d'une Higgsstrahlung
o\`u le Higgs se d\'esint\`egre en 2 $Z$
et les trois $Z$ ainsi produits se d\'esint\`egrent en 2 quarks)
jusqu'\`a des \'ev\`enements \`a 2 muons et \'energie manquante
(\eg\
dans le cas d'une fusion de $W$
o\`u le Higgs se d\'esint\`egre en 2 leptons $\tau$
qui eux-m\^emes se d\'esint\`egrent chacun en un muon et deux neutrinos).
Vous l'avez devin\'e,
il y a beaucoup de possibilit\'es 
et j'en ai repr\'esent\'e un petit \'echantillon
dans la figure~\ref{fhtopo}.

\begin{figure}[htb]
\unitlength1mm
\makebox(0,80)[bl]{\includegraphics{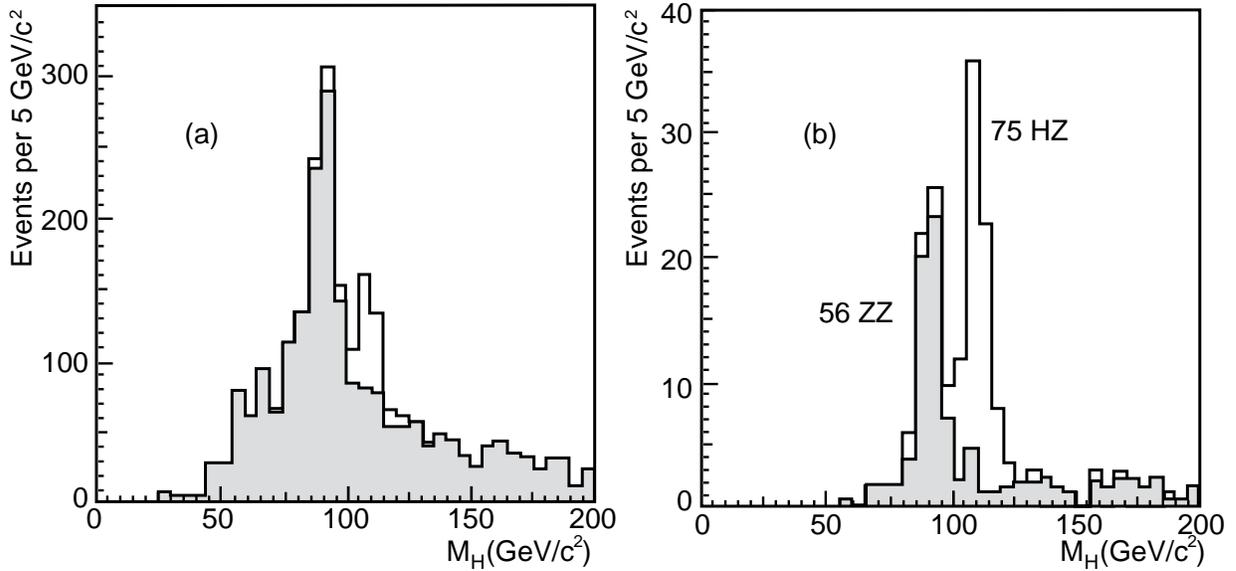}}
\caption{
Distribution de la masse invariante des jets 
issus de la production par Higgsstrahlung d'un boson de Higgs 
de 110 GeV,
et de sa d\'esint\'egration en une paire de quarks $b$.
La surface grise correspond aux bruits de fond
et l'aire transparente au signal d\^u au Higgs.
Les cas (a) et (b) correspondent respectivement aux analyses 
en l'absence et en pr\'esence de $b$-tagging.
Cette figure est adapt\'ee de la r\'ef\'erence~\protect\cite{nlc}.
}
\label{fhbtag}
\end{figure}

La question importante est maintenant la suivante:
quels sont les bruits de fond 
pour ces nombreuses signatures du Higgs?
Il s'av\`ere que si l'on tient compte 
des caract\'eristiques cin\'ematiques de ces signaux,
les bruits de fond peuvent en g\'en\'eral \^etre r\'eduits
\`a un niveau parfaitement inoffensif.
\Eg,
dans le cas o\`u les Higgs se d\'esint\`egrent en une paire de quarks $b$,
il faut que l'on reconnaisse les jets 
qui r\'esultent de l'hadronisation de ces quarks
comme originant en effet de quarks $b$,
et les masses invariantes de ces syst\`emes
doivent \^etre centr\'ees tr\`es pr\'ecis\'ement sur la masse du Higgs.
Cette technique d'\'etiquetage de jets
s'appelle $b$-{\em tagging} dans le jargon,
et pourra se faire au FLC avec au moins
70\% d'efficacit\'e pour une purit\'e de 99\%.

\begin{figure}[htb]
\unitlength1mm
\makebox(0,80)[bl]{\includegraphics{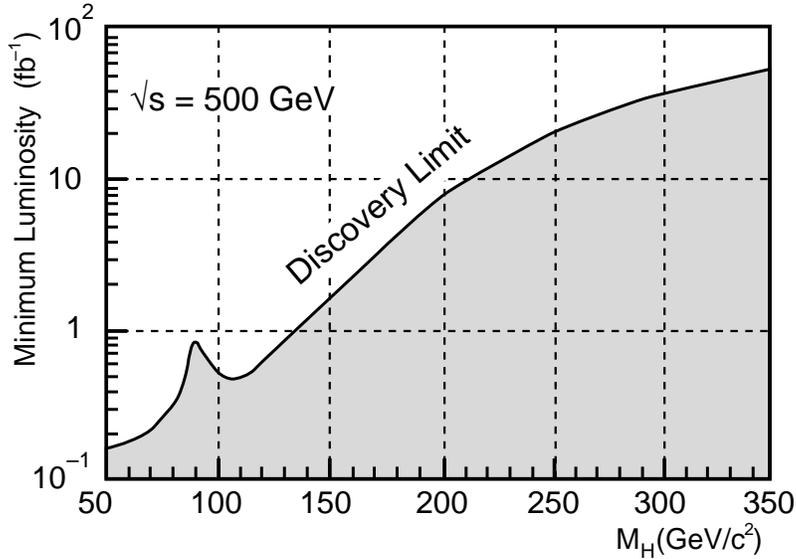}}
\caption{
Luminosit\'e n\'ecessaire au FLC
pour la d\'ecouverte du boson de Higgs,
en fonction de la masse de ce dernier.
Cette figure est adapt\'ee de la r\'ef\'erence~\protect\cite{nlc}.
}
\label{fhlum}
\end{figure}

Un exemple illustrant ce genre de proc\'edure 
est montr\'e dans la figure~\ref{fhbtag}.
Il s'agit d'une simulation d'\'ev\`enements \`a quatre jets
qui prend en compte \`a la fois 
le signal d'un Higgs de 110 GeV
et le bruit de fond.
Le signal provient d'une Higgsstrahlung
suivie de la d\'esint\'egration du Higgs en une paire de quarks $b$,
comme d\'ecrit par la figure~\ref{fhtopo}b.
Le bruit de fond peut \^etre d\^u \`a un grand nombre de processus,
tel que la production d'une paire de $Z$ \eg.
L'abscisse repr\'esente la masse invariante
de la paire de jets
candidate au titre de Higgs.
M\^eme dans le pire des cas,
o\`u la masse du Higgs est \'egale \`a la masse du $Z$,
la d\'ecouverte du Higgs est garantie.

Si l'on consid\`ere tous les m\'ecanismes 
de production et de d\'esint\'egration du Higgs
ainsi que leurs bruits de fond,
un Higgs l\'eger
peut \^etre d\'ecouvert au FLC
au bout d'une semaine
en assumant la luminosit\'e projet\'ee annuelle de 50 fb$^{-1}$ du FLC.
Comme le montre la figure~\ref{fhlum},
il faudra simplement attendre plus longtemps
pour qu'un nombre suffisant d'\'ev\`enements 
originant d'un Higgs plus lourd
soit accumul\'e
et obtienne une statistique suffisante
pour \^etre clairement diff\'erenci\'e du bruit de fond.
Le petit pic \`a $\sim$90 GeV
est d\^u \`a l'exc\`es de bruit de fond originant du $Z$.

En r\'esum\'e
on peut donc affirmer
que le FLC pourra certainement d\'ecouvrir le boson de Higgs
pr\'edit par le \sm,
si sa masse est inf\'erieure \`a 350 GeV.

\section{Extensions du secteur de Higgs}

Bien que nous n'ayons \`a l'heure actuelle
aucune \'evidence exp\'erimentale s\'erieuse
quant \`a l'existence du boson Higgs du \sm,
les extensions du secteur de Higgs
sont soumises \`a des contraintes exp\'erimentales draconiennes!
Ceci en raison des excellentes mesures 
des masses des bosons de jauge \EW s 
dont nous disposons,
et de la fa\c con subtile dont ces masses d\'ependent 
des d\'etails de la \ssb.

Il existe n\'eanmoins une classe d'extensions du secteur de Higgs
qui \'echappe \`a ces contraintes.
J'examinerai en particulier le
{\em mod\`ele \`a deux doublets de Higgs},
qui non seulement repr\'esente l'alternative la plus simple,
mais constitue aussi une pr\'ediction fondamentale de la \susy.

\subsection{Contraintes}

Afin d'analyser ce point plus en d\'etails
calculons ces masses
en pr\'esence d'un multiplet de Higgs $\phi_n$
de dimension $n$ arbitraire.
Pour ce faire,
nous n\'ecessitons la partie cin\'etique du lagrangien de Higgs
en fonction de la d\'eriv\'ee covariante $D_\mu$

\bea
\label{hl}
|D_\mu \phi_n|^2
&=&
\left|
\left( \partial_\mu - igT_n^aW^a_\mu - ig'Y_nB_\mu \right) \phi_n
\right|^2
\\\nonumber
&=&
\left|
\left[ 
\partial_\mu 
- ig \left( T^+_n W^-_\mu + T^-_n W^+_\mu \right)
- i \left( gT^3_n W^3_\mu + g'Y_n B_\mu \right)
\right] \phi_n
\right|^2~
~.
\eea

Pour extraire les masses des bosons de jauge 
il suffit de se concentrer sur les termes
contenant la \vev\ $v_n$ du champ scalaire.
Cette derni\`ere appartient n\'ecessairement 
\`a la composante \'electriquement neutre du multiplet,
qui satisfait la relation
$Q_n=T^3_n+Y_n=0$.
En sommant sur tous les multiplets de Higgs possibles
$n=1,2,3,\dots$,
on obtient donc

\bea
\label{hle}
\sum_n |D_\mu \phi_n|^2
~=~
\cdots
&+&
\sum_n |v_n|^2 g^2 
\underbrace{\left(T^+_nT^-_n + T^-_nT^+_n\right)}_%
{\displaystyle ({\bf T}_n)^2 - (T^3_n)^2 = T_n(T_n+1) - Y_n^2}
~W^+_\mu W^-_\mu
\\\nonumber\\\nonumber
&&
+~~
\sum_n |v_n|^2 Y_n^2
~\underbrace{\Big( g W^3_\mu - g' B_\mu \Big)}_{\makebox[0em]{$
(g\cw+g'\sw)
~Z_\mu
~+~
\underbrace{(g\sw-g'\cw)}_{\displaystyle0}
~A_\mu
$}}
\rule{0em}{2ex}^2
\\\nonumber\\\nonumber
~=~
\cdots
&+&
\underbrace{g^2 \sum_n |v_n|^2 \left[ T_n(T_n+1) - Y_n^2 \right]}_%
{\displaystyle m_W^2}
~W^+_\mu W^-_\mu
\hskip5em\mbox{}
\\\nonumber\\\nonumber
&&
+~~
\underbrace{{g^2\over\cwt} \sum_n |v_n|^2 Y_n^2}_%
{{1\over2} \displaystyle m_Z^2}
~Z_\mu Z_\mu~
~,
\eea

o\`u j'ai utilis\'e la condition de neutralit\'e $T^3_n=-Y_n$
et j'ai ins\'er\'e
les d\'efinitions des \'etats propres de masses
$Z$ et $\gamma$ (\ref{mixing}).
Afin d'\'eviter de g\'en\'erer d'ind\'esirables
masse du photon et m\'elange $Z$-photon,
j'ai d\^u imposer la relation famili\`ere (\ref{cc})
entre les constantes de couplage et l'\wma.
Les masses des bosons de jauge deviennent ainsi

\bea
m_\gamma^2 
&=&
0
\\\nonumber\\
\label{mw}
m_W^2 
&=&
g^2~
\sum_n
|v_n|^2~
\left[ T_n(T_n+1) - Y_n^2 \right]
\\
\label{mz}
m_Z^2 
&=&
{2 g^2 \over \cwt}~
\sum_n
|v_n|^2~
Y_n^2~
~,
\eea

ce qui permet d'introduire
l'important param\`etre ph\'enom\'enologique $\rho$

\beq
\label{rho}
\rho ~=~ {m_W^2 \over m_Z^2 \cwt} ~
=~
{ \sum |v_n|^2\left[T_n(T_n+1)-Y_n^2\right] \over 2\sum |v_n|^2Y_n^2 }~
~,
\eeq

qui quantifie le rapport de force
des courants neutres et charg\'es des interactions \EW s.
La combinaison d'un grand nombre de mesures
a permis de d\'eterminer la valeur de ce param\`etre
de fa\c con tr\`es pr\'ecise~\cite{ll}

\beq
\label{rhoexp}
\rho ~\approx~ 1.001 \pm 0.002~
~.
\eeq

Cette contrainte exp\'erimentale
est donc une puissante ``tueuse de mod\`eles''!

La solution non-triviale la plus simple 
consiste \`a n'avoir qu'un unique doublet de Higgs
d'hypercharge $Y=\pm1/2$.
Le choix du signe positif correspond au \sm\
et permet aussi de g\'en\'erer les masses des fermions
\`a travers des interactions de Yukawa.

Mais la contrainte (\ref{rhoexp}) est automatiquement satisfaite
par un nombre arbitraire de doublets
d'hypercharge $Y=\pm1/2$.
N\'eanmoins,
la somme des carr\'es des \vevs\  de ces doublets
est aussi s\'ev\`erement restreinte
par les \'equations (\ref{mw}) et (\ref{mz}),
qui font intervenir les masses bien connues
des bosons de jauge $W$ et $Z$.
Le mod\`ele qui consiste \`a consid\'erer deux doublets de Higgs
est immens\'ement populaire
\`a l'heure actuelle
et sa ph\'enom\'enologie fait l'objet d'un grand nombre d'\'etudes.
Le fait que cette extension du secteur de Higgs
soit pr\'edite par la \susy\
explique partiellement l'engouement dont il fait l'objet.

\co
En principe,
la contrainte exp\'erimentale (\ref{rhoexp})
permet \`a tout multiplet scalaire
satisfaisant $T(T+1)=3Y^2$
d'aqu\'erir une \vev.
Mais la premi\`ere solution rationnelle au-del\`a de $(T,Y) = (1/2,\pm1/2)$
implique un septuplet $(T,Y) = (3,\pm2)$.
Horrible!
\oc

\co
Rien n'emp\^eche non plus en principe
de contempler des valeurs irrationnelles pour l'hypercharge,
comme \eg\ $(1,\pm\sqrt{2/3})$.
Abstrus!
\oc

\co
On peut aussi imaginer avoir plusieurs repr\'esentations diff\'erentes
dont les \vevs\ s'accordent de mani\`ere suffisamment pr\'ecise
pour satisfaire la contrainte (\ref{rhoexp}).
Miraculeux!
\oc

\co
Finalement,
compte tenu des erreurs exp\'erimentales
sur la mesure de $\rho$ (\ref{rhoexp}),
des repr\'esentations arbitraires
peuvent \^etre consid\'er\'ees,
\`a condition que leurs \vevs\  soient suffisamment petites.
En g\'en\'eral,
ces exigences alambiqu\'ees
sont difficile \`a r\'ealiser de mani\`ere naturelle
et ne jouissent pas de la faveur populaire\dots
\oc

\subsection{Le mod\`ele \`a deux doublets de Higgs}
\label{thdm}

L'alternative au \sm\ la plus simple
qui respecte automatiquement la contrainte (\ref{rhoexp})
consiste \`a consid\'erer un deuxi\`eme doublet scalaire
d'hypercharge $Y=1/2$.
Nous sommes ainsi en pr\'esence de $2\times2\times2=8$ \dofs\
qui correspondent aux deux doublet scalaires complexes:

\beq
\label{thd}
\Phi_i ~\equiv~ {\Phi_i^+ \choose \Phi_i^0}
\qquad
i=1,2~
~.
\eeq

Sous des conditions assez g\'en\'erales,
le potentiel de Higgs
qui implique ces deux champs
provoque une \ssb\
et leur fournit les \vevs\ 

\beq
\label{thdvev}
\langle\Phi_i\rangle ~\equiv~ {0 \choose v_i}
\qquad
i=1,2~
~.
\eeq

En principe,
le potentiel de Higgs peut \^etre tel que
ces deux \vevs\ n'aient pas la m\^eme phase complexe.
Dans ce cas
la sym\'etrie $CP$ n'est pas respect\'ee
et il s'ensuit une ph\'enom\'enologie tr\`es int\'eressante.
En particulier,
il s'agit-l\`a d'une importante alternative
au m\'ecanisme de brisure de $CP$
de Kobayashi-Maskawa.
Toutefois,
par simplicit\'e,
je ne consid\`ere ici 
que la situation 
o\`u cette phase relative est nulle.
Les deux \vevs\ peuvent alors en toute g\'en\'eralit\'e
\^etre rendues r\'eelles
par une transformation unitaire des champs (\ref{thd}).
Ce choix se justifie d'autant plus
qu'il correspond au cas des deux doublets de Higgs
pr\'edits par la \susy.

En l'absence de violation de $CP$,
les parties r\'eelles et imaginaires 
des composantes neutres,
sont des \'etats propre de l'op\'erateur de $CP$
et ne se m\'elangent pas.
Les parties r\'eelles
sont de parit\'e $CP$ positive
et leurs \'etats propres de masse
sont donn\'es par la transformation unitaire

\bea
h
&=&
\sqrt{2}~\left[ \left(\Re{e}\Phi_1^0-v_1\right) \cos\alpha ~-~ \left(\Re{e}\Phi_2^0-v_2\right) \sin\alpha \right]
\nonumber\\\label{thcpe}\\\nonumber
H
&=&
\sqrt{2}~\left[ \left(\Re{e}\Phi_1^0-v_1\right) \sin\alpha ~+~ \left(\Re{e}\Phi_2^0-v_2\right) \cos\alpha \right]~
~,
\eea

o\`u $\alpha$ est un angle 
d\'etermin\'e par les param\`etres du potentiel de Higgs.
Les parties imaginaires
sont de parit\'e $CP$ n\'egative
et leurs \'etats propres de masse
sont donn\'es par la transformation unitaire

\bea
A
&=&
\sqrt{2}~\left( \Im{m}\Phi_1^0 \cos\beta ~-~ \Im{m}\Phi_2^0 \sin\beta \right)
\nonumber\\\label{thcpo}\\\nonumber
G
&=&
\sqrt{2}~\left( \Im{m}\Phi_1^0 \sin\beta ~+~ \Im{m}\Phi_2^0 \cos\beta \right)~
~,
\eea

o\`u $\beta$ est l'angle d\'efini par le rapport des \vevs\

\beq
\label{tanb}
\tan\beta ~=~ {v_2 \over v_1}~
~.
\eeq

Les \'etats propres de masse
des composantes charg\'es 
sont

\bea
H^\pm
&=&
\Phi_1^\pm \cos\beta ~-~ \Phi_2^\pm \sin\beta
\nonumber\\\label{thcha}\\\nonumber
G^\pm
&=&
\Phi_1^\pm \sin\beta ~+~ \Phi_2^\pm \cos\beta~
~,
\eea

Trois des huit \dofs\ scalaires
sont transf\'er\'es par le m\'ecanisme de Higgs
dans les composantes longitudinales des vecteurs de jauge $Z^0$ et $W^\pm$.
Ce sont les modes de bosons de Goldstone $G$ et $G^\pm$.
Les cinq autres \dofs\ scalaires qui subsistent 
sont physiques et peuvent \^etre observ\'es.
On compte ainsi 
deux scalaires neutres $h$ et $H$
de parit\'e $CP$ positive,
deux scalaires charg\'es $H^\pm$
et un scalaire neutre $A$
de parit\'e $CP$ n\'egative.
Par abus de langage,
ce dernier est souvent trait\'e de
{\em pseudo-scalaire}
dans la lit\'erature.

Bien que le mod\`ele \`a deux doublets de Higgs
soit la plus simple extension du secteur scalaire du \sm,
sa ph\'enom\'enologie est assez complexe
car elle d\'epend de six param\`etres,
les quatre masses et les deux angles de m\'elange.
Ceci est a contraster avec le secteur de Higgs du \sm,
qui est enti\`erement d\'etermin\'e par la masse et la \vev\ du scalaire.

\chapter{Le \lrsm}

Une des caract\'eristiques les plus singuli\`eres du \sm,
est le fait que les interactions \EW s
traitent tr\`es diff\'eremment 
les fermions gauches et les fermions droits.
Pour beaucoup
cette violation de sym\'etrie
rel\`eve d'un manque d'esth\'etique insupportable!
C'est aussi pourquoi
avant m\^eme l'av\`enement final du \sm,
Mohapatra et Pati propos\`erent
une extension tr\`es naturelle du groupe \EW~\cite{mp},
qui restore la sym\'etrie de parit\'e,
du moins aux hautes \'energies.
Pour des raisons \'evidentes,
on l'appele le \lrsm.
Il est bas\'e sur la sym\'etrie de jauge

\beq
\label{lrsgr}
G = SU(2)_L \otimes SU(2)_R \otimes U(1)_{B-L}
~.
\eeq

Les cons\'equences du facteur $SU(2)_R$ suppl\'ementaire
sont consid\'erables et profondes.
Le mod\`ele pr\'edit non seulement l'existence 
de nouveaux scalaires, fermions et vecteurs de jauge,
mais il g\'en\`ere aussi des masses de Majorana
pour les neutrinos,
avec tout ce que cela implique 
pour la conservation du nombre leptonique!

Le \lrsm\ rec\`ele donc une ph\'enom\'enologie tr\`es riche,
dont je n'analyserai ici
qu'un aspect tr\`es particulier,
la physique du $Z'$,
et ce de mani\`ere tr\`es incompl\`ete.
Je passerai ainsi sous silence
toutes les autres facettes fascinantes du mod\`ele,
en autres
celles des $W'$,
des Higgs doublement charg\'es
et du m\'ecanisme 
{\em see-saw}
de g\'en\'eration de masses de neutrino.

\section{La sym\'etrie de jauge et sa brisure spontan\'ee}

L'ajoute du facteur $SU(2)_R$ 
au groupe de jauge \EW\
implique l'existence de trois bosons de jauge suppl\'ementaires
correspondants au trois g\'en\'erateurs du groupe.
Par analogie avec les bosons \EW s qui nous sont familiers,
on les appelle commun\'ement $W'$ et $Z'$.
Nous verrons que le facteur ab\'elien $U(1)_{B-L}$
correspond \`a une sym\'etrie
qui conserve la diff\'erence entre 
le nombre baryonique $B$
et le nombre leptonique $L$.

On observe clairement,
jusqu'aux \'energies les plus \'elev\'ees 
que l'on ait pu atteindre jusqu'\`a ce jour,
que les fermions gauches se comportent diff\'eremment des fermions droits.
S'il y a une sym\'etrie gauche-droite,
elle doit donc \^etre bris\'ee
\`a une \'echelle d'\'energie non encore explor\'ee.
Pour obtenir cette brisure de sym\'etrie
on doit invoquer un nouveau secteur de Higgs,
qui fournit aux bosons de jauge $W'$ et $Z'$ une masse suffisamment \'elev\'ee,
pour rendre ces derniers invisibles aux basses \'energies.

La fa\c con la plus simple de briser 
la sym\'etrie gauche-droite du groupe (\ref{lrsgr})
vers la sym\'etrie \EW,
est de laisser en un premier temps la sym\'etrie $SU(2)_L$ intacte
comme suit

\beq
\label{lrsbr}
\hspace{-.5em}
\renewcommand{\arraystretch}{.2}
\begin{array}{ccccccccc}
SU(2)_L & \otimes & {SU(2)_R} & \otimes & {U(1)_{B-L}}
& \stackrel{{\Delta}}{\Black{\longrightarrow}} & 
SU(2)_L & \otimes & U(1)_Y
\\\\\\
\Black{\downarrow} & & \Black{\downarrow} & & \Black{\downarrow} & & \Black{\downarrow} & & \Black{\downarrow}
\\\\
W_{L\mu}^i & & W_{R\mu}^i & & C_\mu & & W_{L\mu}^i & & B_\mu
\\\\\\
g_L & & {g_R} & & {g'} & & g_L & & g_Y
\end{array}
\renewcommand{\arraystretch}{1}
\eeq

Les bosons de jauge associ\'es \`a chaque groupe
sont indiqu\'es sur la deuxi\`eme ligne de l'\'equation (\ref{lrsbr}).
Il s'agit bien des \'etats propres de jauge 
et non n\'ecessairement des \'etats propres de masse!
Les membres \'electriquement neutres de ces derniers
sont le photon $A$, le $Z$
et un $Z'$ lourd.
Les deux bases sont reli\'ees par les transformations unitaires

\beq
\label{m2gbasis}
\left\{
\renewcommand{\arraystretch}{1}
\begin{array}{lcr@{\:}c@{\:}r}
C     & = & {\ct} ~B & - & {\st} ~Z' \\\\
W^3_R & = & {\st} ~B & + & {\ct} ~Z' 
\end{array}
\renewcommand{\arraystretch}{1}
\right.
\qquad
\left\{
\renewcommand{\arraystretch}{1}
\begin{array}{lcr@{\:}c@{\:}r}
B     & = & {\cw} ~A & - & {\sw} ~Z \\\\
W^3_L & = & {\sw} ~A & + & {\cw} ~Z 
\end{array}
\renewcommand{\arraystretch}{1}
\right.
\eeq

ou bien

\beq
\left\{
\renewcommand{\arraystretch}{1}
\begin{array}{lccrcrcr}
W^3_L & = & & & & {\cw} ~Z & + & {\sw} ~A \\
W^3_R & = &   & {\ct} ~Z' & - & {\st} {\sw} ~Z & + & {\st} {\cw} ~A \\
C     & = & - & {\st} ~Z' & - & {\ct} {\sw} ~Z & + & {\ct} {\cw} ~A 
\end{array}
\renewcommand{\arraystretch}{1}
\right.
\eeq

o\`u $\theta_w$ est l'\wma\
et $\theta_s$ est un nouvel angle
qui d\'etermine le m\'elange entre les bosons de jauge 
$SU(2)_R$ and $U(1)_{B-L}$.

\begin{figure}[htb]
\unitlength1mm
\centerline{\makebox(0,100)[bl]{\includegraphics{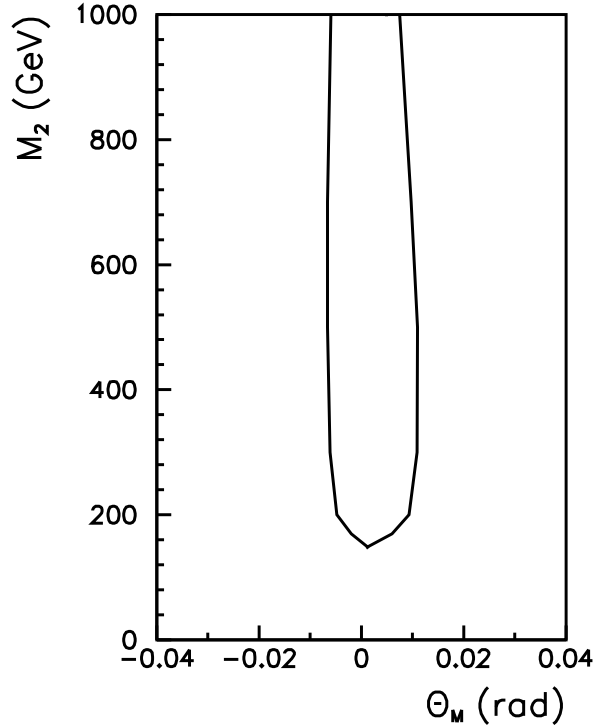}}}
\caption{
Valeurs typiques admissibles pour 
un \'eventuel angle de m\'elange $ZZ'$ ($\Theta_M$)
et la masse du boson de jauge neutre lourd ($M_2$).
Cette figure est adapt\'ee de la r\'ef\'erence~\protect\cite{lr2}.
}
\bigskip
\label{fzpmix}
\end{figure}

\co
Cette s\'eparation de la matrice de transformation unitaire $3\times3$
qui relie la base de jauge \`a la base de masse
en une combinaison de deux transformations $2\times2$
est dict\'ee par la cha\^\i ne de brisure de sym\'etrie (\ref{lrsbr}).
On peut en principe concevoir un m\'ecanisme plus compliqu\'e,
brisant simultan\'ement la sym\'etrie $SU(2)_L$.
Dans ce cas il faut introduire 
un troisi\`eme angle de m\'elange $ZZ'$
ainsi qu'une phase complexe qui brise la sym\'etrie $CP$.
Toutefois,
les r\'esultats obtenus sur la r\'esonance $Z^0$ au LEP1
restreignent s\'ev\`erement 
les valeurs admissibles de ces param\`etres~\cite{lr2},
comme l'illustre la figure~\ref{fzpmix}.
On peut donc affirmer sans crainte
que les \'equations (\ref{m2gbasis}) 
constituent une excellente premi\`ere approximation.
\oc

J'ai sp\'ecifi\'e dans la troisi\`eme ligne de l'\'equation (\ref{lrsbr})
les constantes de couplage associ\'ees \`a chaque groupe de jauge.
Comme nous identifions la sym\'etrie $SU(2)_L$
de l'\'equation (\ref{lrsgr}) 
avec la sym\'etrie $SU(2)_L$ du \sm,
sa constante de couplage est celle des interactions \EW s $g_L=g=e/\sw$,
o\`u $e$ est la charge de l'\'electron.
L'un des deux autres couplages,
$g_R$ ou $g'$,
peut \^etre choisi arbitrairement.
L'autre est alors d\'etermin\'e de mani\`ere unique,
comme nous le verrons plus loin.

\co
En principe,
si $g_R \ne g_L$
on ne peut pas vraiment qualifier ce mod\`ele de
{\em sym\'etrique gauche-droit}.
En pratique personne ne se soucie de cet abus de langage
et le choix $g_R = g_L$
n'est en fait absolument pas l'option pr\'ef\'er\'ee des th\'eoriciens.
En effet,
m\^eme si cette \'egalit\'e est impos\'ee
\`a une quelconque haute \'energie,
les constantes de couplages \'evolueront fatalement diff\'eremment
vers les basses \'energies.
Ceci est in\'eluctable,
puisque deux diff\'erentes \'echelles d'\'energies sont impliqu\'ees,
celle de la brisure de $SU(2)_R \otimes U(1)_{B-L}$
et celle de la brisure de $SU(2)_L \otimes U(1)_Y$.
\oc

Par analogie avec la brisure de la symetrie 
$SU(2)_L \otimes U(1)_Y \to U(1)_{\rm EM}$
qui donne l'expression 
$Q=T_L^3+Y$
pour l'op\'erateur de charge \EM,
la brisure de la symetrie 
$SU(2)_R \otimes U(1)_{B-L} \to U(1)_Y$
donne
$Y=T_R^3+(B-L)/2$
pour l'op\'erateur d'hypercharge\footnote{
Il faudrait en principe \'ecrire 
$U(1)_{(B-L)/2}$ au lieu de $U(1)_{B-L}$,
mais personne ne se donne cette peine.
}.
Dans le cadre du \lrsm,
la l'op\'erateur de charge \EM\ s'\'ecrit donc

\beq
\label{cem}
Q ~=~ T_L^3 ~+~ T_R^3 ~+~ {B-L\over2}~
~.
\eeq

La sym\'etrie $SU(2)_R\otimes U(1)_{B-L}$ est bris\'ee
dans l'\'equation (\ref{lrsbr})
\`a l'aide d'un champ de Higgs $\Delta$

\beq
\label{thdef}
\Delta({\bf1},{\bf3},2) \equiv
\left( 
\begin{array}{c} 
{\delta^{++}} \\ {\delta^+} \\ {\delta^0}
\end{array} 
\right)
\eeq

qui porte la charge $B-L=2$
et se transforme
comme un singulet de $SU(2)_L$
et un triplet de $SU(2)_R$.
L'ensemble de trois \qns\ 
qui sont indiqu\'es entre parenth\`eses
derri\`ere le $\Delta$
permettent une classification commode des particules.
Les deux premiers chiffres en caract\`eres gras
indiquent les dimension des repr\'esentations
$SU(2)_L$ et $SU(2)_R$ 
auxquelles elles appartiennent.
Le troisi\`eme \qn\ est la diff\'erence 
de leurs nombres baryoniques et leptoniques $B-L$.

Lorsque l'\'el\'ement neutre du triplet $\Delta$ (\ref{thdef})
acqui\`ere une \vev\ $v$

\beq
\label{thvev}
\langle \Delta \rangle \equiv
\left( \begin{array}{c} 0 \\ 0 \\ v \end{array} \right)
~.
\eeq

la partie cin\'etique du lagrangien de Higgs
prend la forme suivante
en fonction de la d\'eriv\'ee covariante $D_\mu$

\bea
\label{lrshlag}
\left| D_\mu \Delta \right|^2
&=&
\left|\left( 
\partial_\mu ~-~ ig_R T^a W^a_{R\mu} ~-~ ig' {B-L\over2} C_\mu 
\right) \Delta\right|^2
\\\nonumber\\\nonumber
&=&
\cdots ~+~ 
\Bigg|\Bigg[~
ig_R {1\over\sqrt{2}}
\left( \begin{array}{ccc} 0&1&0 \\ 1&0&1 \\ 0&1&0 \end{array} \right)
W^{1}_{R\mu}~
+~
ig_R {1\over\sqrt{2}}
\left( \begin{array}{ccc} 0&-i&0 \\ i&0&-i \\ 0&i&0 \end{array} \right)
W^{2}_{R\mu}~
\\\nonumber
&&\qquad\qquad
~+~
ig_R \left( \begin{array}{ccc} 1&0&0 \\ 0&0&0 \\ 0&0&-1 \end{array} \right)
W^{3}_{R\mu}
~+~
ig' \left( \begin{array}{ccc} 1&0&0 \\ 0&1&0 \\ 0&0&1 \end{array} \right)
C_\mu
~\Bigg]
\left( \begin{array}{c} 0 \\ 0 \\ v \end{array} \right)
\Bigg|^2
\\\nonumber
&=&
\cdots 
~+~ {1\over2} {|v|^2} g_R^2 ~~ W_\mu^+ W_\mu^-
~+~ {|v|^2} ~
\underbrace{\Big( g_R W_{R\mu}^3 - g'C_\mu \Big)}_{\makebox[0em]{$
\left( g_R \ct + g' \st \right)
~Z'_\mu
-
{\underbrace{{\left( g_R \st - g' \ct \right)}}_{\displaystyle0}}
~B_\mu
$}}
\rule{0em}{2ex}^2
\eea

Je n'ai conserv\'e ici que les termes contenant la \vev\ $v$,
parce que eux seuls sont responsables de la brisure de sym\'etrie,
le sujet qui m'int\'eresse ici.
Comme \`a ce stade la sym\'etrie $SU(2)_L \otimes U(1)_Y$ 
doit rester conserv\'ee,
la masse du boson $B$
(associ\'e \`a la sym\'etrie $U(1)_Y$)
doit \^etre nulle.
Ce fait implique une int\'eressante relation
entre l'angle de m\'elange $\theta_s$
et les constantes de couplage $g_R$ and $g'$.
La masse du $Z'$ est proportionnelle au carr\'e de la norme de la \vev\ $v$
et reste donc essentiellement un param\`etre libre.

\section{Les fermions}

Par d\'efinition du \lrsm,
les quarks et leptons
se transforment comme des doublets de $SU(2)_L$ and $SU(2)_R$.
En utilisant la m\^eme notation qu'auparavent 
pour le triplet de Higgs $\Delta$ 
dans l'\'equation (\ref{thdef}),
nous avons pour chaque g\'en\'eration

\beq
\label{lrsql}
\left\{
\begin{array}{l}
Q_L({\bf2},{\bf1},{1\over3}) \equiv 
\left( \begin{array}{c} u_L \\ d_L \end{array} \right)
\\\\
Q_R({\bf1},{\bf2},{1\over3}) \equiv 
\left( \begin{array}{c} u_R \\ d_R \end{array} \right)
\end{array}
\right.
\qquad\qquad
\left\{
\begin{array}{l}
L_L({\bf2},{\bf1},-1) \equiv 
\left( \begin{array}{c} \nu_L \\ \ell_L \end{array} \right)
\\\\
L_R({\bf1},{\bf2},-1) \equiv 
\left( \begin{array}{c} {\nu_R} \\ \ell_R \end{array} \right)
\end{array}
\right.
\eeq

Vous n'aurez pas manqu\'e de remarquer
que j'ai introduit un neutrino droit $\nu_R$.
Les cons\'equences de ce ``d\'etail''
sont extr\`emement importantes:
les neutrinos peuvent ainsi d\'evelopper une masse
et cela a maintenant un sens
de se demander si ce sont des fermions de Majorana ou de Dirac.

\subsection{Les neutrinos de Majorana}

Les neutrinos droits
peuvent acqu\'erir une masse de Majorana
au travers de la \vev\ du triplet de Higgs $\Delta$ (\ref{thvev}).
En effet,
on peut \'ecrire une interaction de Yukawa 
invariante de jauge
liant le doublet de leptons droits au triplet de Higgs 
sous la forme

\beq
\label{yuk}
{\cal L}_{\rm Y}~
=~
\lambda~
\bar L^c_R~
i\sigma_2\vec\sigma\cdot\vec\Delta~
L_R~
~,
\eeq

o\`u $L^c_R$ repr\'esente les fermions conjugu\'es de charge
et o\`u $\vec\sigma$ est un vecteur
form\'e des trois matrices de Pauli,
dont le produit scalaire avec le triplet de Higgs $\vec\Delta$
donne une matrice $2\times2$.

Le lagrangien (\ref{yuk}) est trivialement invariant
par transormations de jauge gauches,
puisque les trois champs sont des singulets de $SU(2)_L$.
Il est aussi invariant 
par transormations de jauge $U(1)_{B-L}$,
puisque le Higgs porte la charge $B-L=2$
alors que chaque lepton porte la charge $B-L=-1$.

L'invariance par transormations de jauge droite
est moins \'evidente.
Pour la prouver
il faut se rappeler
leurs formes locale
pour une rotation d\'efinies par le vecteur infinit\'esimal $\vec\theta$

\bea
L_R
&\longrightarrow&
\left( 1 - i g_R{1\over2} \vec\theta \cdot \vec\sigma \right)~
L_R
\nonumber\\\label{inft}
\bar L^c_R
&\longrightarrow&
\bar L^c_R~
\left( 1 - i g_R{1\over2} \vec\theta \cdot \vec\sigma^T \right)
\\\nonumber
\vec\Delta
&\longrightarrow&
\vec\Delta - g_R \vec\theta \times \vec\Delta~
~,
\eea

o\`u les matrices de Pauli $\vec\sigma$ 
Le couplage de Yukawa du lagrangien (\ref{yuk}) 
se transforme ainsi comme

\bea
\label{lt}
\makebox[0em][l]{$
\bar L^c_R~
\sigma_2\vec\sigma\cdot\vec\Delta~
L_R~
\longrightarrow
$}
\\\nonumber\\\nonumber
&&
\bar L^c_R~
\left( 1 - i g_R{1\over2} \vec\theta \cdot \vec\sigma^T \right)~
\sigma_2\vec\sigma\cdot
\left( \vec\Delta - g_R \vec\theta \times \vec\Delta \right)~
\left( 1 - i g_R{1\over2} \vec\theta \cdot \vec\sigma \right)~
L_R
\\\nonumber\\\nonumber
&=&
\bar L^c_R~
\Big[~
\sigma_2\vec\sigma\cdot\vec\Delta
~-~
i g_R{1\over2} 
\underbrace{ (\vec\theta \cdot \vec\sigma^T) \sigma_2}_%
{\makebox(0,0)[ct]{$\displaystyle - \sigma_2 (\vec\theta \cdot \vec\sigma)$}}
(\vec\sigma \cdot \vec\Delta)
~-~
i g_R{1\over2} 
\sigma_2 
(\vec\sigma \cdot \vec\Delta)
(\vec\theta \cdot \vec\sigma) 
\\\nonumber
&&
\hskip17em
~-~
g_R 
\sigma_2\vec\sigma\cdot
\left( \vec\theta \times \vec\Delta \right)
~+~
{\cal O}(\theta^2)
~\Big]~
L_R
\\\nonumber\\\nonumber
&=&
\bar L^c_R~
\Big[~
\sigma_2\vec\sigma\cdot\vec\Delta
~+~
i g_R{1\over2} \sigma_2 
\underbrace{
\left[ \vec\sigma \cdot \vec\theta , \vec\sigma \cdot \vec\Delta \right]
}_{\displaystyle
- i \vec\sigma \cdot \left( \vec\theta \times \vec\Delta \right)}
~-~
g_R 
\sigma_2\vec\sigma\cdot
\left( \vec\theta \times \vec\Delta \right)
~+~
{\cal O}(\theta^2)
~\Big]~
L_R
\\\nonumber
&=&
\bar L^c_R~
\sigma_2\vec\sigma\cdot\vec\Delta~
L_R
~+~
{\cal O}(\theta^2)
\eea

En \'ecrivant explicitement les composantes
des diff\'erents multiplets
dans le lagrangien (\ref{yuk}),
et en nous concentrant uniquement sur les termes 
qui impliquent la \vev\ du Higgs,
nous obtenons le lagrangien de masse du neutrino

\bea
\label{majn}
{\cal L}_{\rm Y}
&=&
\lambda~
\left(~ \bar\nu_R^c ~ \bar\ell_R^c ~\right)
\Bigg[
\underbrace{
  \left( {0\atop-1} {1\atop0} \right)
  \left( {0\atop1} {1\atop0} \right)
}_{\displaystyle \left( {1\atop0} {0\atop-1} \right) }~
{ \delta^{++} + \delta^0 \over \sqrt{2} }
\\\nonumber
&&
\hskip5em
+~
\underbrace{
  \left( {0\atop-1} {1\atop0} \right)
  \left( {0\atop i} {-i\atop0} \right)
}_{\displaystyle \left( {i\atop0} {0\atop i} \right) }~
{ \delta^{++} - \delta^0 \over i\sqrt{2} }
~+~
\underbrace{
  \left( {0\atop-1} {1\atop0} \right)
  \left( {1\atop0} {0\atop-1} \right)
}_{\displaystyle \left( {0\atop-1} {-1\atop0} \right) }~
\delta^{+}~
\Bigg]~
{\nu_R \choose \ell_R}
\\\nonumber
&=&
\dots~+~
\lambda~
{v\over\sqrt{2}}~
\left(~ \bar\nu_R^c ~ \bar\ell_R^c ~\right)
\left( {2\atop0} {0\atop0} \right)
{\nu_R \choose \ell_R}
\\\nonumber\\\nonumber
&=&
\dots~+~
\sqrt{2}\lambda v~
\bar\nu_R^c \nu_R~
~.
\eea

Le neutrino droit acquiert ainsi une masse de Majorana 
$m_{\nu_R} = \sqrt{2}\lambda v$,
qui doit \^etre tr\`es lourde
puisqu'elle est proportionelle \`a la \vev\ du Higgs
responsable de la \ssb\ gauche-droite.
Ce terme de masse est tr\`es particulier,
car il ne conserve aucun nombre quantique additif.
En particulier,
puisque la charge \EM\ n'est pas conserv\'ee,
les particules charg\'ees
ne peuvent que d\'eveloper une masse de Dirac
du type
$m(\bar\psi_L\psi_R+\bar\psi_R\psi_L)$.
Par contre,
les fermions neutres ne sont pas soumis \`a cette contrainte.

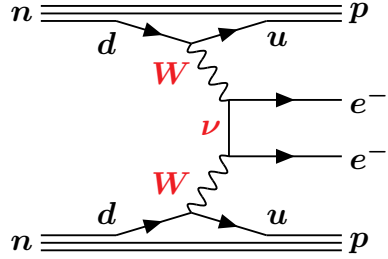
\begin{figure}[htb]
\unitlength.5mm
\SetScale{1.418}
\begin{boldmath}
\begin{center}
\begin{picture}(80,70)(0,0)
\Line(00,00)(80,00)
\Line(00,02)(80,02)
\Line(00,04)(20,04)
\ArrowLine(20,04)(40,10)
\ArrowLine(40,10)(60,4)
\Line(60,04)(80,04)
\Photon(40,10)(50,25){2}{3.5}
\Line(50,25)(50,40)
\ArrowLine(50,25)(80,25)
\ArrowLine(50,40)(80,40)
\Photon(50,40)(40,55){2}{3.5}
\Line(60,61)(80,61)
\ArrowLine(40,55)(60,61)
\ArrowLine(20,61)(40,55)
\Line(00,61)(20,61)
\Line(00,63)(80,63)
\Line(00,65)(80,65)
\Text(-2,02)[r]{$n$}
\Text(-2,63)[r]{$n$}
\Text(82,02)[l]{$p$}
\Text(82,63)[l]{$p$}
\Text(82,25)[l]{$e^-$}
\Text(82,40)[l]{$e^-$}
\Text(20,07)[br]{$d$}
\Text(20,58)[tr]{$d$}
\Text(60,07)[bl]{$u$}
\Text(60,58)[tl]{$u$}
\Text(48,32.5)[r]{$\Red{\nu}$}
\Text(40,15)[br]{$\Red{W}$}
\Text(40,50)[tr]{$\Red{W}$}
\end{picture}
\end{center}
\end{boldmath}
\caption{
Diagramme de Feynman 
responsable de la double d\'esint\'egration $\beta$ sans neutrino.
}
\label{nldbd}
\end{figure}

Le terme de masse de Majorana 
permet toutes sortes de processus
qui ne conservent pas le nombre leptonique,
tel la fameuse
{\em double d\'esint\'egration $\beta$ sans neutrino}
d\'ecrite microscopiquement par le diagramme de Feynman 
de la figure~\ref{nldbd}.
Il permet aussi des oscillations
d'une saveur de neutrino \`a une autre.
Toutes ces implications sont passionantes, 
aussi bien du point th\'eorique que ph\'enom\'enologique,
mais je ne le d\'eveloperai pas plus ici,
mais me concentrerai \`a pr\'esent sur certaines cons\'equences
de l'existence du $Z'$.

\subsection{Les couplages des fermions}

Les couplages des fermions (\ref{lrsql})
aux bosons de jauge
s'obtiennent en \'ecrivant explicitement 
la partie cin\'etique du lagrangien des fermions
en fonction de la d\'eriv\'ee covariante $D_\mu$.
Il est clair que
les couplages aux $W'$ 
sont donn\'es 
par les interactions chirales
$g_R P_R$.
Les couplages aux $Z'$ 
sont d\'etermin\'es par la partie diagonale
des interactions avec les bosons de jauge:

\begin{eqnarray}
\bar\Psi \gamma^\mu D_\mu \Psi
&=&
\bar\Psi \gamma^\mu
\left[
\partial_\mu 
-
ig_L {1\over2}\sigma^a W^a_{L\mu} P_L
-
ig_R {1\over2}\sigma^a W^a_{R\mu} P_R
-
ig' {B-L\over2}
\right]
\Psi
\\\nonumber\\\nonumber&&
\hspace{-6em}=~
\cdots~ - ~
i ~(\bar U ~ \bar D) \gamma^\mu
\left[
g_L {1\over2} \left( \begin{array}{cc} 1&0\\0&-1 \end{array} \right) 
W^3_{L\mu} P_L
+
g_R {1\over2} \left( \begin{array}{cc} 1&0\\0&-1 \end{array} \right) 
W^3_{R\mu} P_R
\right.
\\\nonumber&&\qquad\qquad\qquad\qquad\qquad\qquad\qquad\quad
\left.
+
g' {B-L\over2} \left( \begin{array}{cc} 1&0\\0&1 \end{array} \right) 
C_\mu
\right]
\left( \begin{array}{c} U \\ D \end{array} \right)
\end{eqnarray}

o\`u $P_{R,L} = (1\pm\gamma_5)/2$ sont les op\'erateurs de projection
sur les chiralit\'es gauche et droite
des doublets (\ref{lrsql}).
En ins\'erant les d\'efinitions des \'etats propres de masse (\ref{m2gbasis}),
nous obtenons

\bea
\label{lrsqllag}
&& \hskip-2em \bar\Psi \gamma^\mu D_\mu \Psi~ =~ \cdots \\\nonumber&&\hskip-2em
\renewcommand{\arraystretch}{1}
\begin{array}{lccc@{\:}c@{\:}l@{\:}c@{\:}l@{\:}c@{\:}l@{\:}ccr}
%\multicolumnn{9}{l}{\bar\Psi \gamma^\mu D_\mu \Psi~ =~ \cdots} \\
-~i ~\bar u ~\gamma^\mu
& \Bigg[ & Z'_\mu & ( 
&   &                       
&   & {1\over2} {g_R \ct} \quad~ P_R 
& - & {1\over6} {g'  \st} 
& ) \\
&   & Z_\mu  & ( 
&   & {1\over2} {g_L} \cw P_L 
& - & {1\over2} {g_R \st} \sw P_R 
& - & {1\over6} {g'  \ct} \sw
& ) \\
&   & A_\mu  & ( 
&   & {1\over2} {g_L} \sw P_L 
& + & {1\over2} {g_R \st} \cw P_R 
& + & {1\over6} {g'  \ct} \cw
& ) & \Bigg] & u
\\
&&&
\multicolumn{8}{c}{{\stackrel{\underbrace{\makebox[22em][c]{}}}{{2\over3}e}}}
\end{array}
\eea
$$
\begin{array}{lccc@{\:}c@{\:}l@{\:}c@{\:}l@{\:}c@{\:}l@{\:}ccr}
-~i ~\bar d ~\gamma^\mu
& \Bigg[ & Z'_\mu & ( 
&   &                       
& - & {1\over2} {g_R \ct} \quad~ P_R 
& - & {1\over6} {g'  \st} 
& ) \\
&   & Z_\mu  & ( 
& - & {1\over2} {g_L} \cw P_L 
& + & {1\over2} {g_R \st} \sw P_R 
& - & {1\over6} {g'  \ct} \sw
& ) \\
&   & A_\mu  & ( 
& - & {1\over2} {g_L} \sw P_L 
& - & {1\over2} {g_R \st} \cw P_R 
& + & {1\over6} {g'  \ct} \cw
& ) & \Bigg] & d
\\
&&& \multicolumn{8}{c}{{\stackrel{\underbrace{\makebox[22em][c]{}}}{-{1\over3}e}}}
\end{array}
$$
$$
\begin{array}{lccc@{\:}c@{\:}l@{\:}c@{\:}l@{\:}c@{\:}l@{\:}ccr}
-~i ~\bar \nu ~\gamma^\mu
&\Bigg [ & Z'_\mu & ( 
&   &                       
&   & {1\over2} {g_R \ct} \quad~ P_R 
& + & {1\over2} {g'  \st} 
& ) \\
&   & Z_\mu  & ( 
&   & {1\over2} {g_L} \cw P_L 
& - & {1\over2} {g_R \st} \sw P_R 
& + & {1\over2} {g'  \ct} \sw
& ) \\
&   & A_\mu  & ( 
&   & {1\over2} {g_L} \sw P_L 
& + & {1\over2} {g_R \st} \cw P_R 
& - & {1\over2} {g'  \ct} \cw
& ) & \Bigg] & \nu
\\
&&& \multicolumn{8}{c}{{\stackrel{\underbrace{\makebox[22em][c]{}}}{0}}}
\end{array}
$$
$$
\begin{array}{lccc@{\:}c@{\:}l@{\:}c@{\:}l@{\:}c@{\:}l@{\:}ccr}
-~i ~\bar \ell ~\gamma^\mu
& \Bigg[ & Z'_\mu & ( 
&   &                       
& - & {1\over2} {g_R \ct} \quad~ P_R 
& + & {1\over2} {g'  \st} 
& ) \\
&   & Z_\mu  & ( 
& - & {1\over2} {g_L} \cw P_L 
& + & {1\over2} {g_R \st} \sw P_R 
& + & {1\over2} {g'  \ct} \sw
& ) \\
&   & A_\mu  & ( 
& - & {1\over2} {g_L} \sw P_L 
& - & {1\over2} {g_R \st} \cw P_R 
& - & {1\over2} {g'  \ct} \cw
& ) & \Bigg] & \ell
\\
&&& \multicolumn{8}{c}{{\stackrel{\underbrace{\makebox[22em][c]{}}}{-e}}}
\end{array}
\renewcommand{\arraystretch}{1}
$$

Bien que les couplages des quarks et leptons au $Z'$
ne soient pas connus {\em a priori},
leurs couplages au photon $A$ et au $Z^0$
nous sont connus du \sm.
En guise d'illustration,
j'ai indiqu\'e explicitement dans l'\'equation (\ref{lrsqllag})
les couplages au photon.

Ces huit relations 
(plus celle que nous avions obtenue pr\'ec\'edemment 
en traitant du triplet de Higgs)
fournissent un syst\`eme sur-d\'etermin\'e d'\'equations.
Fort heureusement il a une solution consistante
en fonction de trois param\`etres ind\'ependants du mod\`ele.
Le choix pi\'eton pour ceux-ci
seraient les constantes de couplage associ\'ees au groupe de jauge 
$(g_L,g_R,g')$.
Mais je pr\'ef\`ere faire le choix \'equivalent et plus astucieux
{$(e,\sw,\kappa=g_L/g_R)$},
qui a le grand avantage de contenir d'embl\'ee 
deux quantit\'es bien connues et mesur\'ees,
\`a savoir la charge de l'\'electron $e$
et le sinus de l'\wma\ \sw.

Un manipulation \'el\'ementaire nous donne:

\beq
\label{lrssol}
\left.
\renewcommand{\arraystretch}{1.5}
\begin{array}{l}
e \\ \sw \\ \kappa = \displaystyle{g_R \over g_L}
\end{array}
\right\}
\qquad
\longrightarrow
\qquad
\left\{
\renewcommand{\arraystretch}{2}
\begin{array}{l}
g_L = \displaystyle{e\over\sw} \\
g_R = \displaystyle{e\kappa\over\sw} \\
g' = \displaystyle{e\kappa\over\sqrt{\kappa^2\cwt-\swt}} \\
\st = \displaystyle{\sw\over\kappa\cw} 
\end{array}
\renewcommand{\arraystretch}{1}
\right.
\eeq

Il est aussi utile 
pour simplifier les expressions par la suite,
d'introduite le param\`etre

\beq
\label{beta}
\beta ~=~ \cot\theta_s ~=~ \sqrt{{\kappa^2\cwt\over\swt}-1}~
~.
\eeq

Outre les param\`etres qui interviennent d\'ej\`a dans le \sm.
le \lrsm\ minimal est donc sp\'ecifi\'e 
en fonction de deux nouveaux param\`etres:

\begin{enumerate}
\item 
Le rapport des constantes de couplage droite et gauche
$\kappa = g_R/g_L$,
qui d\'etermine le couplage des bosons de jauge lourds 
$Z'$ et $W'$ 
aux fermions.
\item
La \vev\ $v$ du triplet droit de Higgs $\Delta$,
qui brise la sym\'etrie $SU(2)_R \otimes U(1)_{B-L} \to U(1)_Y$
et d\'etermine les masses des bosons de jauge lourds
$Z'$ et $W'$.
\end{enumerate}

\section[La physique du $Z'$]{La physique du \boldmath$Z'$}

Concentrons nous \`a pr\'esent 
sur une pr\'ediction bien particuli\`ere du \lrsm,
l'existence du boson de jauge massif $Z'$,
et explorons bri\`evement les perspectives pour sa d\'ecouverte.

Comme les masses des $W'$ et $Z'$ 
sont proportionnelles \`a $|v|^2$,
%sont proportionnelles au carr\'e de la \vev\ $|v|^2$,
elles peuvent en principe prendre des valeurs tr\`es \'elev\'ees.
Elles sont d\'etermin\'ees
en ins\'erant les conclusions de la section pr\'ec\'edente (\ref{lrssol})
dans le lagrangien (\ref{lrshlag}):

\bea
m_{W'}^2 &=& e^2 ~ |v|^2 ~ {\kappa^2 \over 2\swt}
\nonumber\\\label{lrsclm}\\\nonumber
m_{Z'}^2 &=& e^2 ~ |v|^2 ~ {2\kappa^4\cwt \over \beta^2\swf}~ 
~.
\eea

Les couplages des quarks et leptons au $W'$ 
sont \'evidemment donn\'es 
par les interactions chirales
$g_R P_R$.
Leurs couplages au au $Z'$
sont d\'etermin\'es
en ins\'erant les conclusions de la section pr\'ec\'edente (\ref{lrssol})
dans le lagrangien (\ref{lrsqllag}).
\Eg,
pour les leptons charg\'es
$\ell=(e,\mu,\tau)$
on obtient les couplages vectoriel et axial
${v_{Z'}}$ et ${a_{Z'}}$ 
normalis\'es \`a la charge de l'\'electron $e$

\bea
\label{lrsclv}
{v_{Z'}} &=& {1 \over 4c_w} ~ {2-\beta^2 \over \beta}
\\
\label{lrscla}
{a_{Z'}} &=& -~{1 \over 4c_w} ~ \beta
~.
\eea

Ces couplages d\'ecrivent en fonction des valeur du rapport $\kappa$
une demi hyperbole
dans l'espace param\'etrique 
$(v_{Z'},a_{Z'})$,
dont les asymptotes sont $a_{Z'}=0$ et $a_{Z'}=v_{Z'}$
avec $a_{Z'}<0$.

\subsection[Autres mod\`eles de $Z'$]{Autres mod\`eles de \boldmath$Z'$}

Comme il existe beaucoup d'autres extensions du \sm\
qui pr\'edisent l'existence d'un $Z'$,
il serait domage de les ignorer
dans une analyse ph\'enom\'enologique.
D'une mani\`ere g\'en\'erale,
les interactions d'un quelconque $Z'$ aux quarks et leptons
peuvent \^etre \'ecrits sous la forme g\'en\'erale

\beq
\label{zpc}
e ~ {Z'_\mu}
~ \bar\psi ~ \gamma^\mu ~ \left( {v_{Z'}} + {a_{Z'}} \gamma_5 \right) ~ \psi
~,
\eeq

o\`u ${v_{Z'}}$ et ${a_{Z'}}$ 
sont les couplages vectoriel et axial
du $Z'$ au lepton $\psi$.
Remarquez que ces couplages sont normalis\'es
\`a la charge de l'\'electron $e$.

\begin{figure}[htb]
\hskip-0em{\input{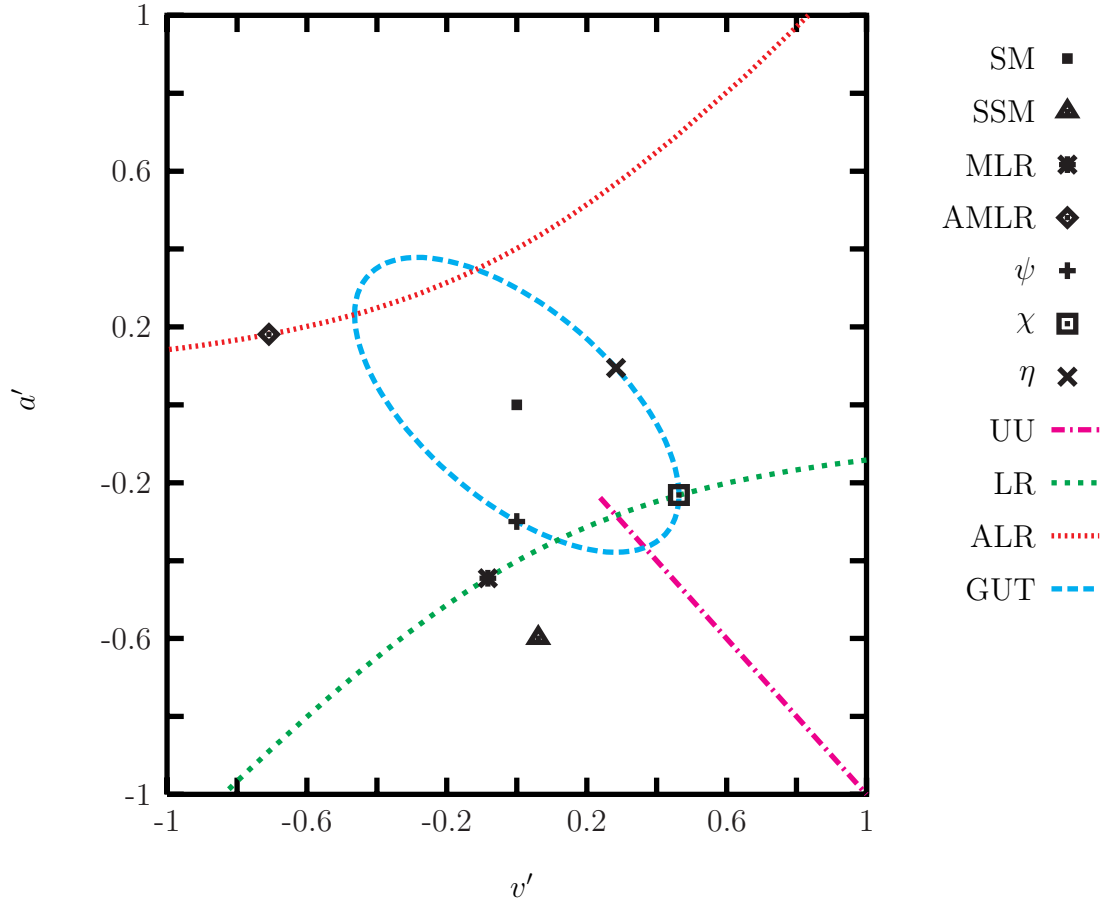}}
\bigskip
\caption{
Valeurs des couplages vectoriels et axiaux
des leptons charg\'es
aux $Z'$ pr\'edits par les mod\`eles
mentionn\'es dans la Table~\protect\ref{tzp}.
%Les courbes UU, LR, ALR et GUT
%correspondent respectivement aux mod\`eles
%standard d\'esunifi\'e,
%sym\'etrique gauche-droite,
%sym\'etrique gauche-droite alternatif
%et de \gu.
%Les points SM, SSM, MLR et AMLR
%repr\'esentent respectivement les mod\`eles
%standard,
%standard s\'equentiel
}
\bigskip
\label{fzpmodels}
\end{figure}

\begin{table}[htb]
\renewcommand{\arraystretch}{2.6}
$$
\begin{array}{||l||c|c|l||}
\hline\hline
\mbox{mod\`ele} 
& v_{Z'}
& a_{Z'}
& \mbox{param\`etres}
\\
\hline\hline
\raisebox{-.5ex}[.5ex]{\shortstack[l]{\sm\\s\'equentiel}}
& \displaystyle{1-4\sin^2\theta_w \over 4\sin\theta_w\cos\theta_w}
& \displaystyle{-1 \over 4\sin\theta_w\cos\theta_w}
& \raisebox{-.5ex}[.5ex]{\shortstack[l]{banc d'essai}}
\\
\hline
\raisebox{-.5ex}[.5ex]{\shortstack[l]{\sm\\d\'esunifi\'e}}
&  \displaystyle{1\over2\sin\theta_w} ~ \tan\phi
& -\displaystyle{1\over2\sin\theta_w} ~ \tan\phi
& \sin\phi \geq \sin\theta_w
\\
\hline
\raisebox{-.5ex}[.5ex]{\shortstack[l]{sym\'etrique\\gauche-droite}}
& \displaystyle{1 \over 4c_w} ~ {2-\beta^2 \over \beta}
& -\displaystyle{1 \over 4c_w} ~ \beta
& \beta > 0
\\
\hline
\raisebox{-1.2ex}[1.2ex]{\shortstack[l]{sym\'etrique\\gauche-droite\\alternatif}}
& \displaystyle{1 \over 4c_w} ~ {1-2\beta^2 \over \beta}
& \displaystyle{1 \over 4c_w} ~ \beta
& \beta > 0
\\
\hline
\raisebox{-.5ex}[.5ex]{\shortstack[l]{th\'eories de\\\gu}}
& \displaystyle{1\over\sqrt{6}\cos\theta_w} ~ \cos\beta
& -\displaystyle{1\over2\sqrt{6}\cos\theta_w} 
\left( \cos\beta + \sqrt{\displaystyle{5\over3}} \sin\beta \right)
& -{\pi} \leq \beta \leq {\pi}
\\
\shortstack[l]{$E_6:~\psi$}
& ``
& ``
& \beta=\displaystyle{\pi\over2}
\\
\shortstack[l]{$SO(10):~\chi$}
& ``
& ``
& \beta=0
\\
\shortstack[l]{inspir\'e des\\supercordes:~$\eta$}
& ``
& ``
& \tan\beta=-\displaystyle\sqrt{5\over3}
\\
\hline\hline
\end{array}
$$
\renewcommand{\arraystretch}{1}
\caption{
Valeurs des couplages vectoriels et axiaux
des leptons charg\'es
aux $Z'$ pr\'edits par diff\'erents mod\`eles populaires.
}
\label{tzp}
\end{table}

Les couplages pr\'edits par les mod\`eles les plus populaires
sont montr\'ees sur la figure~\ref{fzpmodels}
et leurs \'equations param\'etriques sont donn\'ees dans le tableau~\ref{tzp}.
Tous ces mod\`eles supposent l'universalit\'e,
\ie\ que les trois g\'en\'erations sont trait\'ees 
de mani\`ere identique,
mais aucun ne fait de pr\'ediction pr\'ecise 
quant \`a la masse du $Z'$.

\begin{figure}[htb]
\unitlength1mm
\centerline{\makebox(0,120)[bl]{\includegraphics{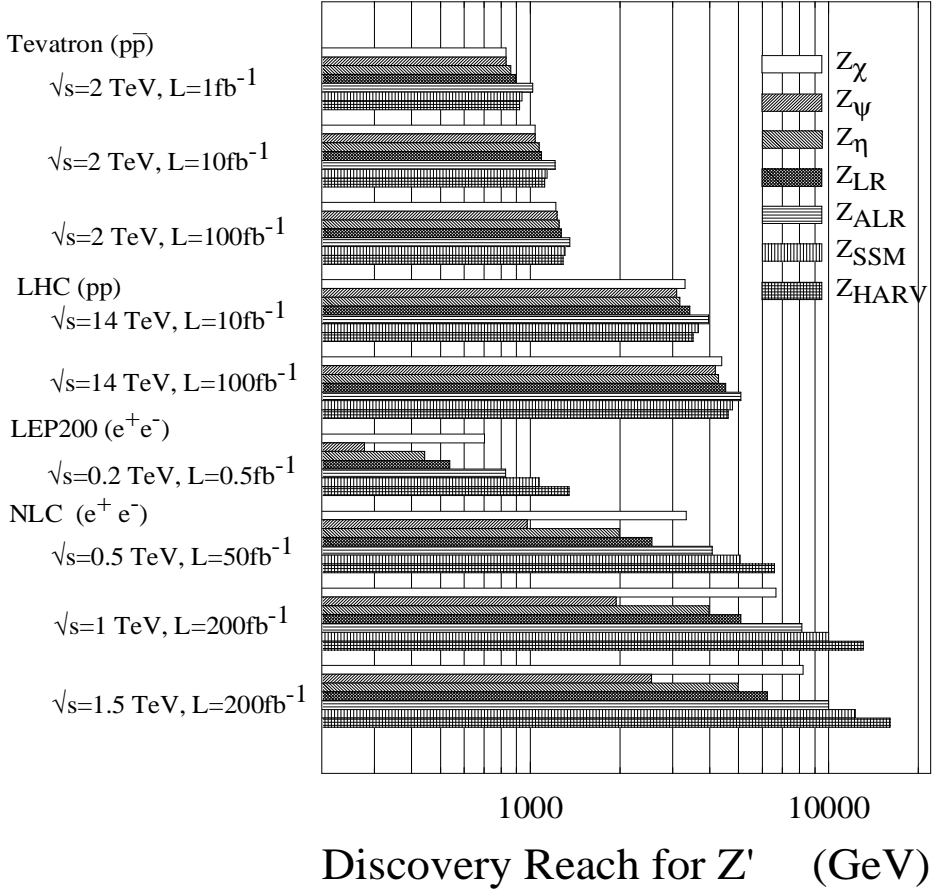}}}
\caption{
Valeurs maximales observables
des masses de $Z'$ pr\'edits par diff\'erents mod\`eles populaires.
Cette figure est adapt\'ee de la r\'ef\'erence~\protect\cite{nlc}.
}
\bigskip
\label{fzlim}
\end{figure}

Examinons \`a pr\'esent les perspectives du LHC et du FLC
quant \`a la d\'ecouverte d'un de ces $Z'$.
Elle sont r\'esum\'ees 
pour les mod\`eles populaires
dans la figure~\ref{fzlim},
qui d\'ecrit aussi les limites sup\'erieures des masses de $Z'$
observables au TEVATRON et au LEP2.

\subsection{Collisions hadroniques}

Le principal m\'ecanisme de production de $Z'$ 
dans les collisions $pp$ ou $p\bar p$
est la production d'une paire de lepton
par une r\'eaction du type Drell-Yan
d\'ecrite dans la figure~\ref{fdy}.
Un $Z'$ y est produit par l'anihilation d'une paire quark anti-quark
et se d\'esint\`egre ensuite en une paire lepton anti-lepton.
Comme ces derniers ont une masse invariante centr\'ee sur la masse du $Z'$,
qui est tr\`es \'elev\'ee,
ce signal ne souffre quasiment d'aucun bruit de fond.
On estime qu'une dizaine d'\'ev\`enement de ce type
suffisent \`a garantir la d\'ecouverte du $Z'$.

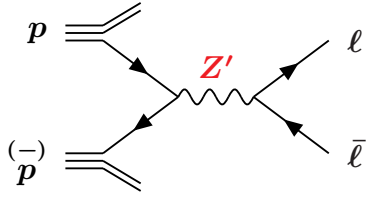
\begin{figure}[htb]
\unitlength.5mm
\SetScale{1.418}
\begin{boldmath}
\begin{center}
\begin{picture}(70,50)(0,0)
\Line(00,10)(10,10)
\Line(00,08)(10,08)
\Line(00,06)(10,06)
%\Vertex(10,08)(5)
\Line(10,08)(20,02)
\Line(10,06)(20,00)
\ArrowLine(30,25)(10,10)
\Line(00,40)(10,40)
\Line(00,42)(10,42)
\Line(00,44)(10,44)
%\Vertex(10,42)(5)
\Line(10,42)(20,48)
\Line(10,44)(20,50)
\ArrowLine(10,40)(30,25)
\Photon(30,25)(50,25){2}{3}
\ArrowLine(70,10)(50,25)
\ArrowLine(50,25)(70,40)
\Text(-5,42)[rc]{$p$}
\Text(-5,08)[rc]{$\stackrel{(-)}{p}$}
\Text(40,30)[cb]{\Red{$Z'$}}
\Text(75,40)[lc]{$\ell$}
\Text(75,10)[lc]{$\bar\ell$}
\end{picture}
\end{center}
\end{boldmath}
\caption{
Diagramme de Feynman 
responsable \`a l'ordre le plus bas
de la production et de la d\'esint\'egration d'un $Z'$
dans les collisions $pp$ ou $p\bar p$.
}
\label{fdy}
\end{figure}

C'est d'ailleurs ce m\^eme m\'ecanisme
qui a permis de d\'ecouvir le boson interm\'ediaire $Z$ 
\`a partir d'une poign\'ee d'\'ev\`enements
dans les collisions $p\bar p$ au SPS du CERN en 1983.

\ex
Il est facile de se faire une id\'ee approximative
de ce que les collisions hadroniques 
peuvent nous apporter comme information sur un \'eventuel $Z'$
en estimant sa \xs\ de production.
La \xs\ microscopique
du sous-proc\`es Drell-Yan
de la figure~\ref{fdy},
est donn\'e par~\cite{iz}

\bea
d\sigma
&=&
\underbrace%
{|{\cal M}|^2}_%
{\mbox{dynamique}}
~\times~
\underbrace%
{1 \over |\vec v_{\bar{q}}-\vec v_{q}|~2E_{\bar{q}}~2E_{q}}_%
{\mbox{flux}}
\\\nonumber
\\\nonumber
&&
\hspace{3em}
~\times~
\underbrace%
{{d^3p_{\bar{\ell}} \over (2\pi)^3~2E_{\bar{\ell}}}~{d^3p_{\ell} \over (2\pi)^3~2E_{\ell}}~(2\pi)^4~\delta^4(p_{\bar{q}}+p_{q}-p_{\bar{\ell}}-p_{\ell})}_%
{\mbox{espace des phases}}
~.
\eea

Dans le \cmf\ et dans la limite ultra-relativiste o\`u
l'\'energie du \cm\ $E_{\rm CM} \gg m_q = m_\ell \approx 0$,
cette \xs\ devient

\beq
d\sigma
~=~
|{\cal M}|^2
~\times~
{1 \over 2 \hat{s}}
~\times~
{d\Omega_{\bar{\ell}} \over 32\pi^2}
~,
\eeq

o\`u par d\'efinition
$\hat{s} = (p_{\bar{q}}+p_{q})^2$.

Puisque les faisceaux de protons au LHC
ne seront pas polaris\'es,
les quarks qu'ils contiennent ne le seront pas non plus
et le carr\'e de la norme de l'\'el\'ement de matrice $|{\cal M}|^2$
ne d\'epend pas de l'angle de diffusion azimuthal $\phi_{\bar{\ell}}$ 
de l'\'el\'ement d'angle solide
$d\Omega_{\bar{\ell}} = d\phi_{\bar{\ell}} d\cos\theta_{\bar{\ell}}$
de l'anti-lepton.
L'int\'egration sur cette variable de l'espace des phase
donne donc trivialement un facteur $2\pi$
et la \xs\ diff\'erentielle de production de muons
est donn\'ee par

\beq
\label{dsig}
{d\sigma \over d\cos\theta} 
~=~
{1 \over 32 \pi \hat{s}}~|{\cal M}|^2
~.
\eeq

Reste maintenant \`a calculer le carr\'e de la norme
de l'\'el\'ement de matrice $|{\cal M}|^2$.
L'application des r\`egles de Feynman
au graphe de la figure~\ref{fpe2mm}
donne pour l'\'el\'ement de matrice

\beq
{\cal M} 
~=~
e^2~
{1 \over \hat{s}-m_{Z'}^2 - im_{Z'}\Gamma_{Z'}}~
[ \bar u_{\bar{\ell}} \gamma^\alpha 
(v_{Z'}^{\ell}+a_{Z'}^{\ell}\gamma_5) 
v_{\ell} ]~
[ \bar v_{q} \gamma^\alpha 
(v_{Z'}^{q}+a_{Z'}^{q}\gamma_5) 
u_{\bar{q}} ]
\eeq

et son hermitien conjugu\'e

\beq
{\cal M}^{\dag}
~=~
e^2~
{1 \over \hat{s}-m_{Z'}^2 + im_{Z'}\Gamma_{Z'}}~
[ \bar v_{\ell} \gamma^\beta 
(v_{Z'}^{\ell}+a_{Z'}^{\ell}\gamma_5) 
u_{\bar{\ell}} ]~
[ \bar u_{\bar{q}} \gamma^\beta 
(v_{Z'}^{q}+a_{Z'}^{q}\gamma_5) 
v_{q} ]
~,
\eeq

o\`u j'ai introduit la largeur de d\'esint\'egration $\Gamma_{Z'}$ du $Z'$
dans son propagateur.
Celle-ci correspond,
en effet,
\`a la partie imaginaire de son \'energie propre,
qui s'ajoute simplement \`a la masse 
par resommation de Dyson.

En sommant sur les polarisations de l'\'etat final
et en moyennant sur les polarisations de l'\'etat initial
(il y en a $2\times2\times3=12$,
chaque quark ayant deux possibilit\'es pour son spin
et trois possibilit\'es pour sa couleur),
on obtient 
dans la limite ultra-relativiste o\`u les masses des quarks et leptons
peuvent \^etre n\'eglig\'ees
pour le carr\'e de la norme
de l'\'el\'ement de matrice

\bea
|{\cal M}|^2
&=&
{1\over12}~
e^4~
{1 \over (\hat{s}-m_{Z'}^2)^2 + m_{Z'}^2\Gamma_{Z'}^2}~
\\\nonumber
&&
\hspace{-5em}
\times
\bigg[~~
({v_{Z'}^{\ell}}^2 + {a_{Z'}^{\ell}}^2)
({v_{Z'}^{q}}^2 + {a_{Z'}^{q}}^2)
\underbrace
{tr[ {p_{\ell}} \gamma^\beta {p_{\bar{\ell}}} \gamma^\alpha ]~
 tr[ {p_{\bar{q}}} \gamma^\beta {p_{q}} \gamma^\alpha ]}_
{\makebox[1em][l]{$\displaystyle
32~( p_{\bar{q}} \cdot p_{\ell} ~ p_{q} \cdot p_{\bar{\ell}} ~+~ p_{\bar{q}} \cdot p_{\bar{\ell}} ~ p_{q} \cdot p_{\ell} )
~=~ 4 \hat{s}^2 (1 + \cos^2\theta)$}}
\\\nonumber
&&
\hspace{-5em}
~~+~
(v_{Z'}^{\ell} a_{Z'}^{\ell} + a_{Z'}^{\ell} v_{Z'}^{\ell})
(v_{Z'}^{q} a_{Z'}^{q} + a_{Z'}^{q} v_{Z'}^{q})
\underbrace
{tr[ \gamma_5 {p_{\ell}} \gamma^\beta {p_{\bar{\ell}}} \gamma^\alpha ]~
 tr[ \gamma_5 {p_{\bar{q}}} \gamma^\beta {p_{q}} \gamma^\alpha ]}_%
{\makebox[1em][l]{$\displaystyle
~32~( p_{\bar{q}} \cdot p_{\ell} ~ p_{q} \cdot p_{\bar{\ell}} ~-~ p_{\bar{q}} \cdot p_{\bar{\ell}} ~ p_{q} \cdot p_{\ell} )
~=~ 8 \hat{s}^2 \cos\theta$}}
\bigg]
\rule{5em}{0ex}
\eea

Il est facile d'exprimer les produits scalaires
des vecteurs energie--moment cin\'etique ultrarelativistes
en fonction de l'angle $\theta$
entre l'\'electron initial et le muon final.
En effet,
dans un \cmf\ ceux-ci peuvent \^etre \'ecrits sous la forme

\bea
\label{kincmh}
p_{q}
\equiv
{\sqrt{\hat{s}}\over2}\left(\begin{array}{c}1\\0\\0\\1\end{array}\right)
\qquad&&\qquad
p_{\ell}
\equiv
{\sqrt{\hat{s}}\over2}\left(\begin{array}{c}1\\0\\\sin\theta\\\cos\theta\end{array}\right)
\\\nonumber
p_{\bar{q}}
\equiv
{\sqrt{\hat{s}}\over2}\left(\begin{array}{c}1\\0\\0\\-1\end{array}\right)
\qquad&&\qquad
p_{\bar{\ell}}
\equiv
{\sqrt{\hat{s}}\over2}\left(\begin{array}{c}1\\0\\-\sin\theta\\-\cos\theta\end{array}\right)
~.
\eea

Si l'on exprime la charge de l'\'electron $e$
en fonction de la constante de structure fine $\alpha = e^2 / 4\pi$,
la \xs\ diff\'erentielle devient

\bea
{d\sigma \over d\cos\theta} 
&=&
{\pi\alpha^2 \over 6}~
{\hat{s} \over (\hat{s}-m_{Z'}^2)^2 + m_{Z'}^2\Gamma_{Z'}^2}~
\times
\\\nonumber\\\nonumber
&&
\hskip5em
\Big[~
({v_{Z'}^{\ell}}^2 + {a_{Z'}^{\ell}}^2)~
({v_{Z'}^{q}}^2 + {a_{Z'}^{q}}^2)~
(1 + \cos^2\theta)
\\\nonumber
&&
\hskip5em
+~
(v_{Z'}^{\ell} a_{Z'}^{\ell} + a_{Z'}^{\ell} v_{Z'}^{\ell})~
(v_{Z'}^{q} a_{Z'}^{q} + a_{Z'}^{q} v_{Z'}^{q})~
2 \cos\theta
~\Big]
~.
\eea

En int\'egrant l'angle polaire $\theta$
de $0$ \`a $\pi$,
le terme proportionel \`a $\cos\theta$ s'\'elimine
et on obtient pour la \xs\ 

\beq
\label{xsmic}
\sigma(\hat{s})~
=~
{4\pi\alpha^2 \over 9}~
({v_{Z'}^{\ell}}^2 + {a_{Z'}^{\ell}}^2)~
({v_{Z'}^{q}}^2 + {a_{Z'}^{q}}^2)~
{\hat{s} \over (\hat{s}-m_{Z'}^2)^2 + m_{Z'}^2\Gamma_{Z'}^2}~
~.
\eeq

Cette \xs\ n'est que la \xs\ microscopique,
qui traite les quarks comme des particules libres.
Il faut encore effectuer \`a pr\'esent 
sa convolution avec les fonctions de structure des protons (\ref{sf})
et sommer sur les diff\'erent types de quarks qu'ils contiennent

\beq
\label{conv}
\sigma~
=~
\sum_q~
\int\limits_0^s d\hat{s} ~ F(\hat{s}) ~ \sigma(\hat{s})~
~,
\eeq

o\`u $s = E_{\rm CM}^2$ est l'\'energie du \cm\ du collisionneur
et $F(\hat{s})$ est la fonction de luminosit\'e,
qui se calcule ais\'ement connaissant les fonctions de structure.
Pour des raisons dimensionelles,
elle est {\em grosso modo} invers\'ement proportionelle \`a son argument,
\ie\
$F(\hat{s}) \propto 1/\hat{s}$.

Cette convolution ne peut se faire que num\'eriquement,
car les fonctions de structure,
ne pouvant pas \^etre d\'etermin\'ees par des m\'ethodes perturbatives,
sont typiquement d\'etermin\'ees 
\`a partir des donn\'ees exp\'erimentales
de diffusion in\'elastique lepton-proton.

Mais l'on peut n\'eanmoins se faire une id\'ee approximative des choses,
en remarquant que la \xs\ (\ref{xsmic})
se comporte comme une distribution de Breit-Wigner.
Pour autant que la largeur $\Gamma_{Z'}$ 
ne soit pas trop grande
(ce qu'elle n'est certainement pas,
si les int\'eractions du $Z'$ sont de l'ordre des interactions \EW s)
elle a donc un maximum tr\`es prononc\'e autour du point
$\hat{s}=m_{Z'}^2$
(dans la limite o\`u la largeur $\Gamma_{Z'}$ est nulle
ce point devient m\^eme un p\^ole non-int\'egrable),
et l'on peut en toute s\'er\'enit\'e 
simplifier l'int\'egrale (\ref{conv})
en donnant aux termes de variation lente
une valeur constante centr\'ee sur $\hat{s}=m_{Z'}^2$
et en poussant les limites d'int\'egration
au-del\`a du domaine physique:

\bea
\label{convsimp}
\sigma
&=&
{4\pi\alpha^2 \over 9}~
({v_{Z'}^{\ell}}^2 + {a_{Z'}^{\ell}}^2)~
\sum_q
({v_{Z'}^{q}}^2 + {a_{Z'}^{q}}^2)~
\\\nonumber
&&
\hskip9em
\times~
\underbrace{F(m_{Z'}^2)}_{\propto\displaystyle{1\over m_{Z'}^2}}~m_{Z'}^2~
\underbrace
{\int\limits_{-\infty}^{+\infty}
{d\hat{s} \over (\hat{s}-m_{Z'}^2)^2 + m_{Z'}^2\Gamma_{Z'}^2}}_
{\displaystyle {\pi \over m_{Z'}\Gamma_{Z'}}}
\\\nonumber
&\propto&
{\alpha^2 \over m_{Z'}\Gamma_{Z'}}~
({v_{Z'}^{\ell}}^2 + {a_{Z'}^{\ell}}^2)~
\sum_q
({v_{Z'}^{q}}^2 + {a_{Z'}^{q}}^2)~
~.
\eea

\bco{
Si le $Z'$ n'a pas de modes exotiques de d\'esint\'egration,
\ie,
s'il ne se d\'esint\`egre qu'en des paires de fermions,
sa largeur est proportionelle \`a
$\Gamma_{Z'} \propto m_{Z'}\alpha\sum_f({v_{Z'}^{f2}} + {a_{Z'}^{f2}})$,
o\`u la somme s'effectue sur tous les quarks et leptons.
En introduisant cette largeur 
dans ce r\'esultat,
nous obtenons donc la \xs\ 

\beq
\label{zphadxs}
\sigma~
\propto~
{\alpha \over m_{Z'}^2}~
{\displaystyle
({v_{Z'}^{\ell}}^2 + {a_{Z'}^{\ell}}^2)~
\sum_q
({v_{Z'}^{q}}^2 + {a_{Z'}^{q}}^2) 
\over
\displaystyle
\sum_f
({v_{Z'}^{f}}^2 + {a_{Z'}^{f}}^2)
}
~.
\eeq
}\eco

La masse et la largeur du $Z'$ 
peuvent \^etre d\'etermin\'ees ind\'ependemment,
en mesurant la masse invariante des paires de leptons.
Ainsi la mesure de la \xs\ 
nous fournit la valeur de la combinaison des couplages 
qui intervient dans les deux derniers facteurs de l'\'equation (\ref{convsimp}).
\xe

\subsection{Collisions \'electrons-positrons}

Comme pour le $Z$
que l'on a pu \'etudier de mani\`ere si pr\'ecise au LEP et au SLC,
la production de $Z'$ r\'eels
dans les collisions \pe\ au FLC
est la r\'eaction id\'eale pour l'\'etude de ces derniers.
Mais encore faut-il que l'\'energie du collisionneur soit suffisante!

\begin{figure}[htb]
\unitlength.5mm
\SetScale{1.418}
\begin{boldmath}
\begin{center}
\begin{picture}(60,40)(0,0)
\ArrowLine(0,0)(15,15)
\ArrowLine(15,15)(0,30)
\Photon(15,15)(45,15){2}{5}
\ArrowLine(60,30)(45,15)
\ArrowLine(45,15)(60,0)
\Text(-2,0)[r]{$e^-$}
\Text(-2,30)[r]{$e^+$}
\Text(62,0)[l]{$\mu^-$}
\Text(62,30)[l]{$\mu^+$}
\Text(30,23)[c]{$\gamma$,${Z^0}$,$\Red{Z'}$}
\end{picture}
\end{center}
\end{boldmath}
\caption{
Diagramme de Feynman 
responsable \`a l'ordre le plus bas
de la production de paires de muons
dans les collisions \pe.
}
\label{fpe2mm}
\end{figure}
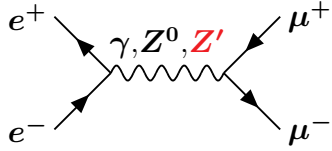

Mais m\^eme si dans un premier stage
l'\'energie nominale du FLC,
\ie\ 500 GeV,
n'atteint pas encore la r\'esonance,
des effets virtuels du $Z'$ pourront peut-\^etre \^etre observ\'es,
\eg\ au travers de la production en paire de muon
d\'ecrite par le diagramme de Feynman de la figure~\ref{fpe2mm}.

\co
Nous serions alors dans une situation analogue 
\`a celle de l'anneau TRISTAN du KEK au Japon.
Ce dernier avait \'et\'e con\c cu pour l'\'etude du p\^ole $Z^0$
\`a une \'epoque 
o\`u l'on croyait encore \`a une masse plus faible de ce dernier.
Bien que cette machine n'ai jamais produit le moindre boson $Z^0$
elle a du moins permis d'observer
au pied de la r\'esonance
un surplus de \xs\ 
par rapport aux pr\'ediction de la QED.
Ceci aurait suffit pour d\'ecouvrir le $Z^0$,
si son existence n'avait pas d\'ej\`a \'et\'e clairement mise en \'evidence
auparavant
dans les collisions $p\bar p$ au SPS du CERN.
\oc

\ex
La \xs\ de production d'une paire muon-antimuon
\`a partir de collisions \'electrons-positrons
est obtenue trivialement
\`a l'aide de la \xs\ (\ref{dsig})

\bco{
est donn\'e par~\cite{iz}

\bea
d\sigma
&=&
\underbrace%
{|{\cal M}|^2}_%
{\mbox{dynamique}}
~\times~
\underbrace%
{1 \over |v_{e^+}-v_{e^-}|~2E_{e^+}~2E_{e^-}}_%
{\mbox{flux}}
\\\nonumber
\\\nonumber
&&
\hspace{3em}
~\times~
\underbrace%
{{d^3p_{\mu^+} \over (2\pi)^3~2E_{\mu^+}}~{d^3p_{\mu^-} \over (2\pi)^3~2E_{\mu^-}}~(2\pi)^4~\delta^4(p_{e^+}+p_{e^-}-p_{\mu^+}-p_{\mu^-})}_%
{\mbox{espace des phases}}
~.
\eea

Dans le \cmf\ et dans la limite ultra-relativiste o\`u
l'\'energie du \cm\ $E_{\rm CM} \gg m_e = m_\mu \approx 0$,
cette \xs\ devient

\beq
d\sigma
~=~
|{\cal M}|^2
~\times~
{1 \over 2s}
~\times~
{d\Omega_{\mu^+} \over 32\pi^2}
~,
\eeq

o\`u par d\'efinition
$s = (p_{e^+}+p_{e^-})^2 = E_{\rm CM}^2$.

En l'absence de polarisation lin\'eaire
des faisceaux d'\'electrons\footnote{
Pour la plupart des applications
la polarisation circulaire est pr\'ef\'er\'ee 
\`a la polarisation lin\'eaire}
le carr\'e de la norme de l'\'el\'ement de matrice $|{\cal M}|^2$
ne d\'epend pas de l'angle de diffusion azimuthal $\phi_{\mu^+}$ 
de l'\'el\'ement d'angle solide
$d\Omega_{\mu^+} = d\phi_{\mu^+} d\cos\theta_{\mu^+}$
de l'antimuon.
L'int\'egration sur cette variable de l'espace des phase
donne donc trivialement un facteur $2\pi$
et la \xs\ diff\'erentielle de production de muons
est donn\'ee par
}\eco

\beq
{d\sigma \over d\cos\theta} 
~=~
{1 \over 32 \pi s}~|{\cal M}|^2~
~,
\eeq

o\`u l'\'el\'ement de matrice ${\cal M}$
s'obtient en applicant les r\`egles de Feynman appropri\'ees
au graphe de la figure~\ref{fpe2mm}

\beq
{\cal M} 
~=~
e^2~
\sum_{i=\gamma,Z,Z'}~
{1 \over s-m_i^2}~
[ \bar u_{\mu^+} \gamma^\alpha (v_i+a_i\gamma_5) v_{\mu^-} ]~
[ \bar v_{e^-} \gamma^\alpha (v_i+a_i\gamma_5) u_{e^+} ]~
~.
\eeq

Son hermitien conjugu\'e 
est donn\'e par

\beq
{\cal M}^{\dag}
~=~
e^2~
\sum_{j=\gamma,Z,Z'}~
{1 \over s-m_j^2}~
[ \bar v_{\mu^-} \gamma^\beta (v_j+a_j\gamma_5) u_{\mu^+} ]~
[ \bar u_{e^+} \gamma^\beta (v_j+a_j\gamma_5) v_{e^-} ]
~,
\eeq

o\`u les couplages vectoriel $v_i$ et axial $a_i$ 
du photon et du $Z^0$ aux leptons
sont donn\'es par

\beq
\left\{
  \begin{array}{l}
  \strut v_\gamma = 1 \\\\
  \strut a_\gamma = 0
  \earr
\right.
\qquad\qquad
\left\{
  \begin{array}{l}
  \strut v_{Z^0} = \displaystyle{1-4\sin^2\theta_w \over 4\sin\theta_w\cos\theta_w} \\\\
  \strut a_{Z^0} = \displaystyle{-1 \over 4\sin\theta_w\cos\theta_w}
  \earr
\right.
\eeq

En sommant sur les polarisations de l'\'etat final
et en moyennant sur les polarisations de l'\'etat initial
(il y en a $2\times2=4$),
on obtient 
dans la limite ultra-relativiste o\`u les masses des \'electrons et muons 
peuvent \^etre n\'eglig\'ees
pour le carr\'e de la norme
de l'\'el\'ement de matrice

\bea
|{\cal M}|^2
&=&
{1\over4}~
e^4~
\sum_{i,j=\gamma,Z,Z'}~
{1 \over s-m_i^2}~
{1 \over s-m_j^2}~
\\\nonumber
&&
\hspace{-5em}
\times
\bigg[~~
(v_i v_j + a_i a_j)^2
\underbrace
{tr[ {p_{\mu^-}} \gamma^\beta {p_{\mu^+}} \gamma^\alpha ]~
 tr[ {p_{e^+}} \gamma^\beta {p_{e^-}} \gamma^\alpha ]}_
{\makebox[1em][l]{$\displaystyle
32~( p_{e^+} \cdot p_{\mu^-} ~ p_{e^-} \cdot p_{\mu^+} ~+~ p_{e^+} \cdot p_{\mu^+} ~ p_{e^-} \cdot p_{\mu^-} )
~=~ 4 s^2 (1 + \cos^2\theta)$}}
\\\nonumber
&&
\hspace{-5em}
~~+~
(v_i a_j + a_i v_j)^2
\underbrace
{tr[ \gamma_5 {p_{\mu^-}} \gamma^\beta {p_{\mu^+}} \gamma^\alpha ]~
 tr[ \gamma_5 {p_{e^+}} \gamma^\beta {p_{e^-}} \gamma^\alpha ]}_
{\makebox[1em][l]{$\displaystyle
~32~( p_{e^+} \cdot p_{\mu^-} ~ p_{e^-} \cdot p_{\mu^+} ~-~ p_{e^+} \cdot p_{\mu^+} ~ p_{e^-} \cdot p_{\mu^-} )
~=~ 8 s^2 \cos\theta$}}
\bigg]
\eea

Il est facile d'exprimer les produits scalaires
des vecteurs energie--moment cin\'etique ultrarelativistes
en fonction de l'angle $\theta$
entre l'\'electron initial et le muon final.
En effet,
dans un \cmf\ ceux-ci peuvent \^etre \'ecrits sous la forme

\bea
\label{kincm}
p_{e^-}
\equiv
{\sqrt{s}\over2}\left(\begin{array}{c}1\\0\\0\\1\end{array}\right)
\qquad&&\qquad
p_{\mu^-}
\equiv
{\sqrt{s}\over2}\left(\begin{array}{c}1\\0\\\sin\theta\\\cos\theta\end{array}\right)
\\\nonumber
p_{e^+}
\equiv
{\sqrt{s}\over2}\left(\begin{array}{c}1\\0\\0\\-1\end{array}\right)
\qquad&&\qquad
p_{\mu^+}
\equiv
{\sqrt{s}\over2}\left(\begin{array}{c}1\\0\\-\sin\theta\\-\cos\theta\end{array}\right)
~.
\eea

Si l'on exprime la charge de l'\'electron $e$
en fonction de la constante de structure fine $\alpha = e^2 / 4\pi$,
la \xs\ diff\'erentielle devient

\bea
{d\sigma \over d\cos\theta} 
&=&
{\pi\alpha^2 s \over 2}~
\sum_{i,j=\gamma,Z,Z'}~
{1 \over s-m_i^2}~
{1 \over s-m_j^2}~
\\\nonumber
&&
\hspace{2em}
\left[~
(v_i v_j + a_i a_j)^2~
(1 + \cos^2\theta)
~+~
(v_i a_j + a_i v_j)^2~
2 \cos\theta
~\right]
~.
\eea

Afin d'obtenir une id\'ee qualitative\footnote{
Qui,
loin du p\^ole $Z'$,
fournit d'ailleurs des r\'esultats parfaitements r\'ealistes!}
de ce qui se passe,
il est pratique de travailler dans la limite o\`u

\beq
\label{as}
m_{Z^0}^2 \ll s \ll m_{Z'}^2
\qquad \mbox{et} \qquad
\sin^2\theta_w = {1\over4}
~ \Rightarrow ~
\left\{
  \begin{array}{l}
  v_{Z^0} = 0 \\\\
  a_{Z^0} = -1/\sqrt{3} 
  \earr
\right.
\eeq

et de n\'egliger la contribution 
du carr\'e du canal $Z'$.\footnote{
Ceci se justifie par le fait que nous cherchons \`a mettre en
\'evidence des effets tr\`es petits,
d\^us \`a un $Z'$ tr\`es lourd ou couplant tr\`es faiblement.
}
La \xs\ diff\'erentielle devient alors

\bea
{d\sigma \over d\cos\theta} 
~=~
{\pi\alpha^2 \over 9}
&\bigg\{&
{1 \over s}
\left( 5 + 6\cos\theta + 5\cos^2\theta \right)
\\\nonumber
&-&
{1 \over m_{Z'}^2}
\Big[~~
   v_{Z'}^2~(3 + 2\cos\theta + 3\cos^2\theta )
\\\nonumber
&&
\hspace{2em}
+~ a_{Z'}^2~(1 + 6\cos\theta +  \cos^2\theta )
~\Big]~\bigg\}~
~.
\eea

On voit donc que 
dans la limite asymptotique (\ref{as})
les couplages et la masse du $Z'$ 
ont un effet corrr\'el\'e:
un $Z'$ l\'eger avec des couplages faibles
est \'equivalent 
\`a un $Z'$ lourd avec des couplages forts.
Pour tenir compte de cette corr\'elation,
il est commode d'utiliser les couplages r\'eduits

\beq
\label{redc}
v' ~=~ v_{Z'} ~ {\sqrt{s} \over m_{Z'}}
\qquad\qquad\qquad
a' ~=~ a_{Z'} ~ {\sqrt{s} \over m_{Z'}}~
~.
\eeq

Il est ais\'e maintenant d'estimer l'effet qu'aura un $Z'$
sur deux observables
ch\'eries des exp\'erimentateurs,
\`a savoir le nombre total d'\'ev\`enements $N$
et l'assym\'etrie avant-arri\`ere $A$:

\bea
N
&=&
{\cal L} ~ \int_{-1}^{+1} d\cos\theta ~ {d\sigma \over d\cos\theta}
\\\nonumber\\\nonumber
&=&
\underbrace{{\cal L} ~ {40\pi\alpha^2 \over 27 s}}_{\displaystyle N_{\rm SM}} ~
\left[ ~ 1 ~-~ {1\over5}  \left(3v'^2 + a'^2\right) ~ \right]
\\\nonumber\\\nonumber\\
A
&=&
\frac%
{ \displaystyle
    \int_{0}^{+1} d\cos\theta ~ {d\sigma \over d\cos\theta}
~-~ \int_{-1}^{0} d\cos\theta ~ {d\sigma \over d\cos\theta} }
{ \displaystyle
    \int_{0}^{+1} d\cos\theta ~ {d\sigma \over d\cos\theta}
~+~ \int_{-1}^{0} d\cos\theta ~ {d\sigma \over d\cos\theta} }
\\\nonumber\\\nonumber
&=&
\underbrace{9\over20}_{\displaystyle A_{\rm SM}} ~ 
\frac%
{\displaystyle 1 - {1\over3} \left(v'^2 + 3a'^2\right)}%
{\displaystyle 1 - {1\over5} \left(3v'^2 + a'^2\right)}
\qquad\simeq\qquad
{9\over20} ~ 
\left[ ~ 1 ~+~ {4\over15} \left(v'^2 - 3a'^2\right) \right]
~.
\eea

Les valeurs pr\'edites par le \sm,
\ie\ en l'absence de $Z'$,
sont donn\'ees par
$N_{\rm SM}$ et $A_{\rm SM}$.
On peut estimer le poids statistique de ces observables
en calculant leurs estimateurs de moindre carr\'es.
Si nous nous concentrons uniquement sur les erreurs statistiques
$\Delta N$ et $\Delta A$
(qui sont donn\'ees par la statistique de Poisson)
et ignorons les erreurs syst\'ematiques
(qui sont donn\'ee par une statistique inconnue
et sont donc beaucoup plus difficile \`a traiter;
heureusement,
dans ce cas-ci il s'av\`ere qu'elles peuvent \^etre n\'eglig\'ees,
car les erreurs statistiques dominent),
nous obtenons

\bea
\label{chin}
\chi^2_N
&=&
\left( {N_{\rm SM}-N \over \Delta N_{\rm SM}} \right)^2
\qquad\qquad\qquad \mbox{avec} \qquad
\Delta N_{\rm SM} = \sqrt{N_{\rm SM}}
\\\nonumber\\\nonumber
&=&
N_{\rm SM} ~ {1\over25} ~ 
{\left( 3v'^2 + a'^2 \right)^{\Black{2}}}
\\\nonumber\\\nonumber\\
\label{chia}
\chi^2_A
&=&
\left( {A_{\rm SM}-A \over \Delta A_{\rm SM}} \right)^2
\qquad\qquad\qquad \mbox{avec} \qquad
\Delta A_{\rm SM} = \sqrt{1-A_{\rm SM}^2 \over N_{\rm SM}}
\\\nonumber\\\nonumber
&=&
N_{\rm SM} ~ {144\over7975} ~ 
{\left( v'^2 - 3a'^2 \right)^{\Black{2}}}
~.
\eea

L'interpr\'etation probabilistique de ces estimateurs est la suivante:
en l'absence de $Z'$,
5\% des fluctuations statistiques 
sont capables de donner lieu \`a une mesure
\'egale \`a la valeur moyenne pr\'edite
en pr\'esence d'un $Z'$
dont les couplages r\'eduits $v'$ et $a'$
fournissent une valeur de $\chi^2$ 
sup\'erieure \`a 6.
On dit dans le jargon,
que la r\'egion de l'espace param\'etrique $(v',a')$
o\`u $\chi^2>6$
peut \^etre explor\'ee \`a 95\%\ de \cl.
Les valeurs de $\chi^2$ correspondant \`a d'autres \cls\
peuvent ais\'ement \^etre obtenues \`a partir de tables 
de la distribution de $\chi^2$~\cite{pdg}.

\begin{figure}[htb]
\unitlength1mm
\qquad\makebox(0,120)[bl]{\includegraphics{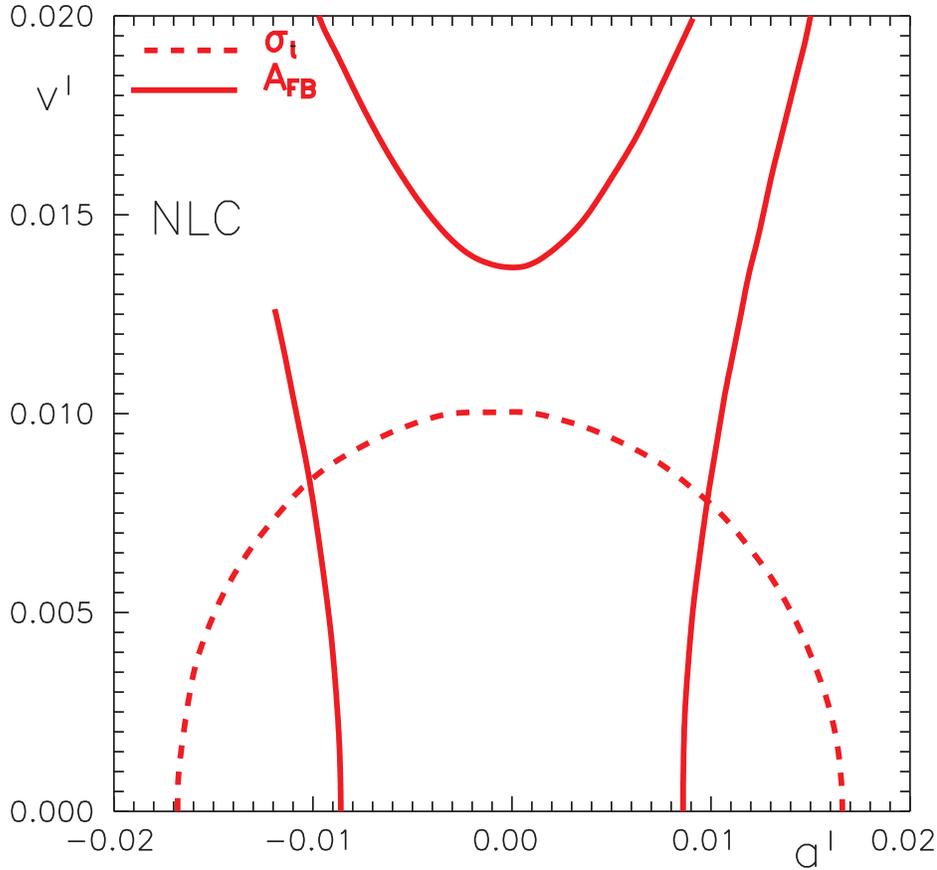}}
\caption{
Valeurs des couplages r\'eduits (\protect\ref{redc})
permettant d'exclure le \sm\ au FLC \`a 95\% de \cl\
en observant le nombre total d'\'ev\`enements ($\sigma_t=N$)
et l'asym\'etrie avant-arri\`ere ($A_{FB}=A$).
Cette figure est adapt\'ee de la r\'ef\'erence~\protect\cite{arnd}.
}
\label{farnd}
\end{figure}

Dans le cas qui nous int\'eresse,
les \'equations (\ref{chin},\ref{chia})
d\'elimitent respectivement une ellipse et deux hyperboles
dans le plan $(v',a')$.
Ce r\'esultat
est confirm\'e 
par une analyse plus compl\`ete,
qui tient compte des effets que j'ai n\'eglig\'es
par mes diverses approximations
et inclut aussi les corrections radiatives 
\`a une boucle~\cite{arnd}.
On peut s'en convaincre \`a l'aide de la figure~\ref{farnd}.
Il s'av\`ere m\^eme que les r\'esultats num\'eriques
de cette analyse plus r\'ealiste
diff\`erent de moins de 20\%
de ceux obtenus plus haut!
\xe

\co
Il est peut-\^etre int\'eressant de remarquer
que les collisions \ee~\cite{ee} peuvent aussi mettre en \'evidence
un $Z'$ virtuel,
en induisant des anomalies dans les distributions angulaires
de la diffusion M\o ller
$\ee\to\ee$.
Il s'av\`ere m\^eme que ce type de r\'eaction
est plus performant que les r\'eactions \pe\
dans la recherche de $Z'$,
si la masse de cux-ci d\'epasse d'environ 20\% 
l'\'energie du collisionneur.
\oc 
\chapter{La \gu}

Le \lrsm\ partage un d\'efaut majeur du \sm,
\`a savoir sa structure de produit de plusieurs groupes simples.
Chacun des facteurs de ces produits directs de groupes simples
est en effet muni de sa propre constante de couplage,
qui n'est pas pr\'edite {\em a priori} par la th\'eorie.
Il ne peut donc \^etre question d'une unification
comme c'est le cas pour l'\'electromagn\'tisme,
o\`u les interactions \'electriques et magn\'etiques
ne sont que deux aspects d'une m\^eme r\'ealit\'e.
Pour qu'il y ai une v\'eritable unification \EW,
voire une \gu\ des interactions fortes et \EW s,
il faut n\'ecessairement avoir au d\'epart une sym\'etrie 
de groupe simple.

Bien s\^ur,
vu le succ\`es du \sm,
il faut aussi imp\'erativement que le groupe simple unifi\'e,
que j'appelerai g\'en\'eriquement $G$,
contienne
$SU(3) \otimes SU(2) \otimes U(1)$
comme sous-groupe.
La sym\'etrie de $G$ doit \^etre bris\'ee
\`a une \'echelle d'\'energies suffisamment \'elev\'ees
$m_{\rm GUT}$,
pour garantir l'absence de ph\'enom\`enes nouveaux
aux \'energies d'ores et d\'ej\`a explor\'ees,
o\`u seule la sym\'etrie 
$SU(3)_c \otimes SU(2)_L \otimes U(1)_Y$
s'est jusqu'\`a pr\'esent manifest\'ee.
La philosophie des \guts\
peut donc se r\'esumer par la cha\^\i ne de \ssb\

\beq
\label{guts}
G~
\stackrel{m_{\rm GUT}}{\longrightarrow}~
SU(3)_c \otimes SU(2)_L \otimes U(1)_Y~
\stackrel{m_W}{\longrightarrow}~
SU(3)_c \otimes U(1)_{\rm EM}~
~.
\eeq

J'examinerai d'abord quelques aspects g\'en\'eraux de la \gu,
partag\'es par toutes les \guts,
et je me sp\'ecialiserai ensuite sur le groupe $SU(5)$,
qui constitue l'exemple le plus simple.

\section{Remarques g\'en\'erales}

Pour qu'un groupe simple $G$ convienne \`a la \gu\
il faut qu'il satisfasse \`a un certain nombre de conditions
qui garantissent la possibilit\'e de d\'ecrire la structure du \sm\
et sa ph\'enom\'enologie.
Ces conditions sont les suivantes:

\begin{itemize}
\item
L'ordre de $G$,
\ie\ son nombre de g\'en\'erateurs,
doit \^etre sup\'erieur \`a $8+3+1=12$,
\ie\ au nombre de g\'en\'erateurs de
$SU(3)_c \otimes SU(2)_L \otimes U(1)_Y$.
\item
Pour la m\^eme raison,
le rang de $G$,
\ie\ son nombre maximum de g\'en\'erateurs
diagonalisable simultan\'ement,
doit \^etre sup\'erieur \`a $2+1+1=4$.
\item
Le groupe $G$
doit admettre des repr\'esentations complexes
afin de permettre de traiter diff\'eremment
les fermions gauches des fermions droits.
\item
Les fermions connus doivent pouvoir s'ins\'erer
dans une repr\'esentation de $G$
qui soit libre d'anomalies.
\end{itemize}

Il s'av\`ere que de la r\'eunion de l'ensemble de ces crit\`eres
r\'esulte une \'elimination draconienne 
des groupes candidats \`a la \gu.
Les principaux survivants sont
$SU(5)$~\cite{gg},
$SO(10)$~\cite{fm} et
$E_6$~\cite{grs},
de rang 4, 5 et 6 respectivement.

\subsection{Quantification de la charge \EM}
\label{chquant}

Dans le cadre du \sm\ 
la charge \EM\ $Q$ d'une particule 
est d\'etermin\'ee par la relation
$Q=T^3+Y$.
Alors que les valeurs de leur projection d'isospin $T^3$ sont quantifi\'ees,
celles de leur hypercharge $Y$,
et donc aussi celles de leur charge $Q$,
ne le sont pas.
En effet,
les valeurs propres des g\'en\'erateurs de groupe simples non-ab\'eliens,
tel le $SU(2)_L$ du \sm,
sont toujours quantifi\'es.
Par contre le g\'en\'erateur
du groupe $U(1)$
admet un continuum de valeurs propres.
La sym\'etrie de jauge du \sm\ ne peut donc pas expliquer
la quantisation de la charge \EM.
Si la perversit\'e vous y pousse,
rien ne vous emp\^eche donc en principe
d'\'elargir la faune du \sm\
en introduisant des particules de charge $\sqrt{2}$ ou $\pi$!

Par contre,
si la sym\'etrie est d\'etermin\'ee au d\'epart 
par un groupe simple
(ce que $SU(2) \otimes U(1)$ n'est pas!),
l'op\'erateur de charge est un de ses g\'en\'erateurs
(autrement ce groupe n'incluerait pas l'\'electromagn\'etisme
et on ne pourrait gu\`ere parler de \gu!)
et doit donc obligatoirement avoir des valeurs propres quantifi\'ees.
Ceci fournit donc 
une explication simple et in\'eluctable
\`a l'incroyable pr\'ecision
avec laquelle les charges de l'\'electron et du proton s'annulent.

Pour cette m\^eme raison
la trace de l'op\'erateur de charge
de charge doit \^etre nulle.
(Comme l'exige la d\'efinition m\^eme
d'un g\'en\'erateur de groupe de Lie simple non-ab\'elien.)
Il en r\'esulte
que pour n'importe quelle repr\'esentation de $G$
la somme des charges 
(et des hypercharges)
de tous ses membres 
doit s'annuler.

\subsection{Unification des constantes de couplage}

Pour chaque repr\'esentation du groupe unifi\'e $G$,
ses g\'en\'erateurs $T_i$
et en particulier les $8+3+1=12$ g\'en\'erateurs des sous-groupes
$SU(3)_c$, $SU(2)_L$ et $U(1)_Y$,
satisfont la relation 

\beq
\label{casimir}
{\rm Tr}(T_iT_j)~ =~ C~\delta_{ij}~
\qquad
\left(i=1\dots n=\mbox{ordre de }G\right)
~,
\eeq

o\`u $C$ est une constante
qui d\'epend
de la repr\'esentation particuli\`ere
sur laquelle est effectu\'ee la trace.
Si nous \'evaluons ce nombre
pour la rep\'esentation qui contient les 15 fermions du \sm\
et pour les g\'en\'erateurs de $SU(3)_c$, $SU(2)_L$, $U(1)_Y$ 
et \Cyan{$U(1)_{EM}$}
(ce dernier, l'op\'erateur de charge \EM,
n'est pas ind\'ependant des autres,
mais je le consid\`ere aussi 
dans l'espoir de rendre les choses plus claires)
nous obtenons:

\bea
\label{casc}
\mbox{}\hspace{-2em}
{\rm Tr}\left(g_s^2\lambda^2\right)
& = &
g_s^2~ 
\underbrace{{\rm tr}\left(\lambda^2\right)}_{\displaystyle2}~
\left[ 0 + 0 + 2 + 1 + 1 \right]~
=~
{2}~ g_s^2
\\\mbox{}\hspace{-2em}
\label{casw}
{\rm Tr}\left(g^2\sigma^2\right)
& = &
g^2~ 
\underbrace{{\rm tr}\left(\sigma^2\right)}_{\displaystyle2}~
\left[ 1 + 0 + 3 + 0 + 0 \right]~
=~
{2}~ g^2
\\\mbox{}\hspace{-2em}
\label{cash}
{\rm Tr}\left(g'^2Y^2\right)
& = &
g'^2~ 
\left[ 2\left(-{1\over2}\right)^2 + \left({1}\right)^2 + 6\left({1\over6}\right)^2 + 3\left(-{2\over3}\right)^2 + 3\left({1\over3}\right)^2 \right]~
=~
{10\over3}~ g'^2\hspace{1em}
\\\nonumber\\\mbox{}\hspace{-2em}
\label{case}
\Cyan{{\rm Tr}\left(e^2Q^2\right)}
& \Cyan{=} &
\Cyan{e^2~ 
\bigg[ 0 + \left(-1\right)^2 + \left(-{1}\right)^2 + 3\left({2\over3}\right)^2}
\\\nonumber&&\hspace{2em}
\Cyan{+ 3\left(-{1\over3}\right)^2 + 3\left(-{2\over3}\right)^2 + 3\left({1\over3}\right)^2 \bigg]~
=~
{16\over3}~ g'^2}
~,
\eea

o\`u $g_s$, $g$, $g'$ et $e$
sont respectivement
les constantes de couplage 
$SU(3)_c$, $SU(2)_L$, $U(1)_Y$
et la charge de l'\'electron.
Nous en d\'eduisons la valeur de la constante (\ref{casimir})
qui correspond \`a la repr\'esentation qui contient les fermions du \sm\

\beq
\label{qc}
C/2~
=~
\underbrace{g_s^2}_{\displaystyle 4\pi\alpha_3}
~=~ 
\underbrace{g^2}_{\displaystyle 4\pi\alpha_2}
~=~ 
\underbrace{{5\over3}g'^2}_{\displaystyle 4\pi\alpha_1}
~=~ 
\Cyan{{8\over3}e^2~}
\eeq

dont nous pouvons d\'eduire
la valeur de l'\wma,
qui est d\'efini par

\beq
\label{swgut}
\swt ~=~ {e^2 \over g^2} ~=~ {3\over8} ~=~ 0.375~
~.
\eeq

\`A premi\`ere vue
il s'agit-l\`a d'une catastrophe pour la \gu!
En effet,
aux \'energies o\`u nous les observons,
les constantes de couplage faibles et fortes
ne sont certainement pas \'egales
et l'\wma,
qui a \'et\'e mesur\'es de mani\`ere tr\`es pr\'ecise
\`a l'\'echelle de la masse du $Z^0$
au LEP et au SLC~\cite{pdg},
a une valeur num\'erique compl\`etement diff\'erente:

\beq
\label{zsw}
\swt(m_Z) ~=~ 0.2315 ~\pm~ 0.0004~
~.
\eeq

Fort heureusement,
les relations (\ref{qc},\ref{swgut})
ne sont valables qu'aux \'echelles d'\'energies $m_{\rm GUT}$
o\`u la sym\'etrie de \gu\ n'est pas bris\'ee.
Mais en-de\c c\`a de ces \'energies
les constantes de couplages \'evoluent diff\'eremment
en fonction de l'\'energie,
comme l'exige le \rg:

\bea
\label{rg}
&&
{1\over\alpha_i(\mu)}~
=~
{1\over\alpha_{\rm GUT}}~
+~
{\beta_i\over2\pi}~
\ln\left({m_{\rm GUT}\over\mu}\right)~
\\\nonumber\\\nonumber\\
&&
\left\{
\begin{array}{rcc@{~}c@{~}c@{~}c@{~}c@{~}ccrcc}
\beta_1&=&&&&\displaystyle{4\over3}~N_g&+&\displaystyle{1\over10}~N_H&=&\displaystyle{41\over10}&>&0 \\\\
\beta_2&=&-&\displaystyle{22\over3}&+&\displaystyle{4\over3}~N_g&+&\displaystyle{1\over6}~N_H&=&-\displaystyle{19\over6}&<&0 \\\\
\beta_2&=&-&{11}&+&\displaystyle{4\over3}~N_g&&&=&-{7~}&<&0 
\end{array}
\right.
\eea

Le premier terme des fonctions beta
(donn\'e en g\'en\'eral par $-11/3N$ pour un groupe $SU(N)$)
est une caract\'eristique unique des groupes de jauge non-ab\'elien.
Les termes suivants font intervenir
le nombre de g\'en\'erations de fermions $N_g$
et le nombre de doublets de Higgs \EW s $N_H$.
Dans le cadre de l'{\em hypoth\`ese du d\'esert},
qui suppose l'absence de ph\'enom\`enes nouveaux
entre les \'echelles \EW\ et de \gu,
ces nombres prennent les valeurs du \sm,
\`a savoir $N_g=3$ et $N_H=1$.
Ce sont les valeurs que j'adopterai dans ce qui suit.
Il est toutefois important de noter
qu'il ne s'agit-l\`a que d'une hypoth\`ese de travail.
Nous verrons au chapitre suivant 
comment la prise en compte de partenaires \susic s
peut avoir des cons\'equences importantes.

La \bc\ (\ref{qc}) est bien r\'ealis\'ee par les \'equations (\ref{rg}),
puisque \`a l'\'echelle de la \gu

\beq
\alpha_1(m_{\rm GUT})
~=~
\alpha_2(m_{\rm GUT})
~=~
\alpha_3(m_{\rm GUT})
~=~
\alpha_{\rm GUT}
~.
\eeq

\co
Au-del\`a de l'\'echelle de la \gu\
les constantes de couplage
$\alpha_1$, $\alpha_2$ et $\alpha_3$
ont toutes trois la m\^eme fonction beta
(d\'etermin\'ee par la structure du groupe de \gu\ $G$)
et \'evoluent donc en unisson.
On ne parle donc plus que d'une seule constante de couplage $\alpha_G$,
celle du groupe de \gu\ $G$,
qui d\'ecroit pour des \'energies croissantes
jusqu'\`a la masse de Planck.
Au-del\`a
plus aucune pr\'ediction n'est possible 
en l'absence d'une th\'eorie quantique de la gravitation.
\oc

Pour r\'esoudre les trois \'equations (\ref{rg})
nous disposons de \bcs\ \`a l'\'echelle de la masse du $Z$

\beq
\mu=m_Z~:
\quad
\left\{
\begin{array}{lccclcl}
\alpha_s & \approx & \displaystyle{1\over10} & = & \alpha_3 \\\\
\alpha & \approx & \displaystyle{1\over128} & = & \alpha_2~\swt 
& = & \displaystyle{3\over5}~\alpha_1~\cwt
\end{array}
\right.
\eeq

Il en r\'esulte pour (\ref{rg})
un groupe de 3 \'equations lin\'eaires \`a 3 inconnues
dont les solutions sont

\bea
m_{\rm GUT} &~\simeq~& 2\cdot10^{14} \mbox{ GeV} \nonumber\\
\alpha_{\rm GUT} &~\simeq~& {1\over42} \label{su5}\\
\swt(m_Z) &~\simeq~& 0.214~\nonumber
~.
\eea

Voil\`a qui est d\'ej\`a nettement mieux que le r\'esultat (\ref{swgut}),
mais cela reste encore toujours incompatible
avec le r\'esultat exp\'erimental (\ref{zsw}).
H\'el\`as,
m\^eme en faisant un calcul plus pr\'ecis 
\`a l'aide de fonctions beta \`a deux boucles
et tenant compte des erreurs 
sur les mesures de $\alpha_s$ et $\alpha$,
la pr\'ediction pour l'\wma\
reste inacceptable face au r\'esultat exp\'erimental (\ref{zsw}).
Comme nous le verrons plus loin,
de telles valeurs de $m_{\rm GUT}$ et $\alpha_{\rm GUT}$
impliquent par ailleurs des taux de d\'esint\'egration du proton
qui auraient d\'ej\`a \'et\'e observ\'es depuis belle lurette!

\co
Sans vouloir trop anticiper sur le prochain chapitre,
une autre fa\c con de pr\'esenter ce d\'esaccord
est expos\'ee dans la figure~\ref{fgut}a.
\oc

Est-ce donc la fin des \guts?
Bien s\^ur que non!
Le calcul pr\'ec\'edant se base enti\`erement sur 
{\bf l'hypoth\`ese du d\'esert},
qui est inh\'erente qu'au mod\`ele minimal de \gu\ $SU(5)$.
En fait,
il est tr\`es facile de peupler ce d\'esert
en rajoutant (arbitrairement) des repr\'esentations exotiques de quarks,
tel le quix {\bf6} ou le queight {\bf8}.
Si en outre on consid\`ere un groupe de \gu\ autre que $SU(5)$,
le m\'echanisme de \ssb\ s'effectue en g\'en\'eral
en cascades successives,
donnant lieu a un o\`u plusieurs
seuils d'\'energie compris entre les \'echelles de \gu\ et \EW s.
\`A chacun de ces seuils des nouvelles particules aqui\`erent leur masse,
et le d\'esert se transforme ainsi en jungle luxuriante\dots

Quel que soit la \gut\ \'etudi\'ee
il est clair que l'\'echelle d'\'energie $m_{\rm GUT}$
\`a laquelle cette sym\'etrie devient manifeste
doit \^etre tr\`es \'elev\'ee,
de l'ordre de $10^{15\pm3}$ GeV.
En effet,
comme les couplages \'evoluent logarithmiquement
en fonction de l'\'energie
dans les \'equations (\ref{rg}),
seule une tr\`es grande valeur de $m_{\rm GUT}$
permet l'unification des couplages.

\subsection{La d\'esint\'egration du proton}

Il est clair
que si des quarks et des leptons
appartiennent \`a une m\^eme \irrep\
du groupe de \gu\ $G$,
il y aura des vecteurs de jauge qui induiront des transitions
des uns aux autres.
\'Evidemment,
puisque ce genre de transition n'est pas observ\'e \`a basses \'energies
ces bosons de jauge doivent correspondre
aux sym\'etries de $G$ qui sont bris\'ees spontan\'ement,
et par cons\'equent
leur masses doivent \^etre de l'ordre de $m_{\rm GUT}$.
Pas \'etonnant qu'on ne les ai pas encore d\'ecouverts!

\begin{figure}[htb]
\unitlength.5mm
\SetScale{1.418}
\begin{boldmath}
\begin{center}
\hfill
\begin{picture}(40,30)(0,0)
\ArrowLine(00,30)(20,30)
\ArrowLine(00,15)(20,15)
\ArrowLine(40,30)(20,30)
\ArrowLine(40,15)(20,15)
\ArrowLine(00,00)(40,00)
\Photon(20,30)(20,15){1}{3}
\Text(-2,30)[r]{$d$}
\Text(-2,15)[r]{$u$}
\Text(-2,00)[r]{$u$}
\Text(42,30)[l]{$e^+$}
\Text(42,15)[l]{$\bar u$}
\Text(42,00)[l]{$u$}
\Text(22,22)[l]{$X$}
\end{picture}
\hfill
\begin{picture}(40,30)(0,0)
\ArrowLine(00,30)(20,30)
\ArrowLine(00,15)(20,15)
\ArrowLine(40,30)(20,30)
\ArrowLine(40,15)(20,15)
\ArrowLine(00,00)(40,00)
\Photon(20,30)(20,15){1}{3}
\Text(-2,30)[r]{$d$}
\Text(-2,15)[r]{$u$}
\Text(-2,00)[r]{$u$}
\Text(42,30)[l]{$\bar u$}
\Text(42,15)[l]{$e^+$}
\Text(42,00)[l]{$u$}
\Text(22,22)[l]{$Y$}
\end{picture}
\hfill
\begin{picture}(40,30)(0,0)
\ArrowLine(00,30)(10,22)
\ArrowLine(00,15)(10,22)
\ArrowLine(40,30)(30,22)
\ArrowLine(40,15)(30,22)
\ArrowLine(00,00)(40,00)
\Photon(10,22)(30,22){1}{4}
\Text(-2,30)[r]{$d$}
\Text(-2,15)[r]{$u$}
\Text(-2,00)[r]{$u$}
\Text(42,30)[l]{$\bar u$}
\Text(42,15)[l]{$e^+$}
\Text(42,00)[l]{$u$}
\Text(20,24)[b]{$Y$}
\end{picture}
\hfill\mbox{}
\end{center}
\end{boldmath}
\caption{
Diagrammes de Feynman 
d\'ecrivant la d\'esint\'egration du proton
dans le cadre de la \gu\ $SU(5)$.
L'\'etat initial est un proton
form\'e (en premi\`ere approximation)
de deux quarks $u$ et d'un quark $d$.
L'\'etat final consiste en un positron
et un m\'eson neutre.
}
\label{fpdec}
\end{figure}
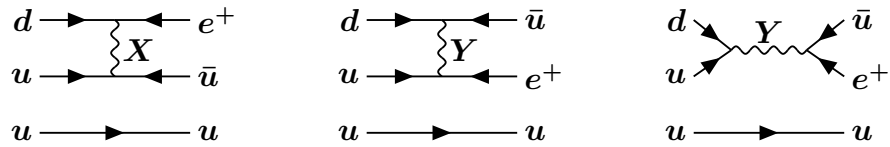

Mais en d\'epit de leur masses inaccessibles
ces bosons de jauge
peuvent se manifester \`a basse \'energie
en induisant la d\'esint\'egration du proton.
Les diagrammes typiques pour la \gu\ $SU(5)$
sont montr\'es dans la figure~\ref{fpdec}.

Il est ais\'e de se faire une id\'ee approximative 
de la largeur de d\'esint\'egration du proton
\`a travers ce m\'ecanisme.
Pour cela il suffit de remarquer
que l'\'echelle des \'energies impliqu\'ees
dans les processus de diffusion de la figure~\ref{fpdec}
est de l'ordre de la masse du proton $m_p$.
Ainsi les propagateurs des bosons \'echang\'es entre les fermions
$1/(p^2-m_{\rm GUT}^2)$,
o\`u $p \sim m_p \ll m_{\rm GUT}$,
peuvent sans danger \^etre remplac\'es par
$1/m_{\rm GUT}^2$.
(On parle alors d'une th\'eorie effective \`a basse \'energie.
Un exemple fameux et encore tr\`es utilis\'e
est la th\'eorie de Fermi des int\'eractions \EW s.)
Comme en outre les processus de la figure~\ref{fpdec}
font intervenir deux fois le couplage
des fermions aux vecteurs de jauge,
leur \'el\'ement de matrice 
est manifestement proportionnel \`a 
$\alpha_{\rm GUT} / m_{\rm GUT}^2$.
La largeur de d\'esint\'egration du proton
\'etant proportionnelle au carr\'e de cet \'el\'ement de matrice,
il est donc clair,
simplement sur la base de consid\'erations dimensionelles,
que l'ordre de grandeur de la dur\'ee de vie du proton 
est donn\'e par

\beq
\label{plt}
\tau_p~
\sim~
{1 \over \alpha_{\rm GUT}^2}~
{m_{\rm GUT}^4 \over m_p^5}~
~.
\eeq

Si l'on utilise les nombres obtenus pr\'ec\'edemment (\ref{su5})
on obtient une dur\'ee de vie du proton
d'environ $10^{29}$ ann\'ees.
De quoi faire p\^alir Mathusalem,
surtout si l'on se souvient que l'\^age de l'univers
exc\`ede \`a peine $10^{10}$ ans\dots

Il n'en reste pas moins que 
nous disposons de pas mal de protons autour de nous.
Ainsi un cube de $5\times5\times5$ m$^3$ de fer 
(\ie\ \`a peu pr\`es 1000 tonnes)
contient plus de $10^{32}$ protons,
ce qui permettrait en principe de d\'etecter
quelques milliers de d\'esint\'egrations par an.
Or les exp\'eriences
visant \`a d\'ecouvrir une d\'esint\'egration \'eventuelle du proton
ont jusqu'\`a pr\'esent fait chou blanc
et ont ainsi permis d'\'etablir une limite inf\'erieure 
de $10^{32}$ ans
pour la dur\'ee de vie du proton.
Bien s\^ur,
le r\'esultat (\ref{plt}) de notre grossi\`ere analyse dimensionelle
ne peut \^etre pris au pied de la lettre,
mais il s'av\`ere que des calculs bien plus sophistiqu\'es
qui tentent de tenir compte des fonctions d'onde hadroniques
confirment cette premi\`ere approximation.
Bien que ces calculs restent ent\^ach\'es d'une grande incertitude,
ils fournissent une \'evidence suppl\'ementaire
qu'une partie du moins du d\'esert 
qui s'\'etend de l'\'echelle \EW\ \`a l'\'echelle de \gu\
est peupl\'ee!

\subsection{Le probl\`eme de la hi\'erarchie}

\`A cause de la lenteur in\'evitable
de l'\'evolution logarithmique des couplages de jauge,
toutes les th\'eories de \gu\
pr\'edisent une \'echelle pour la brisure de la sym\'etrie de \gu\ 
$m_{\rm GUT}$ 
immens\'ement plus \'elev\'ee
que celle de la brisure de la sym\'etrie \EW\
$m_W$.
Ceci implique que les \vevs\ 
(et donc les masses)
des bosons de Higgs responsables des deux \ssb\ de l'\'equation (\ref{guts})
diff\`erent par environ 12 ordres de grandeur.

En principe
il ne s'agit pas-l\`a d'un probl\`eme fondamental,
si ce n'est qu'une telle diff\'erence d'\'echelles d'\'energies
ne peut gu\`ere \^etre qualifi\'ee de {\em naturelle}!
Mais il y a bien pire.
En effet,
les corrections radiatives aux masses des scalaires
contiennent des divergences quadratiques,
ce qui a des cons\'equences d\'esastreuses\dots

Ces corrections radiatives induisent une \'energie propre $\Sigma$
qui renormalise 
(au travers d'une resommation de Dyson)
la masse du scalaire $m_0$,
qui apparait dans le lagrangien.
Comme nous consid\'erons ici des th\'eories de jauge renormalisables,
les divergences contenues dans $\Sigma$ 
doivent \^etre des constantes
qui ne d\'ependent pas du moment du scalaire.
Elles peuvent donc \^etre soustraites et r\'eabsorb\'ees
dans une constante de renormalisation,
le contreterme $\delta m^2_0$.
La masse du scalaire $m$,
mesur\'ee \`a une \'echelle d'\'energie quelconque $\lambda$,
est ainsi donn\'ee par la relation

\beq
\label{se}
m^2(\lambda) ~=~ m^2_0 ~+~ \Sigma(\lambda,\Lambda) ~-~ \delta m^2_0
~,
\eeq

o\`u $\Lambda$ est l'\'echelle d'\'energie
\`a partir de laquelle la th\'eorie consid\'er\'ee devient caduque,
\ie\ $m_{\rm GUT}$ dans le cas de \guts,
ou $m_{\rm Planck}$ en g\'en\'eral.

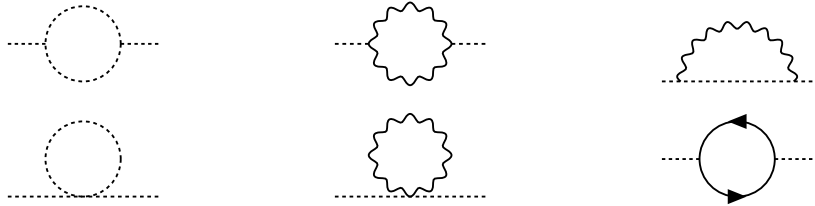
\begin{figure}[htb]
\unitlength.5mm
\SetScale{1.418}
\begin{boldmath}
\begin{center}
\hfill
\begin{picture}(40,30)(0,-10)
\DashLine(00,00)(10,00){1}
\DashLine(30,00)(40,00){1}
\DashCArc(20,00)(10,000,360){1}
\end{picture}
\hfill
\begin{picture}(40,30)(0,-10)
\DashLine(00,00)(09,00){1}
\DashLine(31,00)(40,00){1}
\PhotonArc(20,00)(10,-7.5,352.5){1}{12}
\end{picture}
\hfill
\begin{picture}(40,30)(0,0)
\DashLine(00,00)(40,00){1}
\PhotonArc(20,00)(15,000,180){1}{9.5}
\end{picture}
\hfill\mbox{}
\hfill
\begin{picture}(40,30)(0,0)
\DashLine(00,00)(40,00){1}
\DashCArc(20,10)(10,000,360){1}
\end{picture}
\hfill
\begin{picture}(40,30)(0,0)
\DashLine(00,00)(40,00){1}
\PhotonArc(20,11)(10,-7.5,352.5){1}{12}
\end{picture}
\hfill
\begin{picture}(40,30)(0,-10)
\DashLine(00,00)(10,00){1}
\DashLine(30,00)(40,00){1}
\ArrowArc(20,00)(10,000,180)
\ArrowArc(20,00)(10,180,360)
\end{picture}
\hfill\mbox{}
\end{center}
\end{boldmath}
\caption{
Diagrammes de Feynman typiques
intervenant dans l'\'energie propre
d'un scalaire.
}
\label{fqd}
\end{figure}

\`A une boucle
l'\'energie propre est d\'etermin\'ee
par des diagrammes de Feynman 
du type de ceux repr\'esent\'es dans la figure~\ref{fqd}.
Sa forme peut-\^etre estim\'ee
en appliquant les r\`egles de Feynman usuelles
et en comptant les puissances du moment circulant dans la boucle.
Les quatre derniers diagrammes 
donnent ainsi lieu \`a une correction du type

\bea
\Sigma(\lambda,\Lambda)~
&\sim&
g^2~
\int\limits_{\lambda}^{\Lambda}  d^4p~
{1 \over p^2}~
\\\nonumber
&\sim&
Q~
\int\limits_{\lambda^2}^{\Lambda^2}  dp^2
\quad+\quad
L~
\int\limits_{\lambda^2}^{\Lambda^2}  {dp^2 \over p^2}
\quad+\quad
F~
\int\limits_{\lambda^2}^{\Lambda^2}  {dp^2 \over p^4}
\\\nonumber
&\sim&
Q~
\left( \Lambda^2 - \lambda^2 \right)
\quad+\quad
L~
\ln{\Lambda^2 \over \lambda^2}
\quad+\quad
F~
\left( {1 \over \lambda^2} - {1 \over \Lambda^2} \right)
\\\nonumber\\\nonumber
&\simeq&
Q \Lambda^2
~,
\eea

o\`u $Q$ est un nombre sans dimensions
de l'ordre de grandeur de $g^2$.
Similairement,
par analyse dimensionelle
les coefficients du terme logarithmique et du terme fini
sont de l'ordre de
$L \sim g^2m^2$
et
$F \sim g^2m^4$.
En ins\'erant ce r\'esultat dans l'\'equation (\ref{se})
on obtient pour la masse du Higgs l\'eger
(responsable de la brisure de la sym\'etrie \EW)
mesur\'ee \`a l'\'echelle \EW\ $m_W$

\beq
\label{rm}
m^2(m_W) ~=~ Q ~ m_{\rm GUT}^2 ~-~ \left( \delta m^2_0 - m^2_0 \right)
~.
\eeq

La solution {\em naturelle} \`a cette \'equation
est d'avoir trois termes du m\^eme ordre de grandeur

\beq
\label{nat}
m^2(m_W) ~\sim~ Q ~ m_{\rm GUT}^2 ~\sim~ \left( \delta m^2_0 - m^2_0 \right)~
~.
\eeq

Mais l'existence d'un Higgs $10^{12}$ fois plus l\'eger
que l'\'energie de \gu\
implique une solution tr\`es artificielle

\beq
\label{unnat}
m^2(m_W) ~\ll~ Q ~ m_{\rm GUT}^2 ~\simeq~ \left( \delta m^2_0 - m^2_0 \right)~
~,
\eeq

qui ne peut s'obtenir qu'au prix d'un ajustement incroyablement pr\'ecis,
\`a mieux que douze ordres de grandeurs pr\`es,
entre les trois termes du c\^ot\'e droit de l'\'equation (\ref{rm}).
Ceci est une difficult\'e fondamentale,
commune \`a toutes les th\'eories de \gu,
qu'on l'appelle le {\em probl\`eme de la hi\'erarchie de jauge}.

En principe on peut imaginer 
qu'il est possible d'invoquer une sym\'etrie suppl\'ementaire
du lagrangien\footnote{
Je dis bien {\em en pricipe},
car jusqu'\`a pr\'esent
personne n'a pu impl\'ementer une telle sym\'etrie de mani\`ere convaincante.}
qui explique la quasi annulation 
dont r\'esulte cette hi\'erarchie.
Mais la catastrophe prend sa pleine ampleur
lorsqu'on constate que
les corrections radiatives d'ordre plus \'elev\'e
(\`a deux boucles ou plus)
briserons cette sym\'etrie.
M\^eme si cette brisure est tr\`es faible
et la valeur du coefficient $Q$ de la divergence quadratique
n'est que peu modifi\'ee,
la subtile annulation n\'ecessaire pour maintenir le Higgs \EW\ l\'eger
sera n\'eanmoins compl\`etement d\'etruite.
En fait,
un r\'eajustement du contre-terme $\delta m^2_0$
devrait donc \^etre r\'eeffectu\'e syst\'ematiquement
ordre par ordre en th\'eorie des perturbations
et l'annulation r\'esulterait ainsi d'un hasard
d'une improbabilit\'e inconcevable!

\bco{
\co
Le probl\`eme li\'e aux divergences quadratiques 
est en fait un peu plus subtil
que ce que j'ai d\'ecrit ici.
Apr\`es tout,
la divergence est une constante
qui ne d\'epend pas du moment du scalaire
et peut donc en principe \^etre soustraite
et r\'eabsorb\'ee dans une constante de renormalisation.
\oc

\co
En principe,
si les deux secteurs de Higgs,
responsables respectivement de la \ssb\ de \gu\ et \EW,
sont parfaitement ind\'ependants,
certains esprits tortur\'es pourraient peut-\^etre concevoir 
une telle diff\'erence dans les \'echelles de masses.
Malheureusement,
les deux secteurs de Higgs ne peuvent pas \^etre ind\'ependants,
justement parce que les corrections radiatives 
les recouplent n\'ecessairement.
Il faut donc invoquer
ordre par ordre
une annulation extr\`emement subtile
de diff\'erents termes dans les potentiels des scalaires de Higgs,
pour maintenir la \vev\ responsable de la brisure de la sym\'etrie \EW\
\`a quelques centaines de GeV,
alors que la \vev\ responsable de la brisure de la sym\'etrie de \gu\
doit croiser \`a quelques $10^{15}$ GeV.
\oc
}\eco

\co
Il y a donc en fait deux probl\`emes de hi\'erarchie:
\begin{enumerate}
\item
Le probl\`eme {\bf fondamental} de la hi\'erarchie,
qui est li\'e \`a l'existence de deux \'echelles d'\'energies
s\'epar\'ees par 12 ordres de grandeurs
et dont la solution requi\`ere
un ajustement de param\`etre
d'une pr\'ecision ahurissante.
Jusqu'\`a pr\'esent 
aucune solution satisfaisante \`a cette \'enigme 
n'a \'et\'e trouv\'ee.
\item
Le probl\`eme {\bf technique} de la hi\'erarchie,
qui est li\'e \`a la n\'ecessit\'e
de r\'eeffectuer cet ajustement r\'ep\'etivement,
\`a chaque ordre des perturbations.
Cette difficult\'e est \'evit\'ee par deux th\'eories:
\begin{enumerate}
\item 
La {\bf techni-couleur},
postule l'existence d'une nouvelle interaction forte
dict\'ee par une sym\'etrie de jauge non-ab\'elienne.
Cette force de techni-couleur est asymptotiquement libre,
mais elle confine des {\em techni-fermions}
\`a des \'energies inf\'erieures \`a environ 1 TeV,
\`a l'instar de la QCD
qui confine les quarks en-de\c c\`a de 1 GeV.
Le r\^ole du Higgs est jou\'e
par un \'etat compos\'e de {\em techni-m\'eson}
et l'on parle d'une {\em brisure dynamique} de la sym\'etrie \EW.

Comme la coupure d'\'energie $\Lambda$
devient l'\'echelle de compositivit\'e,
on peut donc contempler la solution naturelle (\ref{nat}),
o\`u $m_{\rm GUT}$ est remplac\'e par $m_{\rm TC}\sim1$ TeV.
En raison des nombreuses complications 
qui ont graduellement du \^etre impl\'ement\'ees
pour ne pas contredire les donn\'ees exp\'erimentales
de plus en plus pr\'ecises,
cette th\'eorie n'est \`a l'heure actuelle 
plus en odeur de saintet\'e.
\item 
La {\bf \susy},
que j'introduirai au chapitre suivant,
a pour cons\'equence d'annuler identiquement 
le coefficient $Q$ des divergences quadratiques.
Comme les corrections logarithmiques qui subsistent
sont beaucoup plus petites,
nous pouvons aussi contempler la solution naturelle (\ref{nat}),
o\`u $Q m_{\rm GUT}^2$ est remplac\'e par 
$L\ln{m_{\rm GUT}^2/m_W^2} \sim m^2$.
\end{enumerate}
\end{enumerate}
\oc

\section[$SU(5)$]{\boldmath$SU(5)$}

Avant de nous embarquer plus avant 
dans la description de quelques aspects plus pr\'ecis
de la \gu\ $SU(5)$
il est peut-\^etre utile de d\'ecrire 
quelques repr\'esentations de ce groupe.
Pour ce faire 
je montre dans le tableau~\ref{tsu5}
les d\'ecompositions 
en repr\'esentations de $SU(3) \otimes SU(2)$
des \irreps\ les plus importantes de $SU(5)$.

\renewcommand{\arraystretch}{1.5}
\begin{table}[htb]
$$
\begin{array}{||c||c|c||}
\hline\hline
\mbox{\irrep} 
& SU(5)
& SU(3) \otimes SU(2)
\\
\hline\hline
\mbox{singulet}
& {\bf1}
& ({\bf1},{\bf1})
\\
\hline
\mbox{fondamentale}
& {\bf5}
& ({\bf3},{\bf1}) \oplus ({\bf1},{\bf2})
\\
\hline
\mbox{fondamentale conjugu\'ee}
& {\bf\bar5}
& ({\bf\bar3},{\bf1}) \oplus ({\bf1},{\bf2})
\\
\hline
\mbox{tenseur antisym\'etrique}
& {\bf10}
& ({\bf\bar3},{\bf1}) \oplus ({\bf3},{\bf2}) \oplus ({\bf1},{\bf1})
\\
\hline
\mbox{tenseur sym\'etrique}
& {\bf15}
& ({\bf6},{\bf1}) \oplus ({\bf3},{\bf2}) \oplus ({\bf1},{\bf3})
\\
\hline
\mbox{adjointe}
& {\bf24}
& ({\bf8},{\bf1}) \oplus ({\bf\bar3},{\bf2}) \oplus ({\bf3},{\bf2})
\oplus ({\bf1},{\bf3}) \oplus ({\bf1},{\bf1})
\\
\hline\hline
\end{array}
$$
\caption{
Quelques \irreps\ du groupe $SU(5)$
et leur d\'ecompositions en repr\'esentations du groupe $SU(3) \otimes SU(2)$.
}
\label{tsu5}
\end{table}
\renewcommand{\arraystretch}{1}

Il est aussi utile de noter
les d\'ecompositions en \irreps\
de quelques produits de \irreps:

\bea
{\bf5} \otimes {\bf\bar5} &=& {\bf1} \oplus {\bf24} \\
\label{5x5}
{\bf\bar5} \otimes {\bf\bar5} &=& {\bf\bar{10}} \oplus {\bf\bar{15}} \\
{\bf\bar5} \otimes {\bf10} &=& {\bf{5}} \oplus {\bf{45}} \\
{\bf10} \otimes {\bf10} &=& {\bf\bar{5}} \oplus {\bf\bar{45}} \oplus {\bf{50}} ~
~.
\eea

\subsection{Les bosons de jauge}

Les $8+3+1=12$ bosons de jauge du \sm\
s'assemblent en 3 \irreps\
de $SU(3)_c \otimes SU(2)_L$,
qui sont univoquement d\'efinies par leurs \qns,
d\'efinis dans le tableau~\ref{tsm}:

\beq
\label{smrepg}
\hspace{-0em}
\begin{array}{c@{\quad\oplus\quad}c@{\quad\oplus\quad}c}
g_a ~ {\scriptstyle(a=1\dots8)}
&
W_i ~ {\scriptstyle(=W^\pm,Z)}
&
B
\\\multicolumn{3}{c}{}\\
({\bf8},{\bf1})
&
({\bf1},{\bf2})
&
({\bf1},{\bf1})
\end{array}
\eeq

Afin d'avoir affaire \`a une th\'eorie de jauge,
ces vecteurs doivent n\'ecessairement s'ins\'erer
dans la repr\'esentation adjointe {\bf24} de $SU(5)$.
Comme on le voit \`a l'aide du tableau~\ref{tsu5},
cela ne pose aucun probl\`eme\footnote{
\'Evidemment,
puisque $SU(3) \otimes SU(2) \subset SU(5)$.
}.
Nous pouvons donc d\'ej\`a affirmer d'embl\'e
qu'en plus des $8+3+1=12$ bosons de jauge du \sm\
il doit y avoir 12 autres bosons de jauge.
Ces derniers,
qu'on d\'enote g\'en\'eralement $X$ et $Y$,
sont inobservables
car ils aqui\`erent une masse de l'ordre de $m_{\rm GUT}$
par \ssb.

En terme de composantes
on repr\'esente g\'en\'eralement 
la repr\'esentation adjointe de la mani\`ere suivante

\bea
\label{adj}
A({\bf24})
&=&
\sum_\alpha {1\over2} \lambda_\alpha A_\alpha~
\\\nonumber
&=&
{1\over\sqrt{2}}
\left(
\begin{array}{ccc|cc} 
&&&    \bar{X_1}& \bar{Y_1} \\
& g_a && \bar{X_2}& \bar{Y_2}\\
&&&    \bar{X_3}& \bar{Y_3}\\
\hline
X_1 & X_2 & X_3 \\
Y_1 & Y_2 & Y_3 & \multicolumn{2}{c}{\raisebox{1.5ex}[-1.5ex]{$W_i$}}\\
\end{array}
\right)~
+~
{1\over2\sqrt{15}}
\left(
\begin{array}{ccc|cc} 
-2&&&& \\ &-2&&& \\ &&-2&& \\
\hline
&&&+3 \\ &&&&+3
\end{array}
\right)~B~
~,
\eea

o\`u les matrices $\lambda$ 
sont les g\'en\'eralisations des matrices de Gell-Mann 
(ou de Pauli)
\`a 5 dimensions.
J'ai utilis\'e des barres horizontales et verticales 
pour aider \`a visualiser les d\'ecompositions
en octet, triplets et doublets de $SU(3) \otimes SU(2)$.

\subsection{Les fermions}

Les 15 fermions du \sm\ s'assemblent en 5 \irreps\
de $SU(3)_c \otimes SU(2)_L$,
qui sont univoquement d\'efinies par leurs \qns,
d\'efinis dans le tableau~\ref{tsm}:

\beq
\label{smrep}
\hspace{-0em}
\begin{array}{c@{\quad\oplus\quad}c@{\quad\oplus\quad}c@{\quad\oplus\quad}c@{\quad\oplus\quad}c}
\left(\begin{array}{c}\nu_{L}\\\ell_L\end{array}\right)
&
\ell_L^c
&
\left(\begin{array}{c}u_{L}\\ d_L\end{array}\right)
&
u_L^c
&
d_L^c
\\\multicolumn{5}{c}{}\\
({\bf1},{\bf2})
&
({\bf1},{\bf1})
&
({\bf3},{\bf2})
&
({\bf\bar3},{\bf1})
&
({\bf\bar3},{\bf1})
\end{array}
\eeq

Au lieu d'utiliser des fermions droits $\psi_R$,
comme il est de coutume dans le cadre du \sm,
j'ai utilis\'e ici des fermions complexes conjugu\'es gauches $\psi_L^c$.
Ceci est uniquement un choix commode
dict\'e principalement par la tradition
et qui ne doit pas vous impressioner.
En effet,
il s'agit du m\^eme \dof\
puisque ces deux types de fermions
sont reli\'es par la bijection

\beq
\psi_L^c~=~C~\gamma_0^T~\psi_R^*~
~,
\eeq

o\`u $C$ est la matrice de conjugaison de charge.
J'ignorerai par la suite l'indice $L$.

Le but du jeu est maintenant d'ins\'erer les fermions (\ref{smrep}) du \sm\
dans des \irreps\ de $SU(5)$
\`a l'aide du tableau~\ref{tsu5}.
L'id\'eal serait \'evidemment d'avoir ces 15 fermions
dans une \irrep.
Malheureusement la \irrep\ ${\bf15}$ ne se pr\^ete absolument pas \`a cela
en raison de sa d\'ecomposition $SU(3) \otimes SU(2)$ incompatible.
La mani\`ere la plus \'economique de repr\'esenter les fermions
est de les inclure dans une 
${\bf\bar5} = ({\bf\bar3},{\bf1}) \oplus ({\bf1},{\bf2})$ 
et une 
${\bf10} = ({\bf\bar3},{\bf1}) \oplus ({\bf3},{\bf2}) \oplus ({\bf1},{\bf1})$,
ce qui permet en effet de reproduire la d\'ecomposition (\ref{smrep}).

En terme de composantes
on repr\'esente g\'en\'eralement ces \irreps\ de la mani\`ere suivante:

\beq
\label{repcomp}
\psi({\bf\bar5})
=
\left(
\begin{array}{c} d_1^c \\ d_2^c \\ d_3^c \\\hline \ell \\ -\nu \end{array}
\right)
\qquad
\chi({\bf10})
=
{1\over\sqrt{2}}
\left(
\begin{array}{ccc|cc} 
0 & u_3^c & -u_2^c & -u_1 & -d_1 \\
-u_3^c & 0 & u_1^c & -u_2 & -d_2 \\
u_2^c & -u_1^c & 0 & -u_3 & -d_3 \\
\hline
u_1 & u_2 & u_3 & 0 & -\ell^c \\
d_1 & d_2 & d_3 & \ell^c & 0 \\
\end{array}
\right)~
~,
\eeq

o\`u j'ai \`a nouveau utilis\'e des barres horizontales et verticales 
pour aider \`a visualiser les d\'ecompositions
en triplets, doublets et singulets de $SU(3) \otimes SU(2)$.
Comme je l'avais indiqu\'e \`a la section \ref{chquant}
en traitant de la quantification de la charge \EM,
on voit bien que la somme des charges de chacun de ces deux multiplets
est nulle.
Le nombre de couleurs de la QCD fournit ainsi une attrayante explication
de la charge $1/3$ des quarks.

\co
En tout il existe trois copies de ces deux multiplets 
qui correspondent aux trois g\'en\'erations de fermions.
\`A ma connaissance
il n'existe actuellement aucune \gut\ attractive
qui contienne ces trois g\'en\'erations
dans une unique \irrep\
et expliquerait ainsi la triplication.
\oc

Le lagrangien fermionique s'\'ecrit

\bea
\label{lf}
{\cal L}
&=&
\bar\psi \gamma^\mu D_\mu \psi~
+~
{\rm Tr} \left( \bar\chi \gamma^\mu D_\mu \chi \right)~
\\\nonumber
&=&
\bar\psi^a \gamma^\mu (D_\mu)^b_a \psi_b~
+~
\bar\chi^{ab} \gamma^\mu (D_\mu)^c_b \chi_{ca} 
\qquad
(a,b,c = 1\dots5)
~,
\eea

o\`u les d\'eriv\'ees covariantes sont donn\'ees par

\bea
\label{cdpsi}
D_\mu \psi
&=&
\partial_\mu \psi~
+~
ig~ \underbrace{A_\mu^\alpha {1\over2}\lambda_\alpha \psi}_{\displaystyle A_\mu \psi}
\\\nonumber\\
\label{cdchi}
D_\mu \chi
&=&
D_\mu (xy-yx)~
\\\nonumber
&=&
(D_\mu x)y + x(D_\mu y) - (D_\mu y)x - y(D_\mu x)
\\\nonumber
&=&
\partial_\mu \chi~
+~
ig~ \underbrace{(A_\mu x)y + x(A_\mu y) - (A_\mu y)x - y(A_\mu x)}_{%
\displaystyle
(A_\mu)^c_d \chi_{cd} - (A_\mu)^c_d \chi_{cd} ~\equiv~ 2 A_\mu \chi
}
~.
\eea

L'expression pour la d\'eriv\'ee covariante de $\psi$ (\ref{cdpsi})
est \'evidente en vertu 
de la d\'efinition de la repr\'esentation fondamentale
et de l'\'equation (\ref{adj}).
Pour trouver la d\'eriv\'ee covariante de $\chi$ (\ref{cdchi})
j'ai exprim\'e ce champ
sous la forme du produit direct antisym\'etris\'e 
de deux repr\'esentations fondamentales $x$ et $y$
(\ref{5x5})
$\chi_{ab} = x_a y_b - x_b y_a$.

En ins\'erant ces d\'eriv\'ees covariantes explicitement 
dans le lagrangien fermionique (\ref{lf})
on retrouve donc automatiquement 
les interactions fortes dans les combinaisons des indices
$a,b,c = 1\dots3$
et les interactions \EW s dans les combinaisons des indices
$a,b,c = 1\dots2$.
Les combinaisons mixtes d'indices
donnent lieu \`a de nouvelles interactions.
\Eg,
des combinaisons $a=1,2,3$ et $b=4$ dans le lagrangien de $\psi$

\beq
ig ~
\left(~
\bar\psi^1 \gamma^\mu (A_\mu)^4_1 ~+~ 
\bar\psi^2 \gamma^\mu (A_\mu)^4_2 ~+~
\bar\psi^3 \gamma^\mu (A_\mu)^4_3 
~\right)~
\psi_4~
=~
{ig\over\sqrt{2}}~
X_\mu~
\bar d^c \gamma^\mu \ell~
\eeq

et des combinaisons $a=4$, $b=1,2,3$ et $c=5$ dans le lagrangien de $\chi$

\beq
2ig ~
\left(~
\bar\chi^{41} \gamma^\mu (A_\mu)^5_1 ~+~ 
\bar\chi^{42} \gamma^\mu (A_\mu)^5_2 ~+~
\bar\chi^{43} \gamma^\mu (A_\mu)^5_3 
~\right)~
\chi_{54}~
=~
-{ig\over\sqrt{2}}~
\bar Y_\mu~
\bar u \gamma^\mu \ell^c~
\eeq

r\'esultent des transitions lepton-quark
du type de celles repr\'esent\'ees dans la figure~\ref{fpdec}.
C'est \`a cause de ce type de transitions
interdite dans le cadre du mod\`ele standard
qu'on appelle les 12 bosons de jauge $X$ et $Y$
des {\bf leptoquarks}.

\subsection{Les bosons de Higgs}

Il y a un grand nombre de possibilit\'es 
pour impl\'ementer le secteur de Higgs
et inclure le doublet de Higgs du \sm\ 
dans une repr\'esentation de $SU(5)$.
Une chose est n\'eanmoins certaine,
au moins deux \irreps\ de scalaires
seront n\'ec\'essaires
pour briser d'abord la sym\'etrie $SU(5)$
et puis la sym\'etrie $SU(2) \otimes U(1)$.
La fa\c con la plus \'economique de faire cela 
est d'invoquer l'existence d'un Higgs $\Phi$ adjoint {\bf24}

\beq
\label{Phi}
\Phi({\bf24})~
=~
\sum_\alpha {1\over{2}} \lambda_\alpha \Phi_\alpha~
\eeq

et d'un Higgs $H$ fondamental {\bf5}

\beq
\label{H}
H({\bf5})
=
\left(
\begin{array}{c} H_1 \\ H_2 \\ H_3 \\\hline \phi^+ \\ \phi^0 \end{array}
\right)~
~.
\eeq

Le lagrangien de Higgs s'\'ecrit donc

\beq
\label{lh}
{\cal L}~
=~
{\rm Tr} |D_\mu \Phi|^2~
+~
|D_\mu H|^2~
-~
V(\Phi,H)~
~,
\eeq

o\`u les d\'eriv\'ees covariantes sont donn\'ees par

\bea
\label{cdH}
D_\mu H
&=&
\partial_\mu H~
+~
ig~ \underbrace{A_\mu^\alpha {1\over2}\lambda_\alpha H}_{\displaystyle A_\mu H}
\\\nonumber\\
\label{cdPhi}
D_\mu \Phi
&=&
\partial_\mu \Phi~
+~
ig~ \underbrace{A_\mu^\alpha T_\alpha \Phi}_{\displaystyle [A_\mu,\Phi]}~
~.
\eea

L'expression pour la d\'eriv\'ees covariante de $H$ (\ref{cdH})
est \'evidente en vertu 
de la d\'efinition de la repr\'esentation fondamentale
et de l'\'equation (\ref{adj}).
Le commutateur dans la d\'eriv\'ees covariante de $\Phi$ (\ref{cdPhi})
s'obtient ais\'ement
si l'on se souvient que le les matrices de la repr\'esentation adjointe
sont donn\'ees par les constantes de structure
$T_\alpha^{\beta\gamma} = i f_{\alpha\gamma\beta}$
qui d\'efinissent l'alg\`ebre de Lie
$[\lambda_\alpha,\lambda_\beta] = 2i f_{\alpha\beta\gamma} \lambda_\gamma$.

Sous des conditions assez g\'en\'erales,
le potentiel de Higgs
prend sa valeur minimale
pour des valeurs non-triviales des champs scalaires.
Il s'av\`ere que leurs \vevs\ 
prennent les formes suivantes

\bea
\label{vevPhi}
\langle \Phi \rangle
&\equiv&
v_\Phi~
{1\over\sqrt{15}}
\left(
\begin{array}{ccc|cc} 
2&&&& \\ &2&&& \\ &&2&& \\
\hline
&&&-3 \\ &&&&-3
\end{array}
\right)
\\
\label{vevH}
\langle H \rangle
&\equiv&
v_H~
\left(
\begin{array}{c} 0\\0\\0\\\hline0\\1 \end{array}
\right)~
~.
\eea

Pour d\'eterminer les masses des vecteurs de jauge
il suffit \`a pr\'esent d'ins\'erer ces \vevs\ 
dans la partie cin\'etique du lagrangien de Higgs (\ref{lh}).
En se concentrant uniquement sur les termes qui impliquent les
carr\'es des \vevs\ 
nous obtenons ainsi pour $\Phi$ 

\bea
{\cal L}
&=&
\dots~+~
g^2~ {\rm Tr}|A\langle \Phi \rangle-\langle \Phi \rangle A|^2
\\\nonumber
&=&
\dots~+~
{5\over6}~ g^2 v_\Phi^2~
\left(~ |X_1|^2 + |X_2|^2 + |X_3|^2 + |Y_1|^2 + |Y_2|^2 + |Y_3|^2 ~\right)~
\eea

et pour $H$ 

\bea
{\cal L}
&=&
\dots~+~
g^2~ |A\langle H \rangle|^2
\\\nonumber
&=&
\dots~+~
{1\over2}~ g^2 v_H^2~
\Bigg[~
|Y_1|^2 + |Y_2|^2 + |Y_3|^2 
+ |W|^2
+ {1\over2}\underbrace{\Big(-W_3 + {3\over\sqrt{15}}B\Big)^2}_%
{\displaystyle {1\over\cwt}Z^2}
~\Bigg]~
~,
\eea

o\`u j'ai fait usage de la valeur de l'\wma\ 
$\swt=3/8$
\`a l'\'echelle de la \gu\ (\ref{swgut})
et de la d\'efinition des \'etats propres de masse
$W$ et $Z$ (\ref{mixing}).
Les masses des vecteurs de jauge sont donc

\bea
\label{mxy}
&& m_Y ~\simeq~ m_X ~=~ \sqrt{5\over6}~ g~ v_\Phi \\
&& m_W ~=~ \cw m_Z ~=~ {1\over\sqrt{2}}~ g~ v_H~
~.
\eea

Comme les vecteurs $W$ et $Z$ 
doivent \^etre environ $10^{12}$ fois plus l\'egers
que les leptoquarks $X$ et $Y$,
nous en d\'eduisons la hi\'erarchie des \vevs\ 

\beq
\label{hier}
{m_X \over m_W} ~\sim~ {v_\Phi \over v_H} ~\sim~ 10^{12}~
~,
\eeq

ce qui justifie donc pleinement l'approximation 
dans l'\'equation (\ref{mxy}).

\chapter{La \susy}

La \susy\ est une sym\'etrie qui \'etablit une relation
entre les fermions et les bosons.
Elle permet de passer des uns aux autres,
comme les transformations de jauge 
permettent de passer d'un champs \`a l'autre
\`a l'int\'erieur d'une m\^eme repr\'esentation du groupe de jauge.
\`A l'instar de ces multiplets de jauge,
on parlera ainsi de super-multiplets de la \susy,
qui contiennent \`a la fois de fermions et des bosons.

J'introduirai ici la \susy\
de la mani\`ere la plus pi\'etonne,
\ie\ \`a partir du lagrangien de Wess-Zumino,
qui est le lagrangien \susic\ le plus simple qui soit.
Il me permettra entre autres
de justifier l'alg\`ebre \susic\
et de montrer explicitement comment disparaissent les divergences quadratiques
dans la renormalisation des masses des scalaires,
qui s'\'etaient montr\'ees si n\'efastes \`a la \gu.
Je d\'ecrirai ensuite bri\`evement
l'extension \susic\ minimale du \sm\
et montrerai comment la \susy\ s'associe bien \`a la \gu.

Toutefois,
avant de nous embarquer plus avant
il est peut-\^etre utile de rappeler 
quelques propri\'et\'es du groupe de Lorentz,
\ie\ du groupe des rotations de l'espace-temps.
Ses \irreps\ sont 
les scalaires
(\`a 1 composante), 
les spineurs
(\`a 2 composantes), 
les vecteurs
(\`a 4 composantes)
\etc.

\co
L'alg\`ebre du groupe de Lorentz $SO(1,3)$ s'av\`ere \^etre identique 
\`a celle du groupe plus familier
$SU(2) \otimes SU(2)$.
Il est tr\`es facile de former ses \irreps\
\`a partir des \qns\ de spin de chaque composante $SU(2)$.
Ainsi $(0,0)$ correspond aux scalaires,
$(1/2,0)$ aux spineurs,
$(0,1/2)$ aux spineurs conjugu\'es complexes,
$(1/2,1/2)$ aux vecteurs
\etc.
Les spineurs de Dirac se transforment comme la repr\'esentation r\'eductible
$(1/2,0) \oplus (0,1/2)$.
\oc

Alors que les scalaires et les vecteurs sont bien connus,
il n'en est peut-\^etre pas de m\^eme pour les spineurs \`a 2 composantes.
En effet les spineurs familiers sont ceux de Dirac,
qui ont 4 composantes
mais ne forment pas une \irrep\ du groupe de Lorentz.
Par contre les spineurs de Majorana,
qui sont d\'efinis comme \'etant leur propre conjugu\'e de charge
(attention, pas leur conjugu\'e complexe!)

\beq
\label{maj}
\psi^c = \psi~
~,
\eeq

n'ont que 2 \dofs\ et forment bien \irrep\ du groupe de Lorentz.
Par la suite j'utiliserai leurs propri\'et\'es suivantes

\bea
\bar\psi_1\psi_2 &=& \bar\psi_2\psi_1 \\
\bar\psi_1\gamma_\mu\psi_2 &=& -\bar\psi_2\gamma_\mu\psi_1 \\
\bar\psi_1\gamma_5\psi_2 &=& \bar\psi_2\gamma_5\psi_1 \\
\bar\psi_1\gamma_\mu\gamma_5\psi_2 &=& \bar\psi_2\gamma_\mu\gamma_5\psi_1 \\
\bar\psi_1\gamma_\mu\gamma_\nu\psi_2 &=& -\bar\psi_2\gamma_\mu\gamma_\nu\psi_1~
~.
\eea

\section{Le mod\`ele de Wess-Zumino}

Consid\'erons le lagrangien libre 
que Wess et Zumino ont introduit en 1974~\cite{wz}

\beq
\label{wzf}
{\cal L}_{WZ}~
=~
{1\over2} \left(\partial^\mu A\right) \left(\partial_\mu A\right)~
+~
{1\over2} \left(\partial^\mu B\right) \left(\partial_\mu B\right)~
+~
{i\over2} \bar\psi \gamma^\mu\partial_\mu \psi~
~,
\eeq

o\`u $A$ est un champ scalaire,
$B$ est un champ pseudo-scalaire
et $\psi$ est un champ spineur de Majorana.
Ceci est la forme forme la plus simple du lagrangien de Wess-Zumino.
Plus loin j'introduirai aussi des masses et des interactions.
Aussi anondin qu'il puisse para\^\i tre,
ce lagrangien poss\`ede des propri\'et\'es remarquables
que je vais d\'ecrire d'ici peu.

\subsection{Les transformations de Poincar\'e}

Le lagrangien (\ref{wzf}) est manifestement un scalaire 
pour les transformations du groupe de Poincar\'e,
\ie\ du groupe des transformations rigides
(rotations et translations) 
d'espace-temps.
Pour ce qui suit
il est n\'eanmoins instructif de v\'erifier ce fait explicitement,
\eg\ dans le cas de translations infinit\'esimales
d'amplitude $\xi^\mu$.
Les variations infinit\'esimales correspondantes des trois champs
s'\'ecrivent en fonction du g\'en\'erateur des translations $P_\mu$

\bea
\label{translf}
\delta A &=& \xi^\mu P_\mu ~ A \\
\delta B &=& \xi^\mu P_\mu ~ B \\
\delta \psi &=& \xi^\mu P_\mu ~ \psi ~
~.
\eea

Comme le g\'en\'erateur des translations
peut s'\'ecrire explicitement 
sous la forme de l' op\'erateur diff\'erentiel
$P_\mu = i\partial_\mu$,
la variation du lagrangien (\ref{wzf}) est donc

\bea
\label{transll}
\delta {\cal L}_{WZ} 
&=&
{1\over2} \left(\partial^\mu \delta A\right) \left(\partial_\mu A\right)~
+~
{1\over2} \left(\partial^\mu A\right) \left(\partial_\mu \delta A\right)~
+~\cdots
\\\nonumber
&=&
{i\over2}\xi^\nu~
\Big[~
\left(\partial^\mu \partial_\nu A\right) \left(\partial_\mu A\right)~
+~
\underbrace{
\left(\partial^\mu A\right) \left(\partial_\mu \partial_\nu A\right)
}_{\makebox[0em][c]{$\partial_\mu\left(\partial^\mu A\partial_\nu A\right)
- \left(\partial_\nu\partial^\mu A\right)\left(\partial_\mu A\right)$}}~
+~\cdots
~\Big]
\\\nonumber
&=&
\partial_\nu\left[i\xi^\nu {\cal L}_{WZ} \right]~
~.
\eea

Puisque $\delta {\cal L}_{WZ}$ est une d\'eriv\'ee totale,
son int\'egrale est nulle
et l'action est donc bien invariante sous l'effet d'une translation
d'espace-temps.
Le m\^eme genre de proc\'edure
est bien s\^ur aussi applicable pour les transformations de Lorentz.

\subsection{Les transformations \susic s}

Tout cela \'etait bien connu.
Mais le lagrangien (\ref{wzf}) est aussi invariant
(\`a une d\'eriv\'ee totale pr\`es)
sous l'effet d'une transformation infinit\'esimales bien moins
\'evidente:

\bea
\label{susyf1}
\delta A &=& \bar\epsilon~\psi \\
\label{susyf2}
\delta B &=& \bar\epsilon~ i\gamma_5~\psi \\
\label{susyf3}
\delta \psi &=& i \gamma^\mu\partial_\mu (A+i\gamma_5 B) ~\epsilon~
~,
\eea

o\`u $\epsilon$ est un fermion de Majorana infinit\'esimal constant.
Le calcul de la variation de ${\cal L}_{WZ}$
est maintenant un peu moins trivial

\bea
\label{susyl}
\delta {\cal L}_{WZ} 
&=&
\left(\partial^\mu A\right) \left(\partial_\mu \delta A\right)~
+~
\left(\partial^\mu B\right) \left(\partial_\mu \delta B\right)~
+~
{i\over2} \left(\delta\bar\psi\right) \partial_\mu\psi~
+~
{i\over2} \bar\psi \partial_\mu\left(\delta\psi\right)~
\\\nonumber
&=&
\bar\epsilon
\left[
\partial^\mu (A+i\gamma_5B) 
~-~ 
{1\over2} \partial_\nu (A+i\gamma_5B) \gamma^\nu\gamma^\mu
\right]
\partial_\mu \psi
\\\nonumber
&&\qquad\qquad\qquad+~
{1\over2} \bar\psi \gamma^\mu\gamma^\nu 
\left[
\partial^\mu \partial^\nu (A+i\gamma_5B) 
\right]
\epsilon
\\\nonumber
&=&
\partial_\mu\{\cdots\} 
~-~ 
\bar\epsilon
\left\{\partial_\mu\left[
\partial^\mu (A+i\gamma_5B) 
~-~ 
{1\over2} \partial_\nu (A+i\gamma_5B) \gamma^\nu\gamma^\mu
\right]\right\}
\psi
\\\nonumber
&&\qquad\qquad\qquad-~
{1\over2} \bar\epsilon 
~\underbrace{\gamma^\mu\gamma^\nu}_%
{\makebox[0em][c]{$2\eta^{\mu\nu}-\gamma^\nu\gamma^\mu$}}~
\left[
\partial^\mu \partial^\nu (A+i\gamma_5B) 
\right]
\psi
\\\nonumber
&=&
\partial_\mu\{\cdots\}~
~.
\eea

Qui l'eut cru?
Les inf\^ames transformations (\ref{susyf1}--\ref{susyf3}})
refl\`etent donc bien une sym\'etrie 
du lagrangien (\ref{wzf})!
Il faut bien admettre
que ces transformations sont pour le moins inhabituelles.

\co
En fait,
si j'avais utilis\'e le formalisme du super-espace~\cite{oto}
(qui me prendrait trop de temps \`a introduire ici)
ces transformations \susic s
d\'ecouleraient de mani\`ere parfaitement logique et \'evidente
de la structure m\^eme de l'espace-temps.
Elles s'\'ecriraient alors sous une forme 
bien plus \'el\'egante et moins myst\'erieuse
que dans les \'equations (\ref{susyf1}--\ref{susyf3}).
\oc

\co
On peut consid\'erer les transformations \susic s 
comme formant le sommet d'une hi\'erarchie
de transformations.
En effet,
alors que les translations
agissent ind\'ependemment 
sur chaque composantes spinorielles ou vectorielles d'un champ,
les transformations de Lorentz
m\'elangent ces diff\'erentes composantes.
Par contre,
les transformations de Lorentz ne m\'elangent jamais diff\'erents champs,
contrairement aux transformations de jauge
qui par essence relient les diff\'erents \'el\'ements d'un m\^eme multiplet,
\ie\ d'un ensemble de champs de spin identique.
Finalement,
les transformations \susic s 
effectuent le dernier pas,
en fournissant un moyen de passer d'un spin \`a un autre!
\oc

\subsection{L'alg\`ebre \susic}

\`A l'instar du g\'en\'erateur des translations $P_\mu$,
il existe aussi un g\'en\'erateur des transformations \susic s $Q$.
\Eg,
la transformation (\ref{susyf1}) du scalaire $A$ 
peut par analogie avec l'\'equation (\ref{translf}) s'\'ecrire

\beq
\label{QA}
\delta A ~=~ \bar\epsilon Q A
~.
\eeq

Mais alors que $P_\mu$,
est un vecteur de Lorentz,
$Q$ est un spineur de Majorana.
Pour en apprendre plus sur lui,
examinons l'effet du commutateur 
de deux transformations (\ref{QA}) successives:

\bea
\label{ssgen1}
[\delta_2,\delta_1] A
&=&
[\underbrace{\bar\epsilon_2 Q}_{\displaystyle \bar Q\epsilon_2}
,\bar\epsilon_1 Q] A
\\\nonumber
&=&
\left(
\bar Q \epsilon_2 \bar\epsilon_1 Q - \bar\epsilon_1 Q \bar Q
\epsilon_2
\right) A
\\\nonumber
&=&
\Big(~
\bar\epsilon_{1\alpha} \epsilon_{2\beta} ~ \bar Q_\beta Q_\alpha
-
\bar\epsilon_{1\alpha} \epsilon_{2\beta} 
\underbrace{Q_\alpha \bar Q_\beta}_{\displaystyle-\bar Q_\beta Q_\alpha}
\Big) A
\\\nonumber
&=&
\bar\epsilon_{1\alpha} \epsilon_{2\beta}
\left\{ Q_\alpha,\bar Q_\beta \right\} A~
~.
\eea

J'ai utilis\'e dans l'avant derni\`ere ligne
la relation d'anticommutation habituelle de deux op\'erateurs fermioniques. 
D'autre part,
si l'on utilise la forme explicite de la transformation (\ref{susyf1})
ce m\^eme commutateur devient

\bea
\label{ssgen2}
[\delta_2,\delta_1] A
&=&
\delta_2 \left(\bar\epsilon_1 \psi\right) 
- 
\delta_1 \left(\bar\epsilon_2 \psi\right)
\\\nonumber
&=&
~-~ i \bar\epsilon_1 \gamma^\mu\partial_\mu(A+i\gamma_5 B) \epsilon_2
~+~ i \bar\epsilon_2 \gamma^\mu\partial_\mu(A+i\gamma_5 B) \epsilon_1
\\\nonumber
&=&
~-~i 
\Big( \bar\epsilon_1 \gamma^\mu \epsilon_2
- \underbrace{\bar\epsilon_2 \gamma^\mu \epsilon_1}_%
{\displaystyle-\bar\epsilon_1 \gamma^\mu \epsilon_2}
\Big)
\partial_\mu A
~-~i 
\Big( \bar\epsilon_1 \gamma^\mu\gamma_5 \epsilon_2
- \underbrace{\bar\epsilon_2 \gamma^\mu\gamma_5 \epsilon_1}_%
{\displaystyle\bar\epsilon_1 \gamma^\mu\gamma_5 \epsilon_2}
\Big)
\partial_\mu B
\\\nonumber
&=&
2i \bar\epsilon_1 \gamma^\mu \epsilon_2 \partial_\mu A
\\\nonumber
&=&
2 \bar\epsilon_{1\alpha} \epsilon_{2\beta}
\gamma^\mu_{\alpha\beta} P_\mu A~
~.
\eea

En comparant les \'equations (\ref{ssgen1}) et (\ref{ssgen2})
On voit donc que les g\'en\'erateurs 
des translations $P_\mu$ et des transformations \susic s $Q$
sont intimement connect\'es
\`a travers l'anticommutateur de ces derniers:

\beq
\label{ssgen}
\left\{ Q_\alpha,\bar Q_\beta \right\}~
=~
2 \gamma^\mu_{\alpha\beta} P_\mu ~
~.
\eeq

On peut ais\'ement
v\'erifier que cette importante relation est aussi valable
pour les autres champs $B$ et $\psi$.
De mani\`ere analogue
on peut aussi \'etablir le reste de l'alg\`ebre de la \susy,
que j'indique ici symboliquement:

\beq
\label{salg}
\mbox{\shortstack{super-alg\`ebre \\ de Poincar\'e}}~
\left\{~
\begin{array}{l}
\multicolumn{1}{l}{
\left.
\begin{array}{@{}l}
\mbox{}[P,P] ~=~ 0 \\
\mbox{}[L,L] ~\propto~ L \\
\mbox{}[P,P] ~\propto~ P
\end{array}~
\right\}~
\mbox{\shortstack{alg\`ebre de Lie \\ de Poincar\'e}}} \\
\mbox{}[P,Q] ~=~ 0 \\
\mbox{}[P,\bar Q] ~=~ 0 \\
\mbox{}[L,Q] ~\propto~ Q \\
\mbox{}[L,\bar Q] ~\propto~ \bar Q \\
\{Q,Q\} ~=~ 0 \\
\{\bar Q,\bar Q\} ~=~ 0 \\
\{Q,\bar Q\} ~\propto~ P
\end{array}
\right.
\eeq

o\`u $L$ repr\'esente les transformations de Lorentz,
\ie\ les rotations d'espace-temps.
Pour simplifier je n'ai pas repr\'esent\'e 
les indices vectoriels et spinoriels
dans les \'equations (\ref{salg}).

L'alg\`ebre (\ref{salg}) est dite {\em \susic} ou {\em gradu\'ee}
en raison de la pr\'esence des anticommutateurs.
Elle contient comme sous-alg\`ebre
la famili\`ere alg\`ebre de Lie
du groupe de Poincar\'e $SO(1,3)$,
qui connecte les translations et rotations spacio-temporelles.
Il est important de remarquer que,
bien que j'ai d\'eduit l'alg\`ebre (\ref{salg})
\`a partir du mod\`ele de Wess-Zumino,
elle ne se restreint pas seulement \`a celui-ci.
Il s'agit de l'alg\`ebre g\'en\'erale de la \susy.

Les \irreps\ de l'alg\`ebre de Poincar\'e
ont toutes un spin bien d\'efini,
$S=0$ pour les singulets scalaires,
$S=1/2$ pour les doublets spineurs,
$S=1$ pour les quadruplets vecteurs,
{\em etc.}.
Par contre,
les \irreps\ de la super-alg\`ebre de Poincar\'e (\ref{salg})
sont des {\em super-multiplets}
qui contiennent des composantes de diff\'erents spins.
Ainsi,
les trois champs $A$, $B$ et $\psi$ 
du lagrangien de Wess-Zumino (\ref{wzf})
forment ce qu'on appelle le {\em super-multiplet chiral}.

N\'ecessairement tout super-multiplet
doit avoir autant de \dofs\ bosoniques que fermioniques,
afin que l'on puisse passer des uns aux autres
par le g\'en\'erateur \susic\ $Q$.
Ainsi la \susy\ offre une explication naturelle de l'existence
simultan\'ee de bosons et de fermions.

\co
Les op\'erateurs \susic s $Q$
g\'en\`erent des translations supl\'ementaires
du super-espace-temps,
qui n'est rien d'autre que l'espace-temps de Minkowski
auquel ont \'et\'e rajout\'e deux dimensions de Grassman,
\ie\ qui ob\'eissent \`a une statistique fermionique.
\oc

\co
On peut en fait rajouter plus que deux dimensions de Grassman.
Cela a pour effet de n'ecessiter des g\'en\'erateurs \susic s additionels
et d'\'elargir l'alg\`ebre \susic\ (\ref{salg}) en cons\'equence.
On parle alors de \susy\ $N=2$, $N=4$ ou $N=8$.
Mais aux basses \'energies
seule la \susy\ $N=1$ que j'ai d\'ecrite ici
\`a l'aide du mod\`ele de Wess-Zumino
est ph\'enom\'enologiquement viable.
\oc

\subsection{La renormalisation des masses scalaires}

Nous avions remarqu\'e au chapitre pr\'ec\'edant,
que la pr\'esence de divergences quadratiques
dans l'\'energie propre des scalaires
est un probl\`eme majeur
auquel sont confront\'ees toutes les \guts.
La \susy\ fournit une solution tr\`es \'el\'egante \`a ce probl\`eme:
puisque les fermions n'ont pas de divervence quadratique
et puisqu'une transformation \susic\
permet de les transformer \`a loisir en scalaires
et {\em vice versa},
il doit y avoir un m\'ecanisme 
dans toute th\'eorie \susic\
qui annule aussi les divergences quadratiques des scalaires.
L'existence de ce m\'ecanisme est garanti
par un th\'eor\`eme fondamental de la \susy,
appel\'e {\em th\'eor\`eme de non-renormalisation}.

Ceci r\'esoud du moins la partie technique du probl\`eme de la hi\'erarchie,
\ie\ qu'une fois que les param\`etres du potentiel des scalaires
sont ajust\'es de mani\`ere \`a reproduire 
la hi\'erarchie des \vevs\ (\ref{hier}),
celle-ci se maintient naturellement 
\`a tous les ordres en th\'eorie de perturbations.
La \susy\ a donc pour vertu de stabiliser les corrections radiatives.
Notons bien,
toutefois,
que le probl\`eme fondamental de la hi\'erarchie
n'en est pas r\'esolu pour autant,
car aucune explication n'est fournie 
quant \`a la raison pour laquelle
les deux \'echelles de masses sont aussi diff\'entes.

\ex
On peut ais\'ement v\'erifier explicitement
l'annulation des divervences quadratiques
dans le cadre du mod\`ele de Wess-Zumino.
Le lagrangien (\ref{wzf})
ne correspond en fait 
qu'\`a la partie libre du lagrangien.
Il s'av\`ere que l'on peut rajouter 
des termes de masses et d'interactions
sans pour autant briser la \susy.
Je me suis abstenu de les inclure pr\'ec\'edemment,
afin de ne pas compliquer inutilement les calculs.
Si vous n'\^etes pas pr\`ets \`a un {\em acte de foi},
il n'est pas trop difficile de v\'erifier
que les transformations (\ref{susyf1}--\ref{susyf3})
laissent aussi invariant
(\`a une d\'eriv\'ee totale pr\`es)
le lagrangien complet de Wess-Zumino suivant:

\bea
\label{wz}
{\cal L}_{WZ}
&=&
~~~ {1\over2} \left(\partial^\mu A\right)^2
~+~ {1\over2} \left(\partial^\mu B\right)^2
~+~ {i\over2} \bar\psi \gamma^\mu\partial_\mu \psi
\\\nonumber
&&
 -~ {1\over2} m^2 A^2
~-~ {1\over2} m^2 B^2
~-~ {1\over2} m \bar\psi\psi
\\\nonumber
&&
 -~ {1\over\sqrt{2}} g \bar\psi (A-i\gamma_5B) \psi
~-~ {1\over\sqrt{2}} m g A(A^2+B^2)
~-~ {1\over4} g^2 (A^2+B^2)^2~
~.
\eea

Il est capital de remarquer
que les trois membres du super-multiplet ont exactement 
{\bf la m\^eme masse \boldmath$m$}
et que leurs interactions sont r\'egies par 
{\bf le m\^eme couplage \boldmath$g$}.
Il s'agit-l\`a d'une propri\'et\'e
que partagent tous les lagrangiens \susic s.

\begin{figure}[htb]
\unitlength.5mm
\SetScale{1.418}
\begin{boldmath}
\begin{center}
\hfill
\begin{picture}(40,30)(0,-10)
\DashLine(00,00)(10,00){1}
\DashLine(30,00)(40,00){1}
\DashCArc(20,00)(10,000,360){1}
\Text(-2,00)[r]{$A$}
\Text(42,00)[l]{$A$}
\Text(20,00)[c]{$A$}
\end{picture}
\hfill
\begin{picture}(40,30)(0,-10)
\DashLine(00,00)(10,00){1}
\DashLine(30,00)(40,00){1}
\DashCArc(20,00)(10,000,360){1}
\Text(-2,00)[r]{$A$}
\Text(42,00)[l]{$A$}
\Text(20,00)[c]{$B$}
\end{picture}
\hfill\mbox{}
\hfill
\begin{picture}(40,30)(0,0)
\DashLine(00,00)(40,00){1}
\DashCArc(20,10)(10,000,360){1}
\Text(-2,00)[r]{$A$}
\Text(42,00)[l]{$A$}
\Text(20,10)[c]{$A$}
\end{picture}
\hfill
\begin{picture}(40,30)(0,0)
\DashLine(00,00)(40,00){1}
\DashCArc(20,10)(10,000,360){1}
\Text(-2,00)[r]{$A$}
\Text(42,00)[l]{$A$}
\Text(20,10)[c]{$B$}
\end{picture}
\hfill
\begin{picture}(40,30)(0,-10)
\DashLine(00,00)(10,00){1}
\DashLine(30,00)(40,00){1}
\CArc(20,00)(10,000,360)
\Text(-2,00)[r]{$A$}
\Text(42,00)[l]{$A$}
\Text(20,00)[c]{$\psi$}
\end{picture}
\hfill\mbox{}
\end{center}
\end{boldmath}
\caption{
Diagrammes de Feynman \`a l'ordre le plus bas
intervenant dans l'\'energie propre
du scalaire $A$.
Les deux premiers diagrammes divergent logarithmiquement
et les trois deniers diagrammes divergent qua\-dra\-ti\-que\-ment.
}
\label{fse}
\end{figure}
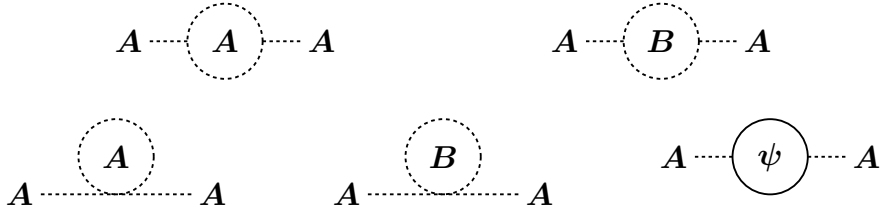

Les interactions du lagrangien de Wess-Zumino
(la troisi\`eme ligne de l'\'equation (\ref{wz}))
renormalisent les masses des champs
(la deuxi\`eme ligne de l'\'equation (\ref{wz})).
Les masses renormalis\'ees sont ainsi d\'etermin\'ees
par les \'energies propres $\Sigma$ (\ref{se}).
Calculons explicitement cette \'energie propre  
\`a une boucle
pour le scalaire $A$.
En vertu du lagrangien (\ref{wz})
elle est d\'etermin\'ee par les cinq graphes de Feynman
de la figure~\ref{fse},
qui donnent les contributions suivantes:

\bea
\unitlength.25mm
\SetScale{.709}
\Sigma\left(
\begin{picture}(40,20)(0,-5)
\DashLine(00,00)(10,00){1}
\DashLine(30,00)(40,00){1}
\DashCArc(20,00)(10,000,360){1}
\Text(20,00)[c]{$\scriptstyle A$}
\end{picture}
\right)
&=&
\left(-{mg\over\sqrt{2}}\right)^2~ (2!)^2 \int\! d^4p ~
{1 \over (p^2-m^2)} ~ {1 \over (p-k)^2-m^2}~
\\\nonumber\\
\unitlength.25mm
\SetScale{.709}
\Sigma\left(
\begin{picture}(40,20)(0,-5)
\DashLine(00,00)(10,00){1}
\DashLine(30,00)(40,00){1}
\DashCArc(20,00)(10,000,360){1}
\Text(20,00)[c]{$\scriptstyle B$}
\end{picture}
\right)
&=&
\left(-{mg\over\sqrt{2}}\right)^2~ 4! \int\! d^4p ~
{1 \over (p^2-m^2)} ~ {1 \over (p-k)^2-m^2}~
\\\nonumber\\
\unitlength.25mm
\SetScale{.709}
\Sigma\left(
\begin{picture}(40,20)(0,0)
\DashLine(00,00)(40,00){1}
\DashCArc(20,10)(10,000,360){1}
\Text(20,10)[c]{$\scriptstyle A$}
\end{picture}
\right)
&=&
-~ {g^2\over4}~ 4!~ \int\! d^4p ~{1 \over p^2-m^2}
\\\nonumber\\
\unitlength.25mm
\SetScale{.709}
\Sigma\left(
\begin{picture}(40,20)(0,0)
\DashLine(00,00)(40,00){1}
\DashCArc(20,10)(10,000,360){1}
\Text(20,10)[c]{$\scriptstyle B$}
\end{picture}
\right)
&=&
-~ {2g^2\over4}~ (2!)^2~ \int\! d^4p ~{1 \over p^2-m^2}
\\\nonumber\\
\unitlength.25mm
\SetScale{.709}
\Sigma\left(
\begin{picture}(40,20)(0,-5)
\DashLine(00,00)(10,00){1}
\DashLine(30,00)(40,00){1}
\CArc(20,00)(10,000,360)
\Text(20,00)[c]{$\scriptstyle \psi$}
\end{picture}
\right)
&=&
\Red{(-)}~ \left(-{g\over\sqrt{2}}\right)^2~ (2!)^2 \int\! d^4p ~
{\rm tr}\left[ 
~ {i \over p\hskip-.5em/-m} ~ {i \over p\hskip-.5em/-k\hskip-.5em/-m} ~
\right]
\\\nonumber
&=&
\Red{(-)}~ -2g^2~ \int\! d^4p ~
{{\rm tr}\left[ 
(p\hskip-.5em/+m) (p\hskip-.5em/-k\hskip-.5em/+m)
\right]
\over
(p^2-m^2) \left[(p-k)^2-m^2\right]}~
~,
\eea

o\`u $k$ est le moment constant 
qui entre dans la boucle.
Manifestement
les deux premiers diagrammes ne donnent lieu
qu'\`a des divergences logarithmiques.
Le signe $\Red{(-)}$ 
qui marque la contribution du fermion,
refl\`ete la statistique de Fermi-Dirac
\`a laquelle il est soumis.
La trace dans la derni\`ere \'equation donne

\bea
{\rm tr}\left[ 
(p\hskip-.5em/+m) (p\hskip-.5em/-k\hskip-.5em/+m)
\right]
&=&
4\left( p^2 -p\cdot k + m^2 \right)
\\\nonumber
&=&
2 \left\{ (p^2-m^2) + \left[(p-k)^2-m^2\right] - (k^2-4m^2) \right\}~
~,
\eea

ce qui permet d'\'ecrire 
la boucle fermionique sous la forme

\bea
\unitlength.25mm
\SetScale{.709}
\Sigma\left(
\begin{picture}(40,20)(0,-5)
\DashLine(00,00)(10,00){1}
\DashLine(30,00)(40,00){1}
\CArc(20,00)(10,000,360)
\Text(20,00)[c]{$\scriptstyle \psi$}
\end{picture}
\right)
~=~
\Red{(-)}~ -4g^2~ \int\! d^4p 
&&\bigg[~
{1 \over (p-k)^2-m^2}
~+~
{1 \over p^2-m^2}
\\\nonumber
&&\hskip2em
-~
{k^2-4m^2 \over (p^2-m^2) \left[(p-k)^2-m^2\right]}
~\bigg]~
~.
\eea

En effectuant le changement de variable
$p \to p+k$
pour le premier terme de l'int\'egrand,
on voit que ce dernier donne la m\^eme contribution que le deuxi\`eme terme.
On observe donc
dans la somme des contributions 
des deux boucles bosoniques et de la boucle fermionique
l'annulation de la divergence quadratique:

\beq
\label{qd}
\Sigma~
=~
\underbrace{(8-6-2)}_{\displaystyle0}~ 
g^2~ 
\underbrace{\int\limits_{\lambda}^{\Lambda} d^4p ~{1 \over p^2-m^2}}_{\displaystyle\sim\Lambda^2-\lambda^2}~
-~
2~ g^2~ (2k^2-15m^2)~ 
\underbrace{\int\limits_{\lambda}^{\Lambda} d^4p ~{1 \over p^4+\cdots}}_{\displaystyle\sim\ln{\Lambda\over\lambda}}~
~.
\eeq

Ainsi le facteur $Q$ du dernier terme de l'\'equation (\ref{rm})
est identiquement nul.
Il en r\'esulte
que la masse du scalaire n'\'evolue donc qu'au plus logarithmiquement
en fonction de l'\'energie,
ce qui permet
m\^eme en pr\'esence d'une profonde hi\'erarchie
d'impl\'ementer une solution {\em naturelle}
\`a l'\'equation (\ref{rm}).
\xe

\section{Le \sm\ \susic\ minimal}

La pr\'ediction fondamentale de la \susy\
est qu'\`a chaque \dof\ de spin $S$
correspond un \dof\ de spin $S+1/2$ ou $S-1/2$.
Or,
dans le cadre du \sm,
on compte $2 \times 3 \time 15 = 90$ \dofs\ fermioniques
(3 g\'en\'erations de 15 quarks et leptons, 
ainsi que leurs anti-particules)
pour seulement $2 \times ( 8 + 3 + 1 ) + 4 = 28$ \dofs\ bosoniques
(12 bosons de jauge \`a 2 projections de spin 
plus le doublet scalaire complexe).
Nous sommes donc loin du compte!

Manifestement,
s'il y a une \susy,
elle doit \^etre bris\'ee \`a basse \'energie.
La question de la brisure de la \susy\
est \`a l'heure actuelle  encore loin d'\^etre r\'esolue,
et,
bien qu'il s'agisse d'un sujet extr\^emement important,
je ne m'y attarderai pas ici.
Je me contenterai de noter que le concensus g\'en\'eral
est d'affirmer (esp\'erer?)
qu'un m\'ecanisme qui \`a lieu \`a des \'energies tr\`es \'elev\'ees
induit des interactions \`a basses \'energies
qui brisent explicitement (par de la magie noire?) la \susy.

Mais alors,
que reste-t-il de l'annulation des divergences quadratiques,
qui sont la raison principale 
de l'engouement pour la \susy?
Pas de danger:
uniquement des termes de {\em brisure molle}
sont rajout\'es au lagrangien \susic.
Ce qu'on veut dire par l\`a,
c'est que ces interactions {\em ad hoc}
sont choisies de mani\`ere \`a ne pas d\'etruire
la subtile annulation des divergences quadratiques.

Ces nouveaux termes mous
l\`event la d\'eg\'en\'erescence de masse 
des divers membres d'un m\^eme supermultiplet
et permettent ainsi aux partenaires \susic s
des particules observ\'ees,
d'aqu\'erir une masse suffisamment \'elev\'ee
pour expliquer leur absence dans les exp\'eriences courrantes.
Toutefois,
ces masses ne peuvent pas \^etre trop \'elev\'ees,
faute de quoi un spectre des divergences quadratiques r\'eapparaitrait,
avec toutes les cons\'equences d\'esastreuses qu'on imagine\dots\
En effet,
il est ais\'e de se convaincre
que l'effet de la lev\'ee de d\'eg\'en\'erescence 
est d'avoir pour l'\'energie propre (\ref{qd})

\beq
\label{q2dm}
\Sigma ~=~ g^2 (\tilde m^2 - m^2)~
~,
\eeq

o\`u $m$ repr\'esente l'\'echelle de masse des particules observ\'ees
et $\tilde m$ repr\'esente l'\'echelle de masse de leurs partenaires \susic s.
La valeur {\em naturelle} de l'\'energie propre 
\'etant de l'ordre de la \vev\ \EW\ 
$v\approx300$ GeV,
les masses des partenaires \susic s 
ne peuvent donc pas d\'epasser {\em grosso modo} 1 TeV.
Or ceci sera le domaine d'\'energie explor\'e par le LHC et le FLC,
ce qui explique en partie le grand int\'er\^et 
que porte la communaut\'e de la physique des particules \`a ces machimes.

On est en droit de se demander 
si certains des bosons et fermions que nous observons 
ne seraient pas d'ores et d\'ej\`a
des partenaires d'un m\^eme multiplet \susic.
H\'el\`as,
pour que des particules appartiennent au m\^eme supermultiplet,
il faut aussi n\'ecessairement 
qu'elles appartiennent 
\`a la m\^eme repr\'esentation du groupe de jauge\footnote{
En vertu d'un th\'eor\`eme fondamental de la \susy,
d\^u \`a Haag, \L opuzanski et Sohnius,
dont il r\'esulte que pour n'importe quelle th\'eorie relativiste des champs,
les g\'en\'erateurs de sym\'etries internes
%(ce qui inclu les sym\'etries de jauge)
sont des scalaires du super-groupe de Lorentz
et commutent donc avec les g\'en\'erateurs \susic s.
}.
Or les bosons de jauge appartiennent aux repr\'esentations adjointes
alors que les fermions appartiennent aux repr\'esentations fondamentales,
ce qui implique qu'ils ne peuvent pas \^etre des partenaires \susic s.
De m\^eme,
bien que le doublet de Higgs appartient \`a la m\^eme repr\'esentation \EW\
qu'un doublet de leptons gauches,
il ne porte pas de nombre leptonique
et ne peut donc pas non plus \^etre un partenaire \susic\ de leptons.

\co
Il existe toutefois un moyen de contourner 
ce dernier argument.
En effet,
il existe des interactions \susic s
qui ne respectent pas la conservation du nombre leptonique.
On ignore celles-ci de mani\`ere {\em ad hoc}
dans le cadre du \sm\ \susic\ minimal,
justement afin d'\'eviter l'apparation de processus
qui brisent le nombre leptonique.
Si de tels termes sont n\'eanmoins introduits dans le lagrangien,
on parle d'une th\'eorie brisant la parit\'e $R$,
et il n'y a plus de diff\'erence fondamentale 
entre les secteurs leptonique et scalaire.
Je reparlerai un peu plus loin de cette importante parit\'e $R$,
mais me confinerai au \sm\ \susic\ minimal,
qui respecte cette sym\'etrie discrette.
\oc

Il faudra donc chercher les partenaires \susic s 
des particules que nous connaissons
\`a plus haute \'energie.
Examinons d'abord cette faune \susic\
et \'etudions ensuite bri\`evement sa ph\'enom\'nologie.

\subsection{La faune \susic}

Le spectre des particules connues
et de leurs partenaires \susic s 
est montr\'e dans le tableau~\ref{tsusy}.
Je n'ai pas \'et\'e trop consistent dans ma nomenclature,
mais l'id\'ee g\'en\'erale est tr\`es simple:
les partenaires fermioniques des bosons recoivent le nom du boson
afubl\'e du sufixe {\em -ino},
alors que les partenaires bosoniques des fermions recoivent le nom du fermions
afubl\'e du pr\'efixe {\em s-}.

\setlength{\arraycolsep}{.5em}
\renewcommand{\arraystretch}{1.5}
\begin{table}[htb]
$$
\begin{array}{||c|c|l||c|c|l||}
\hline\hline
\mbox{symbole} & \mbox{spin} & \mbox{nom} & 
\mbox{symbole} & \mbox{spin} & \mbox{nom}
\\
\hline\hline
H_1 & && 
\tilde{H}_1 & & \\
H_2 & \raisebox{2ex}[-2ex]{$0$} & \raisebox{2ex}[-2ex]{Higgs} & 
\tilde{H}_2 & \raisebox{2ex}[-2ex]{$\displaystyle{1\over2}$} & \raisebox{2ex}[-2ex]{higgsino} \\
\hline\hline
\ell & &&
\tilde{\ell} & & \\
\nu & & \raisebox{2ex}[-2ex]{leptons} & 
\tilde{\nu} & & \raisebox{2ex}[-2ex]{sleptons} \\
\cline{1-1}\cline{3-4}\cline{6-6}
d & \raisebox{2ex}[-2ex]{$\displaystyle{1\over2}$} &&
\tilde{d} & \raisebox{2ex}[-2ex]{$0$} & \\
u & & \raisebox{2ex}[-2ex]{quarks} & 
\tilde{u} & & \raisebox{2ex}[-2ex]{squarks} \\
\hline\hline
\gamma & & \mbox{photon} &
\tilde{\gamma} & & \mbox{photino} \\
\cline{1-1}\cline{3-4}\cline{6-6}
Z & & & 
\tilde{Z} & &  \\
W^\pm & \raisebox{2ex}[-2ex]{$1$} & \sdt{\mbox{bosons}}{\mbox{de jauge}} & 
\tilde{W}^\pm & \raisebox{2ex}[-2ex]{$\displaystyle{1\over2}$} & \raisebox{2ex}[-2ex]{gauginos} \\
\cline{1-1}\cline{3-4}\cline{6-6}
g & & \mbox{gluon} &
\tilde{g} & & \mbox{gluino} \\
\hline\hline
\end{array}
$$
\caption{
Spectre des \'etats propres de jauge du \sm\
et de leurs partenaires \susic s.
}
\label{tsusy}
\end{table}
\renewcommand{\arraystretch}{1}

Vous n'aurez pas manqu\'e de remarquer
que j'ai introduit deux doublets de Higgs.
Ceci est indispensable,
car les higgsinos sont des fermions chiraux,
et il est donc n\'ecessaire d'avoir un deuxi\`eme doublet {\em miroir}
pour annuler les anomalies.
J'ai d\'ej\`a d\'ecrits pr\'ec\'edemment 
au chapitre \ref{thdm}
quelques caract\'eristiques
de ce secteur de Higgs.
Les contraintes impos\'ees par la \susy\
au potentiel de Higgs
sont assez restrictives,
de sorte que celui-ci reste n\'ecessairement invariant 
par transformation de $CP$
et seulement deux param\`etres
suffisent \`a le d\'eterminer
de mani\`ere unique,
au lieu de six
comme c'est le cas pour le mod\`ele g\'en\'eral 
\`a deux doublets de Higgs.

Il y a encore une autre subtilit\'e importante.
Le tableau~\ref{tsusy} indique les \'etats propres de jauge,
mais les termes mous de brisure de \susy\
provoquent un m\'elange higgsinos-gauginos
dans les \'etats propres de masse.
Ces derniers sont appel\'es des charginos et des neutralinos,
et sont donn\'es en fonctions des gauginos et higgsinos
par des transformations unitaires
du type

\bea
\tilde\chi^0_i &=& {a_i}\tilde\gamma + {b_i}\tilde Z + {c_i}\tilde h^0 + {d_i}\tilde H^0 
\qquad (i=1\dots4) 
\nonumber\\\nonumber\\
\label{mix}
\tilde\chi^+_j &=& {e_j}\tilde W^+ + {f_j}\tilde H^+
\qquad\qquad\qquad\quad (j=1,2)~
\\\nonumber\\\nonumber
\tilde\chi^-_j &=& {g_j}\tilde W^- + {h_j}\tilde H^-
\qquad\qquad\qquad\quad (j=1,2)~
~.
\eea

Les coefficients $a,\ b,\ c,\ d,\ e,\ f,\ g$ et $h$
de ces transformations unitaires
d\'ependent des param\`etres qui r\'egissent 
la \ssb\ \EW\ 
et la brisure molle de la \susy.

Le secteur scalaire des sleptons et des squarks
peut aussi impliquer des m\'elanges.
Si l'on fait intervenir tous les termes possibles
de brisure molle de \susy,
on compte en tout 
30 masses,
39 angles de m\'elange
et 41 phases complexes,
\ie\ 110 param\`etres!
Ceci est \'evidemment un d\'efaut majeur,
car la th\'eorie perd ainsi une grande portion 
de son potentiel pr\'edictif.

Une fa\c con tr\`es populaire d'\'echapper \`a cette impasse
est de faire appel \`a la \gu\ 
et d'ainsi contraindre la plupart de ces param\`etres
\`a avoir une valeur commune
\`a l'\'echelle de la \gu.
On parvient ainsi \`a r\'eduire 
le nombre total de param\`etres ind\'ependants
\`a 5!

\subsection{La ph\'enom\'enologie \susic}

Inutile de pr\'eciser 
qu'en pr\'edisant un plus-que-d\'edoublement
du nombre des particules 
observ\'ees jusqu'\`a ce jour,
on obtient une ph\'enom\'enologie riche.
Tr\`es riche!
Une ph\'enom\'enologie si riche,
que je me contenterai ici simplement
de l'illustrer par quelques exemples
et d'\'enoncer deux pr\'edictions fondamentales.
Pour obtenir une vision tr\`es inclusive
(voire presque exhaustive)
je recommande la lecture
(ou du moins le survol)
de l'imposante compilation de Haber et Kane~\cite{hk}.

\subsubsection{La masse du Higgs l\'eger}

En d\'epit du large nombre de param\`etres
dont d\'epend toute extension \susic\ du \sm,
minimale ou non,
il est une cons\'equence de la supersym\'etrisation 
du potentiel de Higgs
\`a laquelle il est pratiquement impossible d'\'echapper:
l'un des scalaires du secteur de Higgs doit \^etre l\'eger!

Si l'on ignore les corrections radiatives,
la masse de ce Higgs doit \^etre inf\'erieure \`a celle du $Z^0$.
Toutefois,
en vertu de la l\'eg\`ere lev\'ee de d\'eg\'en\'erescence 
des masses des fermions et sfermions,
un parfum des divergences quadratiques
dans la renormalisation des masses scalaires 
subsiste,
qui se manifeste principalement par une boucle de quarks top,
ceux qui couplent le plus fortement au Higgs.
Il en r\'esulte
que le carr\'e de la masse du Higgs le plus l\'eger
est donn\'ee par

\beq
\label{shm}
m_h^2~
\simeq~
m_Z^2~\cos2\beta~
+~
{3g^2 \over 4\pi^2}~
{m_t^4 \over m_W^2}~
\ln{m_{\tilde t}^2 \over m_t^2}~
~,
\eeq

o\`u le premier terme,
qui correspond \`a l'approximation d'arbre,
est toujours inf\'erieur \`a la masse du $Z^0$.
Le second terme contient l'effet de la boucle de top
et s'annule bien si le top et le stop sont d\'eg\'en\'er\'es.
Par contre,
si l'on consid\`ere une masse du top de 1 TeV
(la limite sup\'erieure approximative 
pour les masses des particules \susic s,
que j'avais d\'eduite de l'\'equation (\ref{q2dm}))
cette correction radiative
s'\'el\`eve \`a presque $(120 \mbox{ GeV})^2$,
de sorte que 

\beq
\label{shmn}
m_h ~ \lsim ~ 140 \mbox{ GeV}~
~.
\eeq

Cette pr\'ediction s'av\`ere \^etre particuli\`erement robuste,
car elle ne d\'epend pratiquement pas 
des autres param\`etres de la \susy.
En outre,
elle n'est que faiblement corrig\'ee
par les effets de boucles d'ordre sup\'erieur,
en vertu du th\'eor\`eme de non-renormalisation.
On peut donc affirmer sans embage,
que la \susy\ pr\'edit
de mani\`ere cat\'egorique
l'existence d'un Higgs l\'eger.

\subsubsection[La parit\'e $R$ et le LSP]{La parit\'e \boldmath$R$ et le LSP}

La ph\'enom\'enologie du \sm\ \susic\ minimal
est stigmatis\'ee par une importante sym\'etrie discrette,
la parit\'e $R$.
Il s'agit d'un nombre quantique multiplicatif,
qui prend la valeur $R=+1$ pour les particules ordinaires
(celles que nous observons ``quotidiennement'')
et qui prend la valeur $R=-1$ pour leurs partenaires \susic s.
En l'absence de cette sym\'etrie,
il existe des termes \susic s 
que l'on peut rajouter au lagrangien
et qui brisent la conservation du nombre leptonique ou baryonique.
Or comme certains des processus
qui ne conservent pas ces nombres quantiques
(comme la d\'esint\'egration du proton, \eg)
sont soumis \`a des contraintes exp\'erimentales draconiennes,
on pr\'ef\`ere en g\'en\'eral proc\'eder 
\`a l'\'elimination pure et simple
de tous ces termes,
m\^eme ceux qui ne donnent pas lieu 
\`a des effets observables \`a l'heure actuelle.
Ceci est l'unique raison 
qui justifie l'introduction de la parit\'e $R$.

L'existence de cette sym\'etrie
d\'etermine de mani\`ere fondamentale
toute la ph\'enom\'enologie du \sm\ \susic.
En effet,
la conservation de ce nombre quantique
exige que toutes interactions ou r\'eactions
doivent n\'ecessairement impliquer un nombre pair de particules \susic s
(qui peut \^etre nul).
Ceci implique en particulier que
\begin{itemize}
\item
les seuils d'\'energie de production de particules \susic s 
sont \'elev\'es,
car elles ne peuvent \^etre produites que paires;
\item
les effets virtuels d\^us aux d'\'echanges de particules \susic s
sont faibles,
car elles ne peuvent intervenir qu'au travers de diagrammes de boucles;
\item
les produits de d\'esint\'egration de particules \susic s 
contiennent une autre particule \susic\
(ou un nombre impair),
de sorte qu'on a affaire \`a une cascade 
de d\'esint\'egrations successives 
qui se termine toujours par la production
de la particule \susic\ la plus l\'eg\`ere,
le {\em lightest supersymmetric particle} ou LSP.
\end{itemize}

\co
Toute brisure de parit\'e $R$ 
engendre n\'ecessairement des r\'eactions
qui ne conservent pas les nombres leptonique ou baryonique
et ne souffrent donc d'aucun bruit de fond du \sm.
En outre,
comme en l'absence de parit\'e $R$ 
les particules \susic s ne doivent pas n\'ecessairement 
\^etre produites en paires,
leur seuil de production est diminu\'e de moiti\'e,
ce qui permettrait leur observation \`a des \'energies plus basses.
Similairement,
les effets virtuels d'\'echanges de particules \susic s
sont plus importants,
car ils peuvent avoir lieu 
\`a travers des diagrammes d'arbre.
On peut donc affirmer
qu'en d\'epit de son caract\`ere tr\'es {\em ad hoc},
la parit\'e $R$ correspond au cas le plus conservatif,
\ie\ celui o\`u il est le plus difficile 
de d\'etecter la \susy.
\oc

\`A cause de la parit\'e $R$,
le LSP est stable.
Comme il aura en outre \'et\'e produit abondemment 
peu apr\`es le big bang,
il ne peut s'agir que d'une particule neutre
qui interagit faiblement,
faute de quoi nous l'aurions ais\'ement d\'ej\`a d\'etect\'ee autour de nous.
En fait,
le LSP est un candidat id\'eal
pour la mati\`ere sombre!

Les uniques particules \susic s neutres
qui entrent en ligne de compte,
sont un sneutrino $\tilde\nu$
ou le neutralino le plus l\'eger $\tilde\chi^0_1$.
C'est ce dernier qui s'av\`ere \^etre le LSP
dans le cadre des \guts\ \susic s
et c'est aussi l'hypoth\`ese de travail utilis\'ee
dans la vaste majorit\'e des \'etudes ph\'enom\'enologiques
consacr\'ees \`a la \susy.

\co
Il \'etait de coutume anciennement
d'approximer le LSP par un photino $\tilde\gamma$
afin de simplifier les analyses.
On ne peut gu\`ere parler dans ce cas d'une approximation,
car si le neutralino le plus l\'eger 
s'av\`ere \^etre un higgsino $\tilde h$,
ses propri\'et\'es sont totalement diff\'erentes
de celles d'un photino.
De nos jours 
il n'est simplement plus acceptable 
d'effectuer cette simplification.
\oc

Comme le LSP n'interagit que faiblement,
\`a l'instar d'un neutrino
il \'echaperra \`a la d\'etection
en emportant une importante quantit\'e d'\'energie.
Typiquement,
la signature pour la production de particules \susic s
consiste donc en un \'ev\`enement
\`a \'energie ou \'energie transverse manquante.
Par exemple,
le premier graphe de Feynman de la figure~\ref{fsreac}
d\'ecrit la production et d\'esint\'egration d'un squark
par fusion d'un quark et d'un gluon
dans des collisions hadroniques.
On d\'enombre bien dans l'\'etat final
un nombre pair de LSP
qui s'\'echappent du d\'etecteur
sans laisser de traces,
ainsi qu'un jet issu du quark.
Il s'agit-l\`a d'un de ces inf\^ames \'ev\`enements
de {\em monojets},
qu'on avait cru observer en 1985 au CERN
et qu'on avait faussement interpr\`et\'es
comme la d\'ecouverte de la \susy.

\begin{figure}[htb]
\unitlength.5mm
\SetScale{1.418}
\begin{boldmath}
\begin{center}
\begin{picture}(80,60)(0,0)
\ArrowLine(00,00)(20,20)
\Gluon(00,40)(20,20){2}{4.5}
\ArrowLine(20,20)(40,20)
\Line(40,20)(60,00)
\DashLine(40,20)(60,40){1}
\Line(60,40)(80,25)
\ArrowLine(60,40)(80,55)
\Text(-2,00)[r]{$q$}
\Text(-2,40)[r]{$g$}
\Text(62,00)[l]{$\tilde\chi^0_1$}
\Text(82,25)[l]{$\tilde\chi^0_1$}
\Text(82,55)[l]{$q$}
\Text(49,31)[rb]{$\tilde q$}
\end{picture}
\qquad\qquad\qquad\qquad
\begin{picture}(100,70)(0,0)
\ArrowLine(00,00)(20,20)
\ArrowLine(20,20)(00,40)
\Photon(20,20)(40,20){2}{4}
\ArrowLine(40,20)(60,00)
\ArrowLine(60,40)(40,20)
\Line(60,00)(80,-15)
\Photon(60,00)(80,10){2}{4.5}
\ArrowLine(80,30)(60,40)
\DashLine(60,40)(80,55){1}
\Line(80,55)(100,65)
\ArrowLine(80,55)(100,45)
\Text(-2,00)[r]{$e^-$}
\Text(-2,40)[r]{$e^+$}
\Text(102,45)[l]{$\nu$}
\Text(102,65)[l]{$\tilde\chi^0_1$}
\Text(82,30)[l]{$e^+$}
\Text(82,10)[l]{$W^-$}
\Text(82,-15)[l]{$\tilde\chi^0_1$}
\Text(69,48)[rb]{$\tilde\nu$}
\Text(49,31)[rb]{$\tilde\chi^+_1$}
\Text(49,09)[rt]{$\tilde\chi^-_1$}
\end{picture}
\end{center}
\end{boldmath}
\bigskip
\caption{
Diagrammes de Feynman 
d\'ecrivant des typiques m\'ecanismes de productions
de particules \susic s
au LHC
(fusion quark-gluon)
et au FLC
(annihilation \'electron-positron).
}
\label{fsreac}
\end{figure}
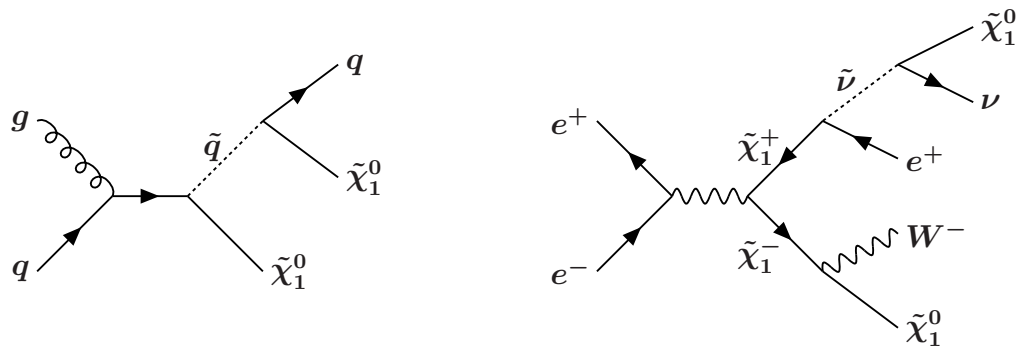

Le deuxi\`eme graphe de Feynman de la figure~\ref{fsreac}
repr\'esente l'annihilations d'\'electrons et positrons
en une paires de charginos,
et la d\'esint\'egration de ces derniers
\`a travers deux canaux diff\'erents.
En g\'en\'eral,
plus les particules \susic s produites sont lourdes,
plus leurs cascades de d\'esint\'egration peuvent \^etre complexes.

\section{La \gu\ \susic}

Puisque la \susy\ annule les divergences quadradratiques
qui apparaissent dans la renormalisation 
des masses des scalaires,
il est bien entendu tentant d'envisager une \gu\ \susic.
Il s'av\`ere que cette audace
est r\'ecompens\'ee au-del\`a de toute esp\'erance.
En effet,
non seulement la \gu\ $SU(5)$ peut-elle ainsi \^etre impl\'ement\'ee
sans plus aucune contradiction avec l'exp\'erience,
mais en plus
la \gu\ \susic\ fournit un m\'ecanisme dynamique 
de brisure de la sym\'etrie \EW.

Une autre r\'ecompense importante 
d\'ej\`a \'evoqu\'ee
de la \gu\ \susic,
consiste en la r\'eduction singuli\`ere
du nombre de param\`etres.
C'est peut-\^etre pour cette derni\`ere raison
(la plus mauvaise!)
que pratiquement toutes les analyses de \susy\
font automatiquement appel \`a la \gu,
souvent sans m\^eme la mentionner\dots

\subsection{L'unification des couplages}

Nous avions vu au chapitre pr\'ec\'edent
que la \gu\ $SU(5)$ sous sa forme la plus simple
n'est pas compatible avec la r\'ealit\'e exp\'erimentale.
Nous en \'etions arriv\'es \`a cette conclusion
en supposant l'unification des trois constantes de couplages
\`a tr\`es haute \'energie,
et en constatant que leur \'evolution vers les \'energies habituelles
donne lieu \`a un \wma\ trop diff\'erent
de celui qui a \'et\'e mesur\'e sur la r\'esonance $Z^0$.
En outre la dur\'ee de vie du proton ainsi d\'eduite
s'av\`ere \^etre trop courte.

Une autre fa\c con de voir les choses
est d'utiliser comme conditions aux limites
les mesures tr\`es pr\'ecises dont nous disposons \`a pr\'esent
\`a l'\'echelle d'\'energie de la masse du $Z^0$ 
et d'\'evoluer les constantes de couplage en sens inverse,
\ie\ vers les hautes \'energies.
C'est ce qui est repr\'esent\'e sur la figure~\ref{fgut}a,
o\`u l'on voit
qu'apr\`es la travers\'e du d\'esert
les trois couplages ne se rencontrent pas au m\^eme point 
comme l'exige la \gu.

\begin{figure}[htb]
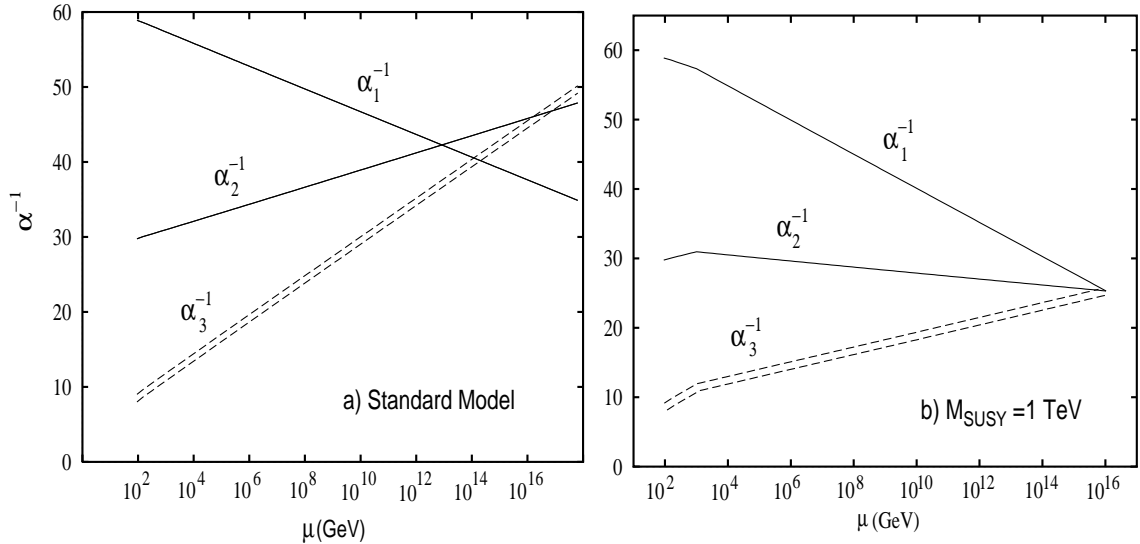

\unitlength1mm
\hspace{10mm}
\makebox(70,80)[bl]{\includegraphics{gutsu5.eps}}
\makebox(0,60)[bl]{\includegraphics{gutsusy.eps}}
%\vspace{5ex}
\caption{
\'Evolution des couplages $SU(3)_c$, $SU(2)_L$ et $U(1)_Y$
en fonction de l'\'energie
\`a partir des donn\'ees exp\'erimentales 
\`a la masse du $Z^0$.
La figure (a) correspond au cas 
o\`u seules les particules connues du \sm\
peuplent les r\'egions d'\'energies
en-de\c c\`a du d\'esert.
La figure (b) suppose en outre 
une population \susic\ aux environs de 1 TeV.
Ces figures sont adapt\'ees de la r\'ef\'erence~\protect\cite{lopez}.
}
\label{fgut}
\end{figure}

Par contre,
comme le montre la figure ~\ref{fgut}b,
l'unification a bien lieu
en pr\'esence d'un seuil \susic\
\`a environ 1 TeV.
Et qui plus est,
elle a lieu a une \'energie suffisamment \'elev\'ee
pour pr\'eclure l'observation de d\'esint\'egrations du proton
dans les exp\'eriences pr\'esentes.
En effet,
\`a cause de la pr\'esence d'un plus grand nombre 
de scalaires et de fermions
les fonctions beta s'accroissent,
d'o\`u il r\'esulte une \'evolution fortement ralentie du couplage fort 
et m\^eme une inversion de la direction
de l'\'evolution du couplage faible.

Si l'on introduit les valeurs de
$m_{\rm GUT} = 10^{16}$ GeV et $\alpha_{\rm GUT} = 1/25$
sugg\'er\'ees par la figures~\ref{fgut}b
dans la formule (\ref{plt})
on obtient un temps de vie du proton d'environ $10^{35}$ ans,
\`a peu pr\`es mille fois plus longue
que la limite impliqu\'ee jusqu'\`a pr\'esent
par l'absence d'observation de ce ph\'enom\`ene.

\subsection{La brisure radiative de la sym\'etrie \EW}

L'\'evolution des masses des scalaires 
pr\'esente aussi un int\'er\^et consid\'erable.
En effet,
au d\'epart d'une masse commune 
de tous les \'el\'ements d'un supermultiplet
\`a l'\'echelle de la \gu,
il s'av\`ere que justement pour l'un des Higgs
le carr\'e de sa masse \'evolue si rapidement vers le bas,
qu'il devient n\'egatif \`a une \'energie apr\'eciable.
Cet effet est montr\'e dans la figures~\ref{fewsb}.
\`A partir de cette \'energie
on ne peut plus traiter de ``masse''
le coefficient du terme quadratique dans le potentiel scalaire,
et ce dernier aqui\`er la typique forme en 
{\em sombr\'ero}.

\begin{figure}[htb]
\unitlength1mm
\hspace{10mm}
\makebox(70,120)[bl]{\includegraphics{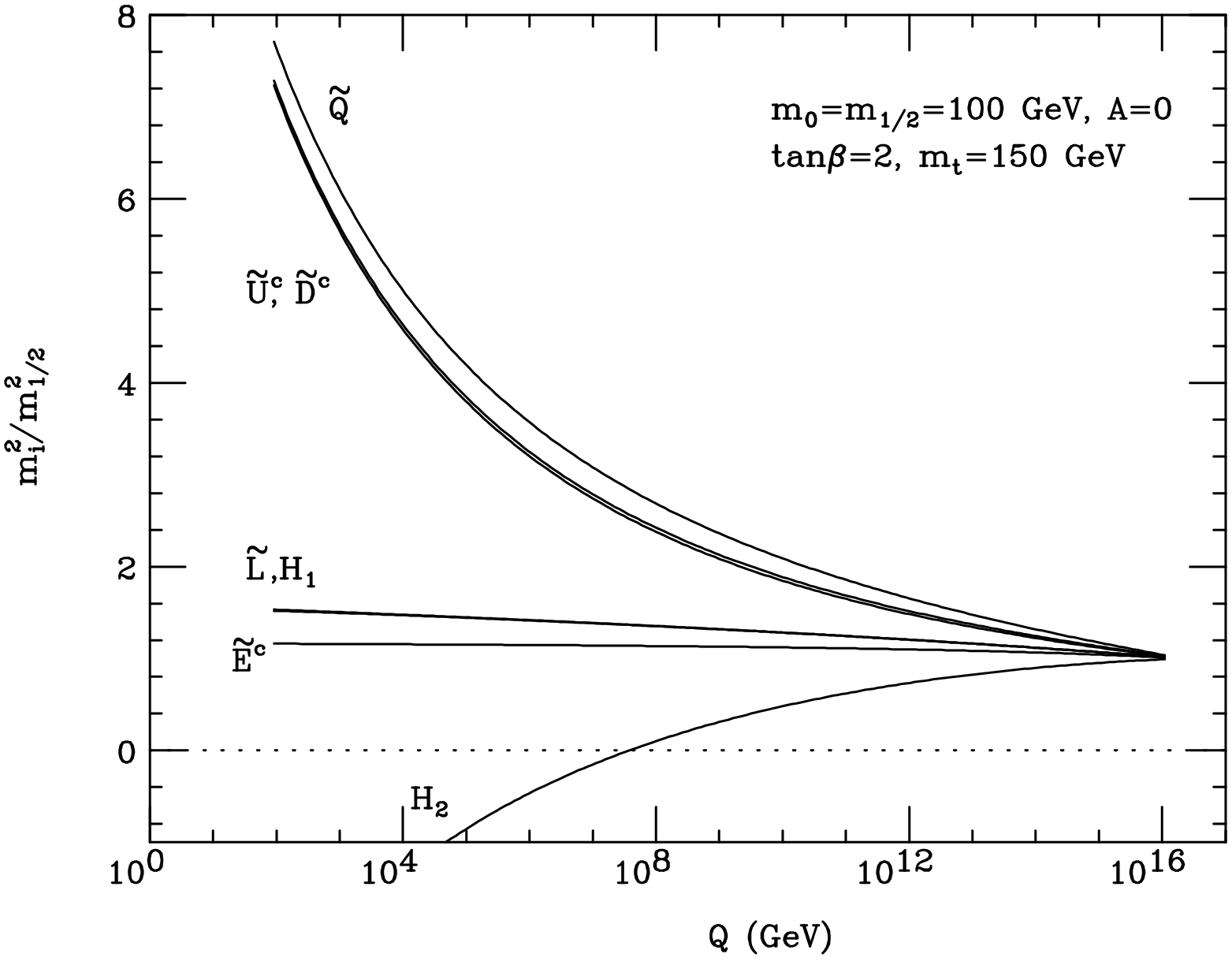}}
\bigskip
\caption{
\'Evolution des masses de divers scalaires
en fonction de l'\'energie.
L'ordonn\'ee repr\'esente le rapport 
des carr\'es des masses des scalaires 
aux carr\'es des masses de leurs partenaires fermioniques.
Cette figure est adapt\'ee de la r\'ef\'erence~\protect\cite{lopez}.
}
\label{fewsb}
\end{figure}

Comme la \ssb\ \EW\
a donc lieu dynamiquement,
on peut \`a premi\`ere vue
\^etre tent\'e de croire avoir ainsi r\'esolu 
le probl\`eme fondamental de la hi\'erarchie.
Il n'en est rien!
En effet,
l'existence d'une hi\'erarchie de jauge
est un ingr\'edient de base de ce r\'esultat,
car les valeurs num\'eriques repr\'esent\'ee 
dans la figures~\ref{fewsb}
d\'ependent tr\`es sensiblememnt 
des valeurs initiales 
utilis\'ees pour les masses des particules observ\'ees\footnote{
En particulier,
l'ingr\'edient majeur est la masse \'elev\'ee du quark top.
}.

\chapter{Ce dont je n'ai pas trait\'e}

Comme je l'avais d\'ej\`a indiqu\'e dans l'introduction,
les quatres th\`emes trait\'es dans ces notes,
\`a savoir 

\begin{itemize}
\item les extensions du secteur de Higgs,
\item la sym\'etrie gauche-droite,
\item la \gu,
\item la \susy,
\end{itemize}

ne forment qu'un sous-ensemble tr\`es restreint
des possibles extensions du \sm.
Bien que cette s\'election soit
tr\`es repr\'esentative
du {\em go\^ut} pr\'evalant \`a l'heure actuelle,
il reste n\'eanmoins important 
de ne pas perdre de vue
qu'il existe un grand nombre de th\'eories et mod\`eles alternatifs.
\bco{
Quoique,
ces derniers ne b\'en\'eficient pas ou plus 
de la faveur populaire,
mis \`a part pour certains {\em aficionados},
il est important de garder son esprit ouvert
\`a de telles \'eventualit\'es.
}\eco
J'en passe bri\`evement en revue
trois des plus importants,
\`a savoir 

\begin{itemize}
\item la compositivit\'e,
\item la technicouleur,
\item la th\'eorie de cordes,
\end{itemize}

en insistant toutefois
qu'il est loin de s'agir-l\`a d'une liste exhaustive.

\section{La compositivit\'e}

En un si\`ecle
nous avons d\'ecouvert
que la mati\`ere peut se d\'ecomposer en atomes,
les atomes en \'electrons et noyaux,
les noyaux en nucl\'eons
et les nucl\'eons en quarks.
Pourquoi n'y aurait-il pas une \'etape supl\'ementaire,
voire plusieurs?
Et aussi les leptons 
pourraient en principe \^etre des \'etats li\'es
de particules plus fondamentales.

Il existe plusieurs mod\`eles
qui invoquent de tels hypoth\'etiques constituants plus \'el\'ementaires,
que ceux que nous connaissons \`a l'heure actuelle,
appel\'es {\em pr\'eons}.
Ces derniers doivent interagir fortement entre eux
par une force dite d'{\em hypercouleur},
qui les confine
\`a l'int\'erieur des particules observ\'ees,
\`a l'instar des quarks 
confin\'es \`a l'int\'erieur des hadrons.

Bien que ces types de mod\`eles
peuvent fournir une explication \'el\'egante
des similarit\'es entre quarks et leptons
et leurs nombres quantiques,
il s'av\`ere ardu
d'\'etablir une base dynamique 
\`a l'origine de leurs confinement.
Toutefois,
cette difficult\'e peut parfaitement \^etre li\'ee 
\`a notre manque d'imagination
quand il s'agit de construire un nouveau m\'ecanisme d'interactions.

Toutefois,
m\^eme en ignorant les d\'etails des interactions entre p\'eons,
ces mod\`eles peuvent toujours \^etre param\'etris\'es
de mani\`ere effective
par leur \'echelle de compositivit\'e $\Lambda$,
\ie\ l'\'energie qui correspond \`a l'inverse du rayon de confinement.
Typiquement,
la compositivit\'e se manifesterait alors 
aux \'energies basses par rapport \`a $\Lambda$,
par des interactions de contacts,
qui permettraient d'observer, \eg,
des moments magn\'etiques anormaux,
des facteurs de forme dans les collisions leptoniques
ou des processus interdits dans le cadre du \sm.
Dans des collisions d'\'energies comparables \`a $\Lambda$,
on pourrait m\^eme observer 
des \'etats excit\'es lourds
des leptons et des quarks.

\section{La technicouleur}

Comme son nom l'indique,
la {\em technicouleur} est une th\'eorie 
essentiellement inspir\'ee de la QCD,
qui suppose l'existence d'une interaction de jauge $SU(N)_T$
dont l'\'echelle d'\'energie est de l'ordre du TeV.
Comme pour le pion dans le cas de la QCD,
la brisure de la sym\'etrie chirale de la technicouleur
implique l'existence de bosons de Goldstone,
les {\em technipions}.
Ces derniers renormalisent les propagateurs des bosons de jauge \EW s
et leur fournissent ainsi leur masses 
par resommation de Dyson.

La technicouleur est donc un m\'ecanisme de g\'en\'eration de masses
alternatif au m\'ecanisme de Higgs.
Il a en outre l'avantage d'\'eviter 
l'\'ecueil des divergences quadratiques.
Ce mod\`ele \'etait \`a sa conception tr\`es simple et \'el\'egant.
Malheureusement,
au fur et \`a mesure
que les contraintes exp\'erimentales
sont devenues plus contraignantes,
il a progressivement d\^u \^etre compliqu\'e,
jusqu'\`a se voir compl\`etement d\'enatur\'e.

N\'eanmoins,
bien qu'il semble \`a pr\'esent
que la technicouleur n'ai plus aucun avenir imm\'ediat,
son id\'ee de base reste tr\`es prometteuse.
Elle a inspir\'e bon nombre d'autres th\'eories de Higgs composites
et pourrait bien un jour resurgir de ses cendres
dans un autre contexte.

\section{La th\'eorie des cordes}

Le fondement de la th\'eorie quantique des champs
est la notion de particule.
On peut abandonner ce paradigme
\`a l'\'echelle de la distance de Planck,
pour consid\'erer des {\em cordes}.

On distingue divers types de cordes, 
entre autres
selon qu'elles sont ouvertes ou ferm\'es,
et
qu'elles n'ont que des \dofs\ bosoniques ou aussi fermioniques.
Dans ce dernier cas
(qui est \'evidemment le cas int\'eressant!)
on parle de {\em supercordes}.

Ces cordes ont diff\'erents modes de vibrations,
comme une corde de violon 
qui peut vibrer dans son mode fondamental 
ou des modes excit\'es.
Pour les cordes quantiques,
il y a beaucoup de diff\'erents modes fondamentaux.
Ils sont d\'eg\'en\'er\'es \`a l'\'energie nulle,
\ie\ qu'ils sont de masse nulle.
Il s'agit-l\`a de toutes les particules que nous observons,
et qui, 
vuent de l'\'echelle de Planck,
ont bien une masse quasiment nulle.\footnote{
Il est ais\'e de g\'en\'erer
par corrections radiatives
les faibles masses observ\'ees
de ces modes fondamentaux.
}
Les \'etats excit\'es de la corde
ont eux des masses 
qui s'accroissent en multiples de la masse de Planck
et ne jouent donc aucun r\^ole \`a nos \'energies.

La th\'eorie des cordes a le grand avantage esth\'etique
d'une \'etonante \'economie de principe:
au d\'epart d'une corde qui peut vibrer de diff\'erente mani\`eres,
on peut en principe reproduire le monde 
tel que nous le connaissons.
H\'el\`as, 
je dis bien ``en principe'',
car les supercordes ont un petit d\'efaut:
elles ne peuvent exister
que dans un espace \`a dix dimensions d'espace-temps!

En soit,
cela n'est pas tr\`es effrayant,
car on peut toujours argumenter\footnote{
Comme Kaluza et Klein au d\'ebut du si\`ecle,
dans le contexte d'une unification 
de la gravitation et de l'\'electromagn\'etisme.}
que six des dimensions spatiales sont recroquevill\'ees sur elles-m\^emes
sur une distances de l'ordre de la longueur de Planck.
Nous n'aurions donc aucune chance de jamais observer
ces dimensions directement.
Voil\`a qui parait bien artificiel \`a premi\`ere vue,
mais devient absolument fascinant
lorsqu'on se rend compte
qu'en se {\em compactifiant},
ces dimensions spatiales g\'en\`erent les interactions de jauge!

Le probl\`eme qui subsiste,
c'est que cette compactification peut s'effectuer
d'un nombre infini de mani\`eres diff\'erentes,
dont chacune d\'etermine 
des interactions
et spectres d'excitations fondamentales
(donc de particules)
diff\'erents.
On est donc loin du r\^eve platonicien
d'une solution unique,
la {\em theory of everything},
qui ber\c cait il y a quinze ans les th\'eoriciens des cordes.

De plus,
comme la structure m\^eme des cordes et de l'espace-temps
ne se manifeste qu'\`a l'\'energie de Planck,
il semble impossible de v\'erifier
l'hypoth\`ese des cordes de mani\`ere directe.
En fait,
il sera peut-\^etre m\^eme impossible
de jamais pouvoir la falsifier,
avec tout ce que cela entraine comme probl\`emes \'epist\'emologiques\dots

Il n'en reste pas moins
que la th\'eorie des supercordes
rec\`ele bon nombre de {\em miracles} math\'ematiques,
qui donnent \`a certains l'impression
que cette th\'eorie est 
{\em trop belle pour ne pas \^etre correcte}.
Ajoutons \`a cela,
qu'une pr\'ediction fondamentale de la th\'eorie des cordes
est l'existence d'exitations de masse nulle
et de spin 2,
des gravitons.
Il s'agit donc peut-\^etre
de la seule th\'eorie quantique de la gravitation
que nous ayons \`a notre disposition.

\end{document}